\documentclass[opre,nonblindrev]{informs3}

\DoubleSpacedXI 

\usepackage{endnotes}
\let\footnote=\endnote

\usepackage{natbib}
 \bibpunct[, ]{(}{)}{,}{a}{}{,}%
 \def\bibfont{\small}%
\usepackage{amsmath,graphicx,amssymb,mathrsfs,bm,bbm}
\usepackage{graphicx}
\usepackage{booktabs}
\usepackage{multirow}
\usepackage{mdwlist}
\usepackage{threeparttable}
\usepackage[top=1in, bottom=1in, left=1in, right=1in]{geometry}
\usepackage{CJK}
\usepackage{soul}
\usepackage[lined,boxruled]{algorithm2e}
\usepackage{colortbl}

\TheoremsNumberedThrough     
\ECRepeatTheorems

\EquationsNumberedThrough    

\begin{document}


\RUNAUTHOR{Li, Lam and Peng}

\RUNTITLE{Efficient Learning for Clustering and Optimizing Context-Dependent Designs}

\TITLE{Efficient Learning for Clustering and Optimizing Context-Dependent Designs}

\ARTICLEAUTHORS{%
\AUTHOR{Haidong Li}
\AFF{Department of Industrial Engineering and Management, College of Engineering, Peking University, Beijing 100871, China, \EMAIL{haidong.li@pku.edu.cn}} 
\AUTHOR{Henry Lam}
\AFF{Department of Industrial Engineering and Operations Research, Columbia University, NY 10027, USA, \EMAIL{henry.lam@columbia.edu}}
\AUTHOR{Yijie Peng}
\AFF{Department of Management Science and Information Systems, Guanghua School of Management, Peking University, Beijing 100871, China, \EMAIL{pengyijie@pku.edu.cn}}
} 

\ABSTRACT{%
We consider a simulation optimization problem for a context-dependent decision-making. A Gaussian mixture model is proposed to capture the performance clustering phenomena of context-dependent designs. Under a Bayesian framework, we develop a dynamic sampling policy to efficiently learn both the global information of each cluster and local information of each design for selecting the best designs in all contexts. The proposed sampling policy is proved to be consistent and achieve the asymptotically optimal sampling ratio. Numerical experiments show that the proposed sampling policy significantly improves the efficiency in context-dependent simulation optimization.
}%



\maketitle

%


\section{Introduction}
Simulation is a powerful tool for optimizing complex stochastic systems. We consider a simulation optimization problem of selecting the best design under different contexts. The mean performances of each design under each context are unknown and can only be estimated via simulation. The performance of each design depends on the contexts, and thus the best design is also context-dependent. For example, in movie recommendation~\citep{liu2009personalized}, movies and users can be regarded as designs and contexts, respectively.  
We aim to recommend the favorite movie for each user.  Other examples include patient-specific treatment regimen-making~\citep{kim2011battle} and automated asset management~\citep{faloon2017individualization}.

For any fixed context, we aim to find the best design among a finite set of alternatives, which is referred to as ranking and selection (R\&S) in the literature. R\&S procedures intelligently allocate simulation replications to efficiently learn the best design. The probability of correct selection (PCS) is used as a measure to evaluate the efficiency of sampling procedure in R\&S. In our problem, the best design is not universal but context-dependent. In this work, the goal is to correctly select all best designs in each context. Besides determining how to allocate simulation replications among different designs in a context, our sampling procedure also needs to consider simulation budget allocation among different contexts since incorrect selection in any context can lead to the failure of our goal. The worst-case notion is used to describe the context with the lowest PCS, and the worst-case probability of correct selection ($\text{PCS}_{\text{W}}$) under all contexts is used to measure the efficiency of sampling procedure in our problem.

In R\&S, the efficiency in sampling for learning the best design is a central issue, because simulation is usually  expensive and there could be a large number of possible designs. In our problem, the learning efficiency is more important because there could also exist a large number of possible contexts, and the number of all design-context pairs is a multiplication of the number of designs and the number of contexts. The complexity of the optimization problem can be substantially reduced if the performance cluster information in designs and contexts can be appropriately utilized. In the movie recommendation example, tremendous movies can be classified into a few categories such as drama, comedy, and action, and countless users can also be characterized by a relatively small number of attributes such as age, gender, and occupation. Users with a common attribute tend to favor movies in certain categories, e.g., arguably young people on average rate action movies higher than senior people, so the performance clustering phenomenon exists in the design-context pairs for the movie recommendation problem. The performance clustering  phenomena are rather common in combinatorial optimization problems with many application backgrounds including manufacturing and healthcare~\citep{peng2019efficient}. 

Accurately identifying the performance clusters would simplify the optimization problem, because designs in a cluster tend to have similar performances while designs in different clusters typically have significant differences in their performances, which provides useful global information for learning the best designs. However, the performances of the designs are estimated by random sampling in simulation, so how to efficiently learn the performance cluster information by sampling is an important issue for our problem. 

To capture the performance clustering phenomena, we use a Gaussian mixture model as the prior distribution for the performance of a design-context pair, and the hyper-parameters in the prior distribution are estimated from sampling information. Under a Bayesian framework, we formulate the sequential sampling decision as a stochastic dynamic programming problem, and provide an efficient scheme to update the posterior information for  each design-context pair. Moreover, we propose a dynamic sampling policy based on the sequentially updated posterior information to efficiently learn  both the global information of each cluster and local information of each design. The proposed sampling policy is proved to be consistent and achieve  the asymptotically optimal sampling ratio. The contribution of our work is threefold. 
\begin{itemize}
	\item We consider performance clustering in design-context pairs to enhance the efficiency for learning the best designs in all contexts.
	\item We provide an efficient scheme to simultaneously learn the global clustering information and local performance information in design-context pairs. 
	\item We propose an efficient dynamic sampling procedure for context-dependent simulation optimization, which is proved to be asymptotically optimal.
\end{itemize}

\subsection{Related Literature}
The R\&S literature consists of the frequentist and the Bayesian branches. See~\cite{kim2006selecting} and~\cite{chen2015ranking} for overviews. Frequentist procedures (e.g.,~\citealp{rinott1978two},~\citealp{kim2001fully},~\citealp{luo2015fully}) allocate simulation replications to guarantee a pre-specified PCS level, whereas Bayesian procedures (e.g.~\citealp{chen2000simulation},~\citealp{chick2012sequential},~\citealp{gao2017new}) aim to either maximize the PCS or minimize the expected opportunity cost subject to a given simulation budget. Bayesian procedures usually achieve better performance than frequentist procedures under a given simulation budget, but they typically do not provide a guaranteed PCS. \cite{peng2019efficient} offered an off-line learning scheme to extract clustering information from auxiliary information of the low-fidelity models in a classic R\&S setting, whereas our work proposes an on-line learning algorithm for simultaneously clustering and optimizing context-dependent designs.

The literature on context-dependent simulation optimization is sparse relative to the actively studied R\&S problem in simulation. Contexts are also known as the covariates, side information, or auxiliary quantities. To the best of our knowledge, the study of~\cite{shen2017ranking} is the first research for this problem. They assume a linear relationship between the response of a design and the contexts, and develop sampling procedures to provide a  guarantee on PCS  for all contexts. \cite{li2018data} further extend the result in~\cite{shen2017ranking} to high-dimensional contexts and general dependence between the mean performance of a design and the contexts. The aforementioned two studies adopt the Indifference Zone paradigm in the frequentist branch. \cite{gao2019selecting} adopted an optimal computing budget allocation (OCBA) approach in R\&S, and solve the problem by identifying the rate-optimal budget allocation rule. None of the existing work formulates the sequential sampling decision as a stochastic dynamic programming problem and considers the performance clustering in design-context pairs.

Our work is related to the literature on contextual multi-arm bandit (MAB) problem in machine learning. In the MAB problem, a fixed amount of samples are allocated to competing alternatives for maximizing their cumulative expected reward \citep{bubeck2012regret}. In our problem, similar to a pure-exploration version of MAB known as the best-arm identification problem, the reward only appears in the final stage for selecting the best designs. \cite{auer2000using} and \cite{hong2011evolutionary} assume a linear dependency between context and the expected reward of an action to provide an approximate solution for the contextual MAB problem. Nonlinear contextual reward functions approximated by nonparametric regression, random forest, and neural network can be found in \cite{rigollet2010nonparametric},~\cite{slivkins2014contextual},~\cite{perchet2013multi},~\cite{allesiardo2014neural},~\cite{feraud2016random}. \cite{han2020sequential} study
sequential batch learning in the adversarial contexts and linear rewards setting. Choosing contexts adversarially allows the decision maker learn knowledge about the rewards as few as possible, which can be considered as the worst case. However, relatively few studies exist on contextual best-arm identification. All analysis of best-arm identification in~\cite{soare2014best},~\cite{xu2018fully}, and~\cite{kazerouni2019best} assume a linear or generalized linear dependence of rewards on the
contexts. None of studies on MAB exploit clustering information in context-dependent designs.

Optimizing the worst-case performance over a range of scenarios when facing model ambiguity is a theme of robust optimization (RO); see~\cite{ben2009robust} and~\cite{bertsimas2011theory} for an introduction. In the simulation literature, the so-call robust simulation applies worst-case analysis on a simulation model when the input distribution is uncertain but postulated to lie within a set, see, e.g.,~\cite{hu2012robust},~\cite{glasserman2014robust},~\cite{Hu2015RobustSO},~\cite{lam2016robust},~\cite{lam2018sensitivity},~\cite{ghosh2019robust}. Focusing on the worst-case calculation, the decision variables in the resulting optimization in these works 
are the unknown input distributions. In contrast, our approach involves optimizing design variables on a criterion that uses the worst-case performance. This is closer to~\cite{hu2013kullback} and~\cite{fan2020distributionally} that consider decision-making over the worst-case scenario in simulation contexts. However, in these works, the performance of each design is context-dependent while the goal is to find a design with the best worst-case performance, whereas in our problem the best design is also context-dependent.

The rest of the paper is organized as follows. In Section~\ref{sec:PD}, we formulate the studied problem and introduce assumptions
of this research. Section~\ref{sec:PE} derives the posterior estimates of parameters in the Gaussian mixture model. In Section~\ref{sec:DSP}, we develop a dynamic sampling procedure. Section~\ref{sec:NE} presents numerical examples and computational results, and Section~\ref{sec:conclusion} concludes the paper and outlines future directions. The proofs of the theorems and propositions in the paper can be found in the online appendix. 

\section{Problem Description}\label{sec:PD}
\begin{figure}[htbp]
	\begin{center}
		\includegraphics[width=3.5in]{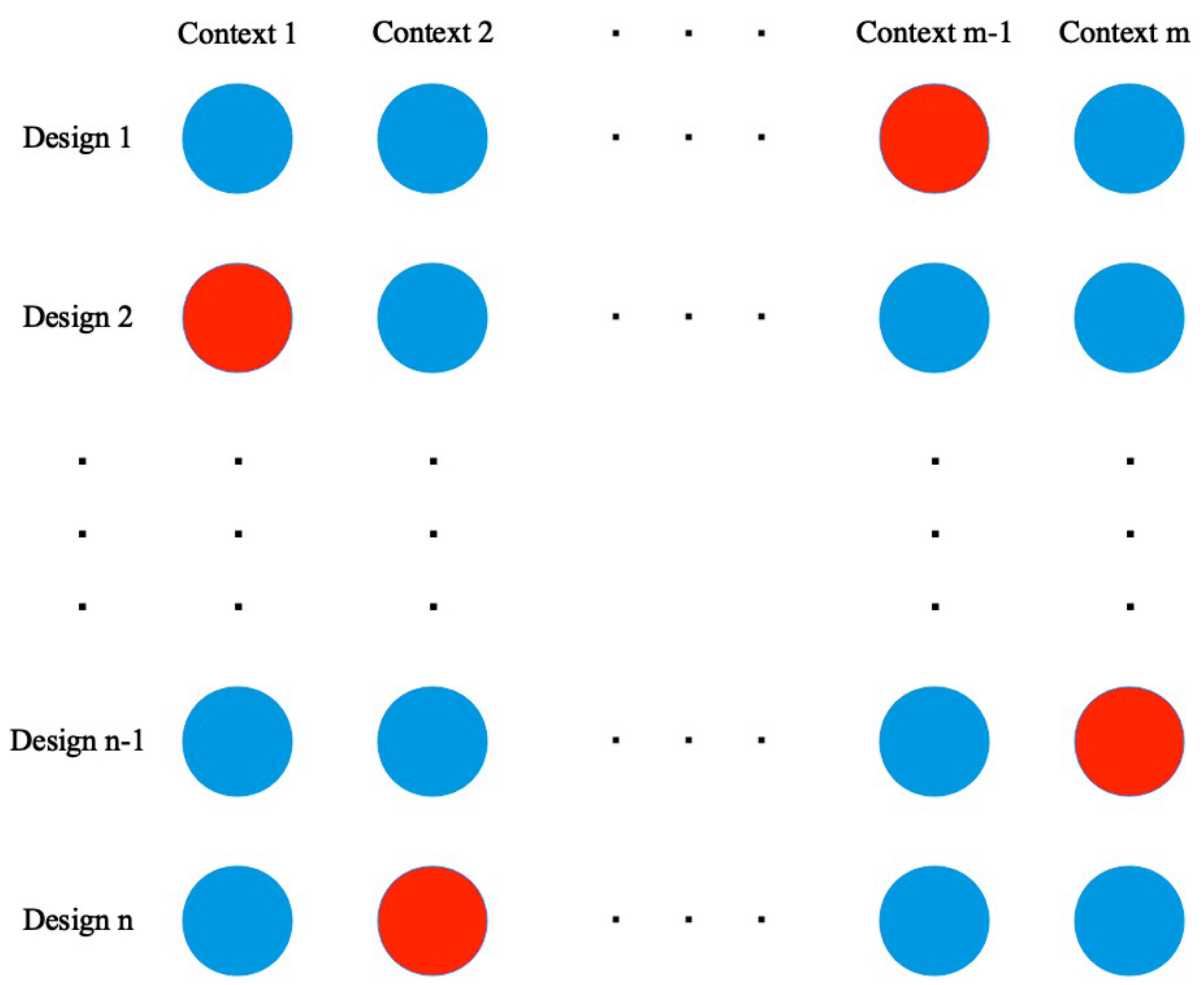}\\
		\caption{Selecting the best (red) design under each context.}\label{illustration_1}
	\end{center}
\end{figure}

Suppose there are $n$ different designs. For $i=1,\ldots,n$, the performance $y_{i}(\bm{x})$ of design $i$ depends on a vector of context $\bm{x}=(x_{1},\ldots,x_{d})^{\top}$ for $\bm{x}\in \mathcal{X}\subseteq\mathbb{R}^{d}$. The performances are unknown and can only be learned via sampling. In this study, we assume that $\mathcal{X}$ contains a finite number of $m$ possible contexts $\bm{x}_{1},\ldots,\bm{x}_{m}$. Our objective is to correctly select the best design for a given value of $\bm{x}$ (see Figure~\ref{illustration_1} for an illustration), i.e., identify ${\arg\max}_{i}\ y_{i}(\bm{x})$. For example, in personalized movie recommendation, we aim to recommend the most favorite movie (design) for the corresponding user (context). Since sampling could be expensive, the total number of samples is usually limited. Moreover, when either $n$ or $m$ is relatively large, it would be practically infeasible to estimate all performances accurately for each design $i$ and each context $\bm{x}_{j}$.

Under a fixed context $\bm{x}$, the quality of the selection for the best design is measured by the probability of correct selection (PCS),
$$\text{PCS}(\bm{x})=\mathbb{P}\left(\bigcap_{i\neq\widehat{i}^{*}(\bm{x})}\left(\widehat{y}_{i}(\bm{x})<\widehat{y}_{\widehat{i}^{*}(\bm{x})}(\bm{x})\right)\right),$$
where $\widehat{i}^{*}(\bm{x})$ is the estimated best design and $\widehat{y}_{i}(\bm{x})$ is the posterior performance for design $i$ and context $\bm{x}$. In this study, we aim to provide the best design for all the $\bm{x}$ that might possibly appear, and therefore need a measure for evaluating the quality of the selection over the entire context space $\mathcal{X}$. Specifically, we adopt the worst-case probability of correct selection over $\mathcal{X}$:
$$\text{PCS}_{\text{W}}=\underset{\bm{x}\in\mathcal{X}}{\min}\ \text{PCS}(\bm{x}).$$
This measure has been used in contextual R\&S~\citep{gao2019selecting}, and uses the worst-case notion in robust optimization~\citep{bertsimas2011theory} and R\&S with input uncertainty~\citep{gao2017robust}.

\subsection{Assumption}
We assume that for each design and context, the simulation observations are i.i.d. normally distributed, i.e., $Y_{i,t}(\bm{x})\sim N(y_{i}(\bm{x}),\sigma_{i}^{2}(\bm{x}))$, $i=1,\ldots,n$, $t\in\mathbb{Z}^{+}$, $\bm{x}\in \mathcal{X}$, and the replications among different designs and different contexts are independent. The variance $\sigma_{i}^{2}(\bm{x})$ in the sampling distribution is assumed to be known in this study and use the sample estimate as a plug-in for the true value in practice. We use $\phi(\cdot|\mu,\sigma^{2})$ to denote the density of a normal distribution with mean $\mu$ and variance $\sigma^{2}$. The normal assumption is the most common assumption in R\&S research.  
For non-normal sampling distributions, the normal assumption is justified by the use of batching \citep{kim2006selecting}. 

\begin{figure}[htbp]
	\begin{center}
		\includegraphics[width=3.8in]{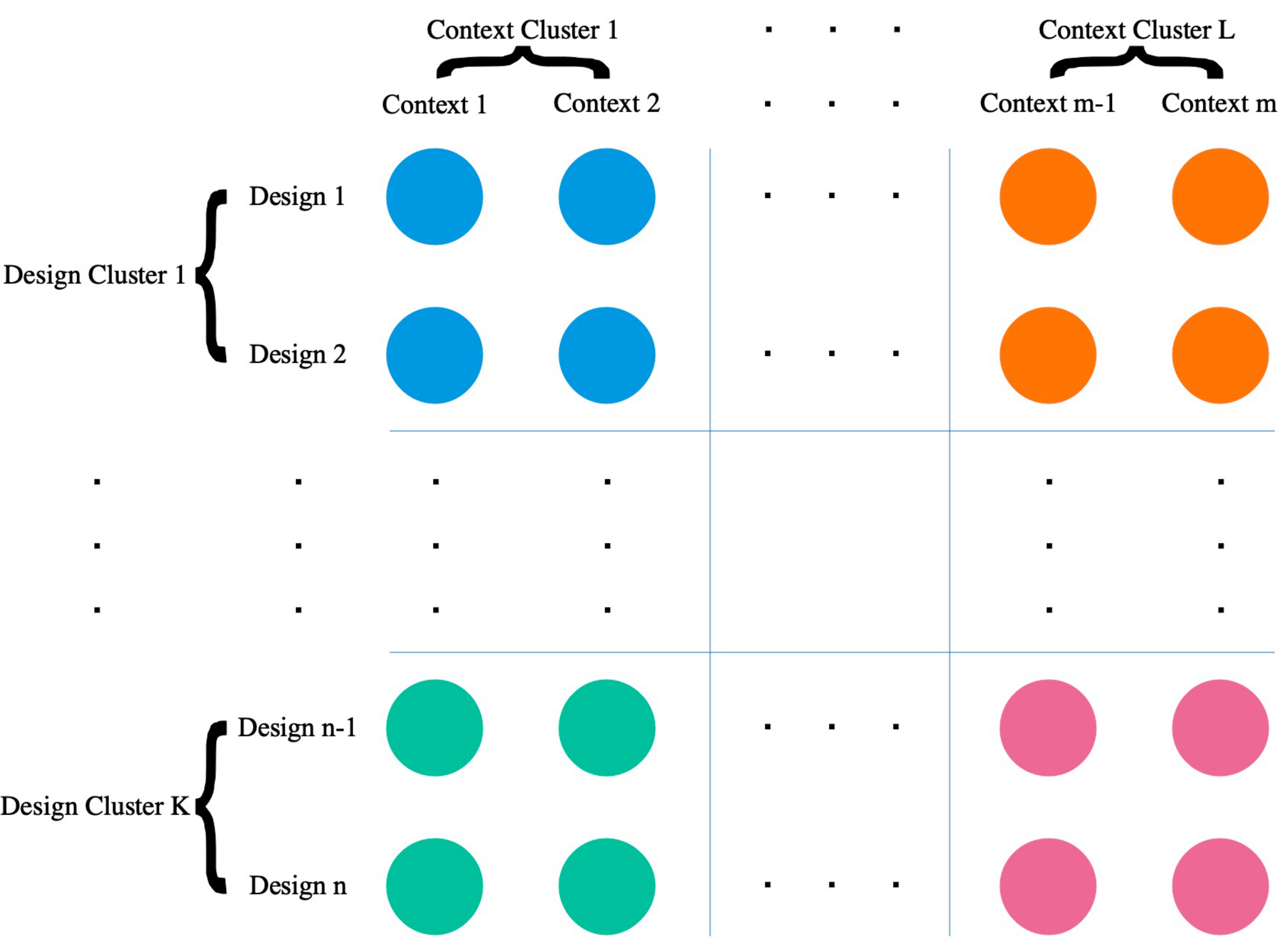}\\
		\caption{The performance clustering phenomenon in designs and contexts.}\label{illustration_2}
	\end{center}
\end{figure}
A Bayesian framework is introduced in learning the unknown performances of different designs under different contexts, and the prior distribution of $y_{i}(\bm{x})$ is assumed to be a Gaussian mixture distribution:
$$y_{i}(\bm{x})\sim \sum_{k=1}^{K}\sum_{\ell=1}^{L}\tau_{k}\omega_{\ell}\phi(\cdot|\mu_{k\ell},\sigma_{k\ell}^{2}),$$
where $K$ is the (unknown) number of clusters in design dimension, $L$ is the (unknown) number of clusters in context dimension, $\tau_{k}$ is the unknown (probability) weight of the $k$-th cluster in design dimension, $\omega_{\ell}$ is the unknown (probability) weight of the $\ell$-th cluster in context dimension, and $\phi(\cdot|\mu_{k\ell},\sigma_{k\ell}^{2})$ is the density of each mixture component. The Gaussian mixture model prior distribution reflects the performance clustering phenomenon in designs and contexts. For example, design performances are similar in each block as shown in Figure~\ref{illustration_2}, that is to say, the users in the same cluster have similar preferences on the movies in the same cluster. In cases without a clear performance clustering structure, our prior distribution would then have only one component and reduce to the widely used normal prior distribution in Bayesian R\&S literature. When there is only one context, we simply consider the performance clustering phenomenon in designs, and our prior distribution reduces to the special case in~\cite{peng2019efficient}.

\section{Parameter Estimation}\label{sec:PE}
To extract the information of performance clustering from observations, we introduce two hidden state random variables $z_{i}=(z_{i,1},\ldots,z_{i,K})$ and $v_{j}=(v_{j,1},\ldots,v_{j,L})$, which follow multinomial distributions:
\begin{gather*}
p(z_{i})=\prod_{k=1}^{K}\tau_{k}^{z_{i,k}},\\
p(v_{j})=\prod_{\ell=1}^{L}\omega_{\ell}^{v_{j,\ell}},\\
s.t.\ \sum_{k=1}^{K}z_{i,k}=1,\ z_{i,k}\in\{0,1\},\ k=1,\ldots,K,\\
\ \ \ \ \sum_{\ell=1}^{L}v_{j,\ell}=1,\ v_{j,\ell}\in\{0,1\},\ \ell=1,\ldots,L.
\end{gather*}
The hidden state random variable $z_{i,k}$ assigns design $i$ to cluster $k$ if $z_{i,k}=1$ while the hidden state random variable $v_{j,\ell}$ assigns context $j$ to cluster $\ell$ if $v_{j,\ell}=1$, which means $y_{i}(\bm{x}_{j})$ comes from a realization of distribution $\phi(\cdot|\mu_{k\ell},\sigma_{k\ell}^{2})$.

Let $t_{ij}$ be the number of simulation replications allocated to design $i$ in context $j$ after allocating a total amount of $t$ simulation replications. To obtain the posterior statistics of the unobservable performance $\bm{y}=[y_{i}(\bm{x}_{j})]_{n\times m}$, we introduce the likelihoods of observations $\mathcal{E}_{t}=\{Y_{i,h}(\bm{x}_{j})\}_{i=1,j=1,h=1}^{n,m,t_{ij}}$ and unobservable $Z=[z_{i,k}]_{n\times K}$ and $V=[v_{j,\ell}]_{m\times L}$. Given parameter $\theta=\{\tau_{k},\omega_{\ell},\mu_{k\ell},\sigma_{k\ell}\}_{k=1,\ell=1}^{K,L}$,  $$\mathcal{L}(Z;\theta)=\prod_{i=1}^{n}p(z_{i})=\prod_{i=1}^{n}\prod_{k=1}^{K}\tau_{k}^{z_{i,k}}$$ and $$\mathcal{L}(V;\theta)=\prod_{j=1}^{m}p(v_{j})=\prod_{j=1}^{m}\prod_{\ell=1}^{L}\omega_{\ell}^{v_{j,\ell}}$$ are the likelihood of design clustering state variable $Z$ and context clustering state variable $V$, respectively, the likelihood of true performance $\bm{y}$ given the clustering state variables $Z$ and $V$ is $$\mathcal{L}(\bm{y}|Z,V;\theta)=\prod_{i=1}^{n}\prod_{j=1}^{m}\left(\prod_{k=1}^{K}\prod_{\ell=1}^{L}\left(\phi\left(y_{i}(\bm{x}_{j})|\mu_{k\ell},\sigma_{k\ell}^{2}\right)\right)^{z_{i,k}v_{j,\ell}}\right),$$
and the likelihood of samples given the true performance $\bm{y}$  is 
\begin{eqnarray*}
\mathcal{L}(\mathcal{E}_{t}|\bm{y};\theta)=\prod_{i=1}^{n}\prod_{j=1}^{m}\left(\prod_{h=1}^{t_{ij}}\phi\left(Y_{i,h}(\bm{x}_{j})|y_{i}(\bm{x}_{j}),\sigma_{i}^{2}(\bm{x}_{j})\right)\right).
\end{eqnarray*}
The likelihood of complete state variables, i.e., $\mathcal{E}_{t}$, $\bm{y}$, $Z$, and $V$, is
\begin{eqnarray*}
\mathcal{L}(\mathcal{E}_{t},\bm{y},Z,V;\theta)=\mathcal{L}(Z;\theta)\mathcal{L}(V;\theta)\mathcal{L}(\bm{y}|Z,V;\theta)\mathcal{L}(\mathcal{E}_{t}|\bm{y};\theta).
\end{eqnarray*}
The likelihood of observations $\mathcal{E}_{t}$ is obtained by integrating out the unobservable state $\bm{y}$, $Z$, and $V$, which is given by
\begin{eqnarray*}
\mathcal{L}(\mathcal{E}_{t};\theta)=\sum_{k_{1:n}\in\mathcal{K}}\sum_{\ell_{1:m}\in\mathcal{L}}f_{\tau}(k_{1:n})f_{\omega}(\ell_{1:m})f_{Y}(\mathcal{E}_{t}|k_{1:n},\ell_{1:m}),
\end{eqnarray*}
where $k_{1:n}$ is the design clustering index set such that $z_{i,k_{i}}=1,\ k_{i}\in\mathcal{K}\triangleq\{1,\ldots,K\}$, $\ell_{1:m}$ is the context clustering index set such that $v_{j,\ell_{j}}=1,\ \ell_{j}\in\mathcal{L}\triangleq\{1,\ldots,L\}$, the probability of design clustering situation~$k_{1:n}$ is
$$f_{\tau}(k_{1:n})\triangleq\prod_{i=1}^{n}\tau_{k_{i}},$$
the probability of context clustering situation~$\ell_{1:m}$ is
$$f_{\omega}(\ell_{1:m})\triangleq\prod_{j=1}^{m}\omega_{\ell_{j}},$$ 
and the probability density of samples given $k_{1:n}$ and $\ell_{1:m}$ is
\begin{equation*}
f_{Y}(\mathcal{E}_{t}|k_{1:n},\ell_{1:m})\triangleq\prod_{i=1}^{n}\prod_{j=1}^{m}C_{ij,k_{i}\ell_{j}},
\end{equation*}
with
\begin{align*}
C_{ij,k_{i}\ell_{j}}\triangleq \int_{\mathbb{R}}\phi\left(y_{i}(\bm{x}_{j})|\mu_{k_{i}\ell_{j}},\sigma_{k_{i}\ell_{j}}^{2}\right)\prod_{h=1}^{t_{ij}}\phi\left(Y_{i,h}(\bm{x}_{j})|y_{i}(\bm{x_{j}}),\sigma_{i}^{2}(\bm{x_{j}})\right)dy_{i}(\bm{x}_{j}).
\end{align*}

\subsection{Number of Clusters}
First, we need to determine numbers of clusters ($K$ and $L$) in the mixture model. Specifically, $K$ and $L$ can be determined by using the Bayesian information criterion (BIC):
\begin{gather}
(\widehat{K},\widehat{L})=\underset{K,L}{\arg\max}\ \left[2\log\mathcal{L}(\mathcal{E}_{t};\widehat{\theta}_{K,L})-(2KL+K+L)\log (nm)\right],\label{eq.BIC}
\end{gather}
where
$$\widehat{\theta}_{K,L}=\underset{\theta}{\arg\max}\ \mathcal{L}(\mathcal{E}_{t};\theta).$$
BIC includes a penalty term $(2KL+K+L)\log (nm)$ for the number of estimated parameters in the model to discount the log-likelihood which captures the statistical fitness. It is always possible to improve the statistical fitness with the data by choosing a more complex model with more parameters, but increased complexity in modeling may result in overfitting. For the Gaussian mixture model, the EM algorithm is one of the most popular methods to efficiently compute the MLE. Given an arbitrary initial value $\theta^{(0)}$, the EM algorithm iteratively executes the following two steps:
\begin{itemize}
\item Expectation step (E-step): given $\mathcal{E}_{t}$ under the current parameter estimate $\theta^{(s)}$, calculate 
\begin{gather}
\mathcal{Q}(\theta|\theta^{(s)})\triangleq\mathbb{E}\left[\log\mathcal{L}(\mathcal{E}_{t},\bm{y},Z,V;\theta)|\mathcal{E}_{t},\theta^{(s)}\right];\label{eq.E-step}
\end{gather}
\item Maximization step (M-step): maximize function $\mathcal{Q}(\cdot|\theta^{(s)})$ to update the parameter estimate 
\begin{gather}
\theta^{(s+1)}=\underset{\theta}{\arg\max} \mathcal{Q}(\theta|\theta^{(s)}).\label{eq.M-step}
\end{gather}
\end{itemize}

\subsection{Posterior Estimates}
We provide a theorem for updating the clustering statistics and the posterior parameter estimates based on the EM algorithm. To start with, we provide a list of the notations. 
\begin{basedescript}{\desclabelstyle{\pushlabel}\desclabelwidth{8em}}
\item[$s$] the iteration number of the EM algorithm;
\item[$\widehat{\theta}^{(t,s)}$] the parameter estimates in the $s$-th iteration of the EM algorithm conditional on $\mathcal{E}_{t}$, which includes elements  $\widehat{\tau}_{k}^{(t,s)}$, $\widehat{\omega}_{\ell}^{(t,s)}$,  $\widehat{\mu}_{k\ell}^{(t,s)}$, and $(\widehat{\sigma}_{k\ell}^{2})^{(t,s)}$;
\item[$\mu_{ij,k\ell}^{(t,s)}$] the posterior mean of $y_{i}(\bm{x}_{j})$ conditional on $\{z_{i,k}=1\}$, $\{v_{j,\ell}=1\}$, $\mathcal{E}_{t}$, and given $\widehat{\theta}^{(t,s)}$;
\item[$(\sigma_{ij,k\ell}^{2})^{(t,s)}$] the posterior variance of $y_{i}(\bm{x}_{j})$ conditional on $\{z_{i,k}=1\}$, $\{v_{j,\ell}=1\}$, $\mathcal{E}_{t}$, and given $\widehat{\theta}^{(t,s)}$.
\end{basedescript}

\begin{theorem}~\label{thm_EM}
The posterior distribution of $y_{i}(\bm{x}_{j})$ conditional on $\{z_{i,k}=1\}$, $\{v_{j,\ell}=1\}$, $\mathcal{E}_{t}$, and given $\widehat{\theta}^{(t,s)}$ is
$$\phi(y_{i}(\bm{x}_{j})|\mu_{ij,k\ell}^{(t,s)},(\sigma_{ij,k\ell}^{2})^{(t,s)}),$$
where 
\begin{gather}
(\sigma_{ij,k_{i}\ell_{j}}^{2})^{(t,s)}=1\Big/\left[\frac{t_{ij}}{\sigma_{i}^{2}(\bm{x}_{j})}+\frac{1}{(\widehat{\sigma}_{k_{i}\ell_{j}}^{2})^{(t,s)}}\right],\label{thm1.1}
\end{gather}
\begin{gather}
\mu_{ij,k_{i}\ell_{j}}^{(t,s)}=(\sigma_{ij,k_{i}\ell_{j}}^{2})^{(t,s)}\left[\frac{\sum_{h=1}^{t_{ij}}Y_{i,h}(\bm{x}_{j})}{\sigma_{i}^{2}(\bm{x}_{j})}+\frac{\widehat{\mu}_{k_{i}\ell_{j}}^{(t,s)}}{(\widehat{\sigma}_{k_{i}\ell_{j}}^{2})^{(t,s)}}\right];\label{thm1.2}
\end{gather}
the posterior probability of $\{z_{i,k}=1\}$ conditional on $\mathcal{E}_{t}$ and given $\widehat{\theta}^{(t,s)}$ is
\begin{gather}
\widehat{z}_{i,k}^{(t,s)}=\frac{\sum_{k_{1:n}\in\mathcal{K},k_{i}=k}\sum_{\ell_{1:m}\in\mathcal{L}}f_{\tau}^{(t,s)}(k_{1:n})f_{\omega}^{(t,s)}(\ell_{1:m})f_{Y}^{(t,s)}(\mathcal{E}_{t}|k_{1:n},\ell_{1:m})}{\sum_{k_{1:n}\in\mathcal{K}}\sum_{\ell_{1:m}\in\mathcal{L}}f_{\tau}^{(t,s)}(k_{1:n})f_{\omega}^{(t,s)}(\ell_{1:m})f_{Y}^{(t,s)}(\mathcal{E}_{t}|k_{1:n},\ell_{1:m})},\label{thm1.3}
\end{gather}
the posterior probability of $\{v_{j,\ell}=1\}$ conditional on $\mathcal{E}_{t}$ and given $\widehat{\theta}^{(t,s)}$ is
\begin{gather}
\widehat{v}_{j,\ell}^{(t,s)}=\frac{\sum_{k_{1:n}\in\mathcal{K}}\sum_{\ell_{1:m}\in\mathcal{L},\ell_{j}=\ell}f_{\tau}^{(t,s)}(k_{1:n})f_{\omega}^{(t,s)}(\ell_{1:m})f_{Y}^{(t,s)}(\mathcal{E}_{t}|k_{1:n},\ell_{1:m})}{\sum_{k_{1:n}\in\mathcal{K}}\sum_{\ell_{1:m}\in\mathcal{L}}f_{\tau}^{(t,s)}(k_{1:n})f_{\omega}^{(t,s)}(\ell_{1:m})f_{Y}^{(t,s)}(\mathcal{E}_{t}|k_{1:n},\ell_{1:m})},\label{thm1.4}
\end{gather}
where $$f_{\tau}^{(t,s)}(k_{1:n})\triangleq\prod_{i=1}^{n}\widehat{\tau}_{k_{i}}^{(t,s)}, \quad f_{\omega}^{(t,s)}(\ell_{1:m})\triangleq\prod_{j=1}^{m}\widehat{\omega}_{\ell_{j}}^{(t,s)},$$
and 
$$f_{Y}^{(t,s)}(\mathcal{E}_{t}|k_{1:n},\ell_{1:m})\triangleq\prod_{i=1}^{n}\prod_{j=1}^{m}C_{ij,k_{i}\ell_{j}}^{(t,s)}$$
with  
$$C_{ij,k_{i}\ell_{j}}^{(t,s)}\triangleq\left(\frac{1}{2\pi\sigma_{i}^{2}(\bm{x}_{j})}\right)^{\frac{t_{ij}}{2}}\sqrt{\frac{(\sigma_{ij,k_{i}\ell_{j}}^{2})^{(t,s)}}{(\widehat{\sigma}_{k_{i}\ell_{j}}^{2})^{(t,s)}}}\exp\left\{\frac{1}{2}\left[\frac{(\mu_{ij,k_{i}\ell_{j}}^{(t,s)})^{2}}{(\sigma_{ij,k_{i}\ell_{j}}^{2})^{(t,s)}}-\frac{\sum_{h=1}^{t_{ij}}Y_{i,h}^{2}(\bm{x}_{j})}{\sigma_{i}^{2}(\bm{x}_{j})}-\frac{(\widehat{\mu}_{k_{i}\ell_{j}}^{(t,s)})^{2}}{(\widehat{\sigma}_{k_{i}\ell_{j}}^{2})^{(t,s)}}\right]\right\}.$$ 
The estimates of the parameters in the $(s+1)$-th iteration of the EM algorithm are given by
\begin{gather}
\widehat{\tau}_{k}^{(t,s+1)}=\frac{\sum_{i=1}^{n}\widehat{z}_{i,k}^{(t,s)}}{n},\ \widehat{\omega}_{\ell}^{(t,s+1)}=\frac{\sum_{j=1}^{m}\widehat{v}_{j,\ell}^{(t,s)}}{m},\label{thm1.5}
\end{gather}
\begin{gather}
\widehat{\mu}_{k\ell}^{(t,s+1)}=\frac{\sum_{i=1}^{n}\sum_{j=1}^{m}\widehat{z}_{i,k}^{(t,s)}\widehat{v}_{j,\ell}^{(t,s)}\mu_{ij,k\ell}^{(t,s)}}{\sum_{i=1}^{n}\sum_{j=1}^{m}\widehat{z}_{i,k}^{(t,s)}\widehat{v}_{j,\ell}^{(t,s)}},\label{thm1.6}
\end{gather}
and
\begin{gather}
(\widehat{\sigma}_{k\ell}^{2})^{(t,s+1)}=\frac{\sum_{i=1}^{n}\sum_{j=1}^{m}\widehat{z}_{i,k}^{(t,s)}\widehat{v}_{j,\ell}^{(t,s)}\left[(\sigma_{ij,k\ell}^{2})^{(t,s)}+\left(\mu_{ij,k\ell}^{(t,s)}-\widehat{\mu}_{k\ell}^{(t,s+1)}\right)^{2}\right]}{\sum_{i=1}^{n}\sum_{j=1}^{m}\widehat{z}_{i,k}^{(t,s)}\widehat{v}_{j,\ell}^{(t,s)}}.\label{thm1.7}
\end{gather}
\end{theorem}
The proof can be found in the e-companion to this paper.

The posterior estimates are the output of the final iteration of the EM algorithm, and we denote these posterior estimates as $\widehat{z}_{i,k}^{(t)}$, $\widehat{v}_{j,\ell}^{(t)}$, $\mu_{ij,k\ell}^{(t)}$, and $(\sigma_{ij,k\ell}^{2})^{(t)}$. Moreover, we have
$$\widehat{\tau}_{k}^{(t)}=\frac{\sum_{i=1}^{n}\widehat{z}_{i,k}^{(t)}}{n},\ \widehat{\omega}_{\ell}^{(t)}=\frac{\sum_{j=1}^{m}\widehat{v}_{j,\ell}^{(t)}}{m},$$
$$\widehat{\mu}_{k\ell}^{(t)}=\frac{\sum_{i=1}^{n}\sum_{j=1}^{m}\widehat{z}_{i,k}^{(t)}\widehat{v}_{j,\ell}^{(t)}\mu_{ij,k\ell}^{(t)}}{\sum_{i=1}^{n}\sum_{j=1}^{m}\widehat{z}_{i,k}^{(t)}\widehat{v}_{j,\ell}^{(t)}},$$
and
$$(\widehat{\sigma}_{k\ell}^{2})^{(t)}=\frac{\sum_{i=1}^{n}\sum_{j=1}^{m}\widehat{z}_{i,k}^{(t)}\widehat{v}_{j,\ell}^{(t)}\left[(\sigma_{ij,k\ell}^{2})^{(t)}+\left(\mu_{ij,k\ell}^{(t)}-\widehat{\mu}_{k\ell}^{(t)}\right)^{2}\right]}{\sum_{i=1}^{n}\sum_{j=1}^{m}\widehat{z}_{i,k}^{(t)}\widehat{v}_{j,\ell}^{(t)}}.$$
We can see that $\widehat{\tau}_{k}^{(t)}$ is the proportion of designs belonging to design cluster $k$, $\widehat{\omega}_{\ell}^{(t)}$ is the proportion of contexts belonging to context cluster $\ell$, cluster mean $\widehat{\mu}_{k\ell}^{(t)}$ is the weighted average of posterior means of design-context pairs belonging to cluster pair ($k,\ell$), and cluster variance $(\widehat{\sigma}_{k\ell}^{2})^{(t)}$ includes the weighted average of posterior variances of design-context pairs belonging to cluster pair ($k,\ell$) and the weighted average of bias with respect to cluster mean $\widehat{\mu}_{k\ell}^{(t)}$.

Given the complexity of the formulas in Theorem~\ref{thm_EM}, it is helpful to examine the limiting case when $t\to+\infty$ such that $y_{i}(\bm{x}_{j})$'s can be estimated accurately. In this case, from classic model-based clustering analysis~\citep{dempster1977maximum,fraley2002model}, we have Proposition~\ref{proposition_lim} on the clustering statistics and the parameter estimates obtained by applying the EM algorithm when observing the true performance $\bm{y}$. In limiting case, $(\sigma_{ij,k\ell}^{2})^{(t)}$ is zero and $\mu_{ij,k\ell}^{(t)}$ is reduced to $y_{i}(\bm{x}_{j})$. Moreover, $C_{ij,k_{i}\ell_{j}}^{(t,s)}$ in Theorem~\ref{thm_EM} is the probability density of samples for design-context pair ($i,j$) given cluster pair ($k_{i},\ell_{j}$), which is replaced by $\phi\left(y_{i}(\bm{x}_{j})|\widehat{\mu}_{k_{i}\ell_{j}}^{(s)},(\widehat{\sigma}_{k_{i}\ell_{j}}^{2})^{(s)}\right)$ in the classic results. The asymptotic result between $C_{ij,k_{i}\ell_{j}}^{(t,s)}$ and $\phi\left(y_{i}(\bm{x}_{j})|\widehat{\mu}_{k_{i}\ell_{j}}^{(s)},(\widehat{\sigma}_{k_{i}\ell_{j}}^{2})^{(s)}\right)$ is shown in Proposition~\ref{proposition_con}, which also concludes that our clustering results given infinite samples are consistent with those when the true performance~$\bm{y}$ is observed, i.e., Corollary~\ref{corollary_con}. All the above observations indicate that the results in Theorem 1 are consistent with the results in Proposition~\ref{proposition_lim} in limiting case.

With regard to practical computation, we note that all $\widehat{\tau}_{k}^{(t,s)}$ and $\widehat{\omega}_{\ell}^{(t,s)}$ are between 0 and 1. In addition, as shown in Proposition~\ref{proposition_zero}, when $\sigma_{i}^{2}(x_{j})$ is no less than certain threshold, $C_{ij,k_{i}\ell_{j}}^{(t,s)}$approaches zero exponentially as the number of allocated samples goes to infinity; otherwise $C_{ij,k_{i}\ell_{j}}^{(t,s)}$ goes to infinity. Both cases will lead to a computational issue that $C_{ij,k_{i}\ell_{j}}^{(t,s)}$ could be smaller or larger than the precision of the computer when the number of allocated samples grows large so that the expressions of $\widehat{z}_{i,k}^{(t,s)}$ and $\widehat{v}_{j,\ell}^{(t,s)}$ become $0/0$ or $\infty/\infty$.

To deal with this computational issue, we provide an equivalent transformation (Algorithm~\ref{Algo_trans} in the e-companion) for the expression of $\widehat{z}_{i,k}^{(t,s)}$ and $\widehat{v}_{j,\ell}^{(t,s)}$. The key idea is to magnify both the numerator and the denominator by the same factor. Specifically, we denote
$$f^{(t,s)}(k_{1:n},\ell_{1:m})\triangleq f_{\tau}^{(t,s)}(k_{1:n})f_{\omega}^{(t,s)}(\ell_{1:m})f_{Y}^{(t,s)}(\mathcal{E}_{t}|k_{1:n},\ell_{1:m}).$$ Given that each $f^{(t,s)}(k_{1:n},\ell_{1:m})$ is too small or too large, we perform a log transformation on $f^{(t,s)}(k_{1:n},\ell_{1:m})$ to scale the value to a suitable range. Furthermore, we denote $$g^{(t,s)}(k_{1:n},\ell_{1:m})\triangleq\log f^{(t,s)}(k_{1:n},\ell_{1:m})-\underset{k_{1:n}\in\mathcal{K},\ell_{1:m}\in\mathcal{L}}{\max}\log f^{(t,s)}(k_{1:n},\ell_{1:m}).$$ Then the expressions of $\widehat{z}_{i,k}^{(t,s)}$ and $\widehat{v}_{j,\ell}^{(t,s)}$ can be rewritten as
$$\widehat{z}_{i,k}^{(t,s)}=\frac{\sum_{k_{1:n}\in\mathcal{K},k_{i}=k}\sum_{\ell_{1:m}\in\mathcal{L}}\exp(g^{(t,s)}(k_{1:n},\ell_{1:m}))}{\sum_{k_{1:n}\in\mathcal{K}}\sum_{\ell_{1:m}\in\mathcal{L}}\exp(g^{(t,s)}(k_{1:n},\ell_{1:m}))},$$
and
$$\widehat{v}_{j,\ell}^{(t,s)}=\frac{\sum_{k_{1:n}\in\mathcal{K}}\sum_{\ell_{1:m}\in\mathcal{L},\ell_{j}=\ell}\exp(g^{(t,s)}(k_{1:n},\ell_{1:m}))}{\sum_{k_{1:n}\in\mathcal{K}}\sum_{\ell_{1:m}\in\mathcal{L}}\exp(g^{(t,s)}(k_{1:n},\ell_{1:m}))}.$$
As shown in Proposition~\ref{proposition_trans}, the denominator of the rewritten $\widehat{z}_{i,k}^{(t,s)}$ and $\widehat{v}_{j,\ell}^{(t,s)}$ is bounded, and thus the computational issue can be addressed by the proposed transformation.

\section{Dynamic Sampling Policy}\label{sec:DSP}
We aim to provide a dynamic sampling policy $\mathcal{A}_{t}$ to maximize the $\text{PCS}_{\text{W}}$. The dynamic sampling policy $\mathcal{A}_{t}$ is a sequence of maps $\mathcal{A}_{t}(\cdot)=(A_{1}(\cdot),\ldots,A_{t}(\cdot))$. Based on sampling observations $\mathcal{E}_{h-1}$, $A_{h}(\mathcal{E}_{h-1})\in\{(i,j):1\leq i\leq n,\ 1\leq j\leq m\}$ allocates the $h$-th sample to estimate the performance of design $i$ in context $j$. Given the information of $t$ allocated samples, we let $k_{i}^{*}=\underset{k=1,\ldots,K}{\arg\max}\ \widehat{z}_{i,k}^{(t)}$ and $\ell_{j}^{*}=\underset{\ell=1,\ldots,L}{\arg\max}\ \widehat{v}_{j,\ell}^{(t)}$ denote the indices of the optimal posterior probabilities of clustering for each design and context, and the selection for context $\bm{x}_{j}$ is to pick the design with the largest
posterior mean, i.e., $\langle1\rangle_{jt}$, where notations $\langle i\rangle_{jt},\ i=1,\ldots,n$, are the ranking indices for context $\bm{x}_{j}$ such that
$$\mu_{\langle1\rangle_{jt}j,k_{\langle1\rangle_{jt}}^{*}\ell_{j}^{*}}^{(t)}>\cdots>\mu_{\langle n\rangle_{jt}j,k_{\langle n\rangle_{jt}}^{*}\ell_{j}^{*}}^{(t)}.$$ Similar to that in~\cite{peng2016dynamic} and~\cite{peng2018ranking}, the sequential sampling decision can be formulated as a stochastic dynamic programming problem. The expected payoff for a sampling policy $\mathcal{A}_{t}$ can be defined recursively by
\begin{gather*}
\mathcal{V}_{t}(\mathcal{E}_{t};\mathcal{A}_{t})\triangleq\underset{j=1,\ldots,m}{\min}\mathbb{P}\left(y_{\langle1\rangle_{jt}}(\bm{x}_{j})>y_{\langle i\rangle_{jt}}(\bm{x}_{j}),i\neq1|\mathcal{E}_{t}\right),
\end{gather*}
and for $0\leq h<t$,
\begin{gather*}
\mathcal{V}_{h}(\mathcal{E}_{h};\mathcal{A}_{t})\triangleq \mathbb{E}\left[\mathcal{V}_{h+1}(\mathcal{E}_{h}\cup\{Y_{i,h+1}(\bm{x}_{j})\};\mathcal{A}_{t})\Big|\mathcal{E}_{h}\right]\Big|_{(i,j)=A_{h+1}(\mathcal{E}_{h})}.
\end{gather*}
Then, the optimal sampling policy is well defined by
\begin{gather*}
\mathcal{A}_{t}^{*}\triangleq \underset{\mathcal{A}_{t}}{\arg\max}\ V(\theta_{0};\mathcal{A}_{t}),
\end{gather*}
where $\theta_{0}$ contains prior hyper-parameters. In principle, the backward induction can be used to solve the stochastic dynamic programming problem, but it suffers from curse-of-dimensionality~\citep{peng2018ranking}. To derive a dynamic sampling policy with an analytical form, we adopt approximate dynamic programming (ADP) schemes which make dynamic decision based on a value function approximation (VFA) and keep learning the VFA with decisions moving forward.

\subsection{Value Function Approximation}
\begin{figure}[htbp]
	\begin{center}
		\includegraphics[width=2.6in]{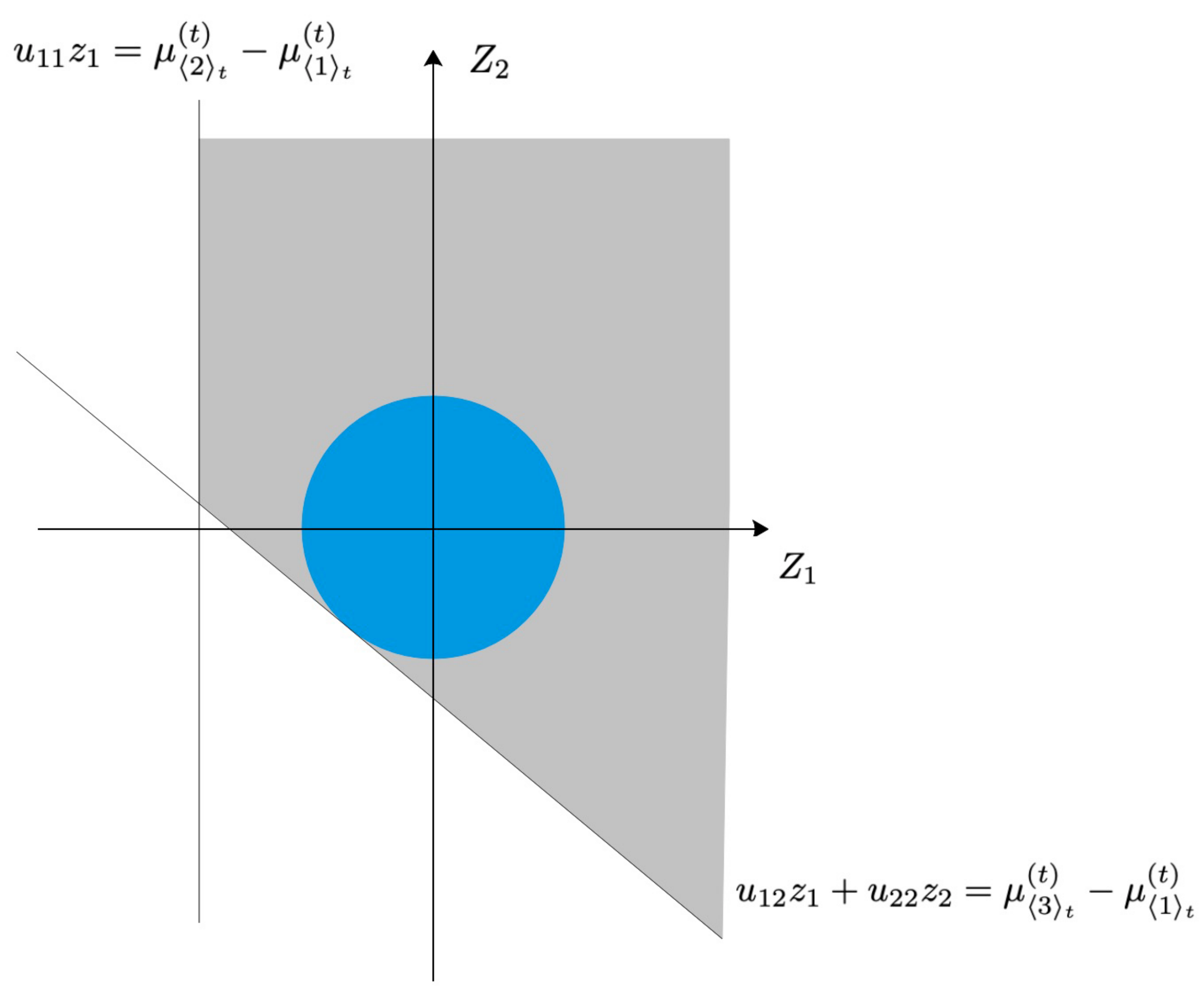}\\
		\caption{Approximation of PCS in~(\ref{VF}).}\label{figure_VFA}
	\end{center}
\end{figure}

The posterior worst-case probability of correct selection can be defined by
\begin{eqnarray*}
&&\underset{j=1,\ldots,m}{\min}\mathbb{P}\left(y_{\langle1\rangle_{jt}}(\bm{x}_{j})>y_{\langle i\rangle_{jt}}(\bm{x}_{j}),i\neq1|\mathcal{E}_{t}\right)\\
&=&\underset{j=1,\ldots,m}{\min}\sum_{k_{1:n}\in\mathcal{K},\ell_{j}\in\mathcal{L}}\left[ \mathbb{P}(\bigcap_{i=1}^{n}\{z_{i,k_{i}}=1\},\{v_{j,\ell_{j}}=1\}\Big|\mathcal{E}_{t})\right.\\
&&\left.\times\mathbb{P}\left(y_{\langle1\rangle_{jt}}(\bm{x}_{j})>y_{\langle i\rangle_{jt}}(\bm{x}_{j}),i\neq1\Bigg|\bigcap_{i=1}^{n}\{z_{i,k_{i}}=1\},\{v_{j,\ell_{j}}=1\},\mathcal{E}_{t}\right)\right].
\end{eqnarray*}
Under a given context and clustering situation, the corresponding probability of correct selection 
\begin{gather}
\mathbb{P}\left(y_{\langle1\rangle_{jt}}(\bm{x}_{j})>y_{\langle i\rangle_{jt}}(\bm{x}_{j}),i\neq1\Bigg|\bigcap_{i=1}^{n}\{z_{i,k_{i}}=1\},\{v_{j,\ell_{j}}=1\},\mathcal{E}_{t}\right)\label{VF}
\end{gather}
is consistent with that in R\&S literature. Then, we use a similar approximation developed in~\cite{peng2018ranking}, i.e.,
\begin{eqnarray*}
V(\mathcal{E}_{t})\triangleq\underset{j=1,\ldots,m}{\min}\sum_{k_{1:n}\in\mathcal{K},\ell_{j}\in\mathcal{L}}p_{z}(k_{1:n},\mathcal{E}_{t})p_{v}(\ell_{j},\mathcal{E}_{t})\text{APCS}(k_{1:n},\ell_{j},\mathcal{E}_{t}),
\end{eqnarray*}
where $p_{z}(k_{1:n},\mathcal{E}_{t})\triangleq \prod_{i=1}^{n}\widehat{z}_{i,k_{i}}^{(t)}$ is the posterior probability of clustering situation $k_{1:n}$ for all designs, $p_{v}(\ell_{j},\mathcal{E}_{t})\triangleq\widehat{v}_{j,\ell_{j}}^{(t)}$ is the posterior probability of clustering situation $\ell_{j}$ for context $\bm{x}_{j}$, and $$\text{APCS}(k_{1:n},\ell_{j},\mathcal{E}_{t})\triangleq\underset{i\neq 1}{\min}\frac{\left(\mu_{\langle1\rangle_{jt}j,k_{\langle1\rangle_{jt}}\ell_{j}}^{(t)}-\mu_{\langle i\rangle_{jt}j,k_{\langle i\rangle_{jt}}\ell_{j}}^{(t)}\right)^{2}}{\left(\sigma_{\langle1\rangle_{jt}j,k_{\langle1\rangle_{jt}}\ell_{j}}^{2}\right)^{(t)}+\left(\sigma_{\langle i\rangle_{jt}j,k_{\langle i\rangle_{jt}}\ell_{j}}^{2}\right)^{(t)}}$$ is an approximation of the  PCS in~(\ref{VF}). Note that the PCS in~(\ref{VF}) is an integral of the multivariate standard normal density over a region encompassed by some hyperplanes. As shown in Figure~\ref{figure_VFA}, the integral over a maximum tangent inner ball in the shadowed region can capture the main body of the integral over entire region due to exponential decay of the normal density. Rigorously, Proposition~\ref{pro_exp} provides an upper bound of the error generated by using the inner ball as an approximation, and shows this upper bound decreases to zero exponentially as the radius of the ball goes to infinity. Therefore, we use the volume of the ball as an approximation for the PCS in~(\ref{VF}).

After an additional sample is allocated to design $r$ and context $q$, we apply a certainty equivalent approximation~\citep{bertsekas1995dynamic} to the value function looking one-step ahead: 
\begin{eqnarray}
&&\mathbb{E}\left[V\left(\mathcal{E}_{t}\cup Y_{r,t_{rq}+1}(\bm{x}_{q})\right)\Big|\mathcal{E}_{t}\right]\nonumber\approx V\left(\mathcal{E}_{t}\cup\mathbb{E}\left[Y_{r,t_{rq}+1}(\bm{x}_{q})\Big|\mathcal{E}_{t}\right]\right)\nonumber\\
&=&\underset{j=1,\ldots,m}{\min}\sum_{k_{1:n}\in\mathcal{K},\ell_{j}\in\mathcal{L}}p_{z}(k_{1:n},\mathcal{E}_{(t;(r,q)_{E})})p_{v}(\ell_{j},\mathcal{E}_{(t;(r,q)_{E})})\text{APCS}(k_{1:n},\ell_{j},\mathcal{E}_{(t;(r,q)_{E})}),\label{VFA_onestep}
\end{eqnarray}
where $\mathcal{E}_{(t;(r,q)_{E})}\triangleq \mathcal{E}_{t}\cup\mathbb{E}\left[Y_{r,t_{rq}+1}(\bm{x}_{q})\Big|\mathcal{E}_{t}\right]$, $p_{z}(k_{1:n},\mathcal{E}_{(t;(r,q)_{E})})\triangleq \prod_{i=1}^{n}\widehat{z}_{i,k_{i}}^{(t,(r,q)_{E})}$ is the one-step-ahead posterior probability of clustering situation $k_{1:n}$ for all designs, $p_{v}(\ell_{j},\mathcal{E}_{(t;(r,q)_{E})})\triangleq\widehat{v}_{j,\ell_{j}}^{(t,(r,q)_{E})}$ the one-step-ahead posterior probability of clustering situation $\ell_{j}$ for context $\bm{x}_{j}$, and $$\text{APCS}(k_{1:n},\ell_{j},\mathcal{E}_{(t;(r,q)_{E})})\triangleq\underset{i\neq 1}{\min}\frac{\left(\mu_{\langle1\rangle_{jt}j,k_{\langle1\rangle_{jt}}\ell_{j}}^{(t)}-\mu_{\langle i\rangle_{jt}j,k_{\langle i\rangle_{jt}}\ell_{j}}^{(t)}\right)^{2}}{\left(\sigma_{\langle1\rangle_{jt}j,k_{\langle1\rangle_{jt}}\ell_{j}}^{2}\right)^{(t;(r,q)_{E})}+\left(\sigma_{\langle i\rangle_{jt}j,k_{\langle i\rangle_{jt}}\ell_{j}}^{2}\right)^{(t;(r,q)_{E})}}.$$
Here for brevity we use a statistic with superscript $(t;(r,q)_{E})$ to denote its corresponding estimate conditional on $\mathcal{E}_{(t;(r,q)_{E})}$.

When the additional sample $Y_{r,t_{rq}+1}(\bm{x}_{q})$ takes a value of its posterior mean $\mathbb{E}\left[Y_{r,t_{rq}+1}(\bm{x}_{q})\Big|\mathcal{E}_{t}\right]$, the posterior mean of $y_{i}(\bm{x}_{j})$ does not change but the posterior variance of $y_{i}(\bm{x}_{j})$ is updated as follows:
\begin{gather}
(\sigma_{ij,k_{i}\ell_{j}}^{2})^{(t;(r,q)_{E})}=\ \left\{
\begin{array}{rcl}
1\Big/\left[\dfrac{t_{ij}+1}{\sigma_{i}^{2}(\bm{x}_{j})}+\dfrac{1}{(\widehat{\sigma}_{k_{i}\ell_{j}}^{2})^{(t;(r,q)_{E})}}\right]&, &(i,j)=(r,q);\\
1\Big/\left[\dfrac{t_{ij}}{\sigma_{i}^{2}(\bm{x}_{j})}+\dfrac{1}{(\widehat{\sigma}_{k_{i}\ell_{j}}^{2})^{(t;(r,q)_{E})}}\right]&, &(i,j)\neq(r,q),(k_{i},\ell_{j})=(k_{r},\ell_{q});\\
(\sigma_{ij,k_{i}\ell_{j}}^{2})^{(t)}&, &\text{otherwise}.
\end{array}
\right.\label{onestep_sigma}
\end{gather}
\begin{eqnarray}
(\widehat{\sigma}_{k\ell}^{2})^{(t;(r,q)_{E})}&=&\frac{\sum_{i=1}^{n}\sum_{j=1}^{m}\widehat{z}_{i,k}^{(t;(r,q)_{E})}\widehat{v}_{j,\ell}^{(t;(r,q)_{E})}\left[1\Big/\left[\dfrac{t_{ij}+\mathbbm{1}\left\{(i,j)=(r,q)\right\}}{\sigma_{i}^{2}(\bm{x}_{j})}+\dfrac{1}{(\widehat{\sigma}_{k\ell}^{2})^{(t)}}\right]+\left(\mu_{ij,k\ell}^{(t)}-\widehat{\mu}_{k\ell}^{(t)}\right)^{2}\right]}{\sum_{i=1}^{n}\sum_{j=1}^{m}\widehat{z}_{i,k}^{(t;(r,q)_{E})}\widehat{v}_{j,\ell}^{(t;(r,q)_{E})}}\nonumber\\
&\approx&\frac{\sum_{i=1}^{n}\sum_{j=1}^{m}\widehat{z}_{i,k}^{(t)}\widehat{v}_{j,\ell}^{(t)}\left[1\Big/\left[\dfrac{t_{ij}+\mathbbm{1}\left\{(i,j)=(r,q)\right\}}{\sigma_{i}^{2}(\bm{x}_{j})}+\dfrac{1}{(\widehat{\sigma}_{k\ell}^{2})^{(t)}}\right]+\left(\mu_{ij,k\ell}^{(t)}-\widehat{\mu}_{k\ell}^{(t)}\right)^{2}\right]}{\sum_{i=1}^{n}\sum_{j=1}^{m}\widehat{z}_{i,k}^{(t)}\widehat{v}_{j,\ell}^{(t)}}.\label{onestep_clustersigma}
\end{eqnarray}
After allocating one more sample to design $r$ and context~$q$, $(\sigma_{ij,k_{i}\ell_{j}}^{2})^{(t;(r,q)_{E})}$ and $(\widehat{\sigma}_{k\ell}^{2})^{(t;(r,q)_{E})}$ denote the posterior variance of $y_{i}(\bm{x}_{j})$ and the cluster variance of cluster pair ($k,\ell$), respectively. In order to simplify $(\widehat{\sigma}_{k\ell}^{2})^{(t;(r,q)_{E})}$, we replace $\widehat{z}_{i,k}^{(t;(r,q)_{E})}$ and $\widehat{v}_{j,\ell}^{(t;(r,q)_{E})}$ with $\widehat{z}_{i,k}^{(t)}$ and $\widehat{v}_{j,\ell}^{(t)}$, which leaves a simplified posterior variance estimate mainly capturing noise reduction caused by sampling. The validation of such replacement is supported by Corollary~\ref{corollary_con}, and thus this approximation will be tight as $t$ goes to infinity. Allocating a sample to design $r$ and context $q$ can reduce both sample variance $\sigma_{r}^{2}(\bm{x}_{q})/t_{rq}$ and cluster variance $(\widehat{\sigma}_{k_{r}\ell_{q}}^{2})^{(t)}$, and thus reduce the posterior variance of $y_{r}(\bm{x}_{q})$. In addition, sampling design-context pair ($r,q$) can also reduce the posterior variance of other $y_{i}(\bm{x}_{j})$ in the same cluster by deceasing the corresponding cluster variance, which increases the confidence for belonging to a cluster. 

\subsubsection{Further Efficiency Enhancement}
\begin{figure}[htbp]
	\begin{center}
		\includegraphics[width=4in]{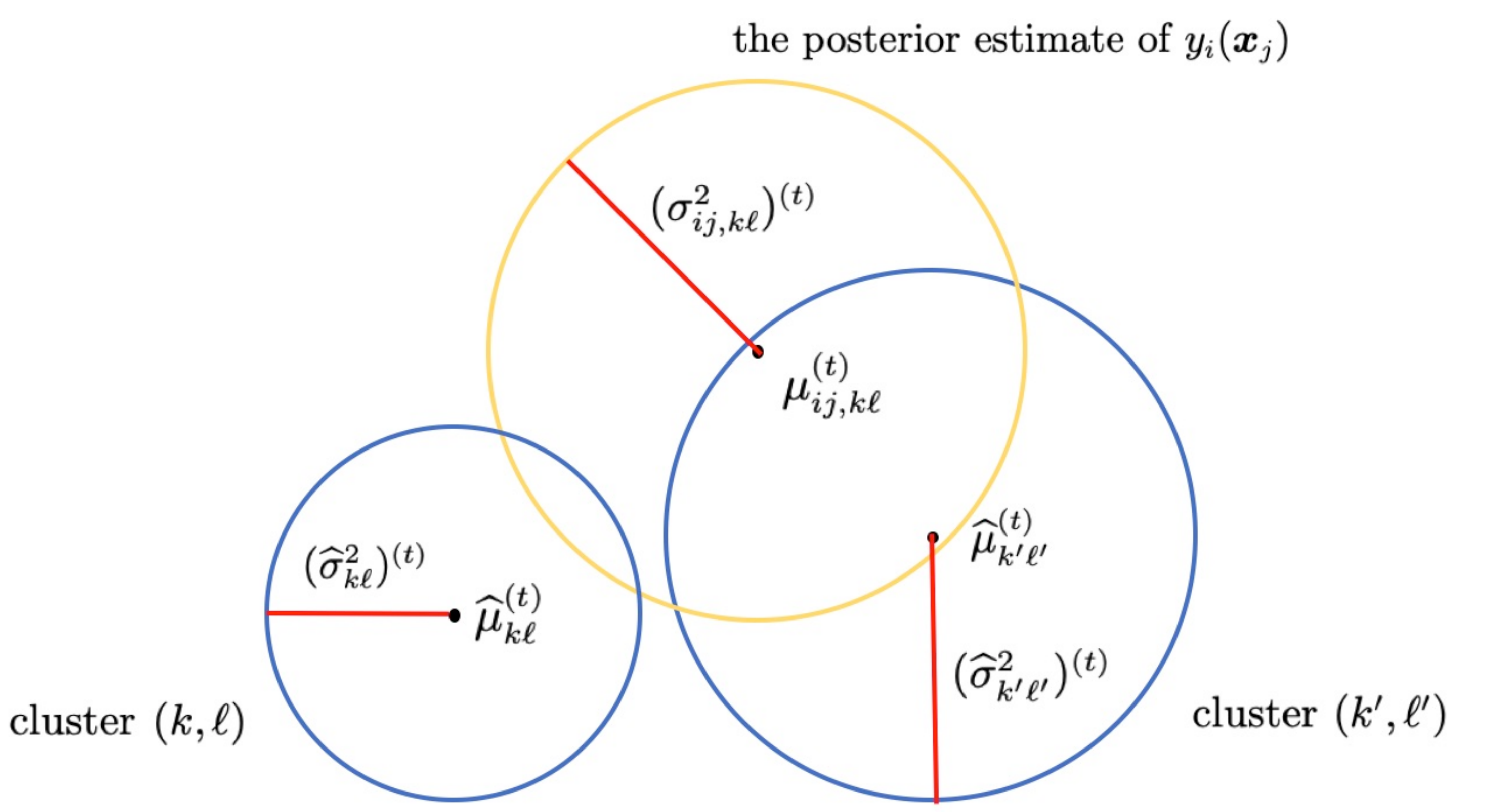}\\
		\caption{Insights in approximations of $\widehat{z}_{i,k}^{(t;(r,q)_{E})}$ and $\widehat{v}_{j,\ell}^{(t;(r,q)_{E})}$.}\label{figure_zv}
	\end{center}
\end{figure}

We need to estimate the one-step-ahead looking posterior probability of the hidden state for determining the cluster in calculating the right hand side of (\ref{VFA_onestep}), i.e., $\widehat{z}_{i,k}^{(t;(r,q)_{E})}$ and $\widehat{v}_{j,\ell}^{(t;(r,q)_{E})}$. In order to balance estimation accuracy and computational efficiency, we make the following approximations:
\begin{gather}
\widehat{z}_{i,k}^{(t;(r,q)_{E})}\approx\ \left\{
\begin{array}{rcl}
\frac{\widehat{z}_{i,k}^{(t)}\exp\left(-\dfrac{(\mu_{iq,k\ell_{q}^{*}}^{(t)}-\widehat{\mu}_{k\ell_{q}^{*}}^{(t)})^{2}(\sigma_{iq,k\ell_{q}^{*}}^{4})^{(t)}}{2\sigma_{i}^{2}(\bm{x}_{q})(\widehat{\sigma}^{4}_{k\ell_{q}^{*}})^{(t)}}\right)}{\sum_{k=1}^{K}\widehat{z}_{i,k}^{(t)}\exp\left(-\dfrac{(\mu_{iq,k\ell_{q}^{*}}^{(t)}-\widehat{\mu}_{k\ell_{q}^{*}}^{(t)})^{2}(\sigma_{iq,k\ell_{q}^{*}}^{4})^{(t)}}{2\sigma_{i}^{2}(\bm{x}_{q})(\widehat{\sigma}^{4}_{k\ell_{q}^{*}})^{(t)}}\right)}&, &i=r;\\
\widehat{z}_{i,k}^{(t)}&, &\text{otherwise}.
\end{array}
\right.\label{onestep_z}
\end{gather}
\begin{gather}
\widehat{v}_{j,\ell}^{(t;(r,q)_{E})}\approx\ \left\{
\begin{array}{rcl}
\frac{\widehat{v}_{j,\ell}^{(t)}\exp\left(-\dfrac{(\mu_{rj,k_{r}^{*}\ell}^{(t)}-\widehat{\mu}_{k_{r}^{*}\ell}^{(t)})^{2}(\sigma_{rj,k_{r}^{*}\ell}^{4})^{(t)}}{2\sigma_{r}^{2}(\bm{x}_{j})(\widehat{\sigma}^{4}_{k_{r}^{*}\ell})^{(t)}}\right)}{\sum_{k=1}^{K}\widehat{v}_{j,\ell}^{(t)}\exp\left(-\dfrac{(\mu_{rj,k_{r}^{*}\ell}^{(t)}-\widehat{\mu}_{k_{r}^{*}\ell}^{(t)})^{2}(\sigma_{rj,k_{r}^{*}\ell}^{4})^{(t)}}{2\sigma_{r}^{2}(\bm{x}_{j})(\widehat{\sigma}^{4}_{k_{r}^{*}\ell})^{(t)}}\right)}&, &j=q;\\
\widehat{v}_{j,\ell}^{(t)}&, &\text{otherwise}.
\end{array}
\right.\label{onestep_v}
\end{gather}
The detailed theoretical supports for the above approximations are provided in the e-companion. Observed from~\eqref{onestep_z} and \eqref{onestep_v}, here we give some insights on how sampling affects the clustering results. As shown in Figure~\ref{figure_zv}, we use circle center to represent posterior mean $\mu_{ij,k\ell}^{(t)}$ and cluster mean $\widehat{\mu}_{k\ell}^{(t)}$, and use radius to represent posterior variance $(\sigma^{2}_{ij,k\ell})^{(t)}$ and cluster variance $(\widehat{\sigma}_{k\ell}^{2})^{(t)}$. Therefore, the yellow circle reflects the posterior estimate of $y_{i}(\bm{x}_{j})$, and it shrinks as more samples are allocated to design $i$ or context $\bm{x}_{j}$; blue circles reflect the scope of each cluster. Large $(\mu_{ij,k\ell}^{(t)}-\widehat{\mu}_{k\ell}^{(t)})^{2}$ or small $(\widehat{\sigma}_{k\ell}^{2})^{(t)}$ for a cluster implies design $i$ or context $j$ is likely to be an outlier in design cluster $k$ or context cluster~$\ell$, and thus as the yellow circle shrinks it will separate with this cluster early, which indicates allocating more samples to estimate $y_{i}(\bm{x}_{j})$ could decrease $\widehat{z}_{i,k}^{(t)}$ and $\widehat{v}_{j,\ell}^{(t)}$. All of these insights are reflected in~\eqref{onestep_z} and \eqref{onestep_v}, that is, $(\mu_{ij,k\ell}^{(t)}-\widehat{\mu}_{k\ell}^{(t)})^{2}$ is on the numerator of the exponential rate while $(\widehat{\sigma}_{k\ell}^{2})^{(t)}$ is on the denominator. Given the approximations of $\widehat{z}_{i,k}^{(t;(r,q)_{E})}$ and $\widehat{v}_{j,\ell}^{(t;(r,q)_{E})}$, the VFA looking one-step ahead can be calculated by
\begin{eqnarray}
&&V(\mathcal{E}_{t};(r,q))\nonumber\\
&\triangleq&\underset{j=1,\ldots,m}{\min}\sum_{k_{1:n}\in\mathcal{K},\ell_{j}\in\mathcal{L}}p_{z}(k_{1:n},\mathcal{E}_{(t;(r,q)_{E})})p_{v}(\ell_{j},\mathcal{E}_{(t;(r,q)_{E})})\text{APCS}(k_{1:n},\ell_{j},\mathcal{E}_{(t;(r,q)_{E})})\label{final_VFA}\\
&=&\underset{j=1,\ldots,m}{\min}\sum_{k_{1:n}\in\mathcal{K},\ell_{j}\in\mathcal{L}}\prod_{i=1}^{n}\widehat{z}_{i,k_{i}}^{(t,(r,q)_{E})}\widehat{v}_{j,\ell_{j}}^{(t,(r,q)_{E})}\underset{i\neq 1}{\min}\frac{\left(\mu_{\langle1\rangle_{jt}j,k_{\langle1\rangle_{jt}}\ell_{j}}^{(t)}-\mu_{\langle i\rangle_{jt}j,k_{\langle i\rangle_{jt}}\ell_{j}}^{(t)}\right)^{2}}{\left(\sigma_{\langle1\rangle_{jt}j,k_{\langle1\rangle_{jt}}\ell_{j}}^{2}\right)^{(t;(r,q)_{E})}+\left(\sigma_{\langle i\rangle_{jt}j,k_{\langle i\rangle_{jt}}\ell_{j}}^{2}\right)^{(t;(r,q)_{E})}}.\nonumber
\end{eqnarray}

\subsection{Dynamic Sampling Policy for Context-Dependent Optimization}
We propose the following dynamic sampling policy for context-dependent optimization (DSCO):
\begin{gather}\label{normal policy}
A_{t+1}(\mathcal{E}_{t})=\left\{(r^{*},q^{*})\Bigg|V(\mathcal{E}_{t};(r^{*},q^{*}))=\underset{(r,q)}{\max}\ V(\mathcal{E}_{t};(r,q))\right\},
\end{gather}
which maximizes VFA looking one-step ahead. As shown in equation~(\ref{final_VFA}), $V(\mathcal{E}_{t};(r,q))$ is determined by the posterior probability of clustering and PCS under each clustering situation, and thus captures the worst-case probability of correct selection over all contexts. Therefore, the sampling rule in equation~(\ref{normal policy}) considers not only correct clustering but also correct selection for improving $\text{PCS}_{\text{W}}$. Moreover, if the decision in equation~(\ref{normal policy}) can increase the VFA almost surely as the number of samples goes to infinity, i.e., 
\begin{gather}\label{condition}
\underset{t\to+\infty}{\lim}V(\mathcal{E}_{t};(r^{*},q^{*}))-V(\mathcal{E}_{t})>0,\ a.s.
\end{gather}
the proposed DSCO is proved to be consistent and achieve the asymptotically optimal sampling ratio.
\begin{remark}\label{special}
Note that VFA in equation (\ref{final_VFA}) captures clustering probability and PCS. However, allocating a sample may not necessarily improve clustering probability and PCS simultaneously so that VFA may decrease. If condition~(\ref{condition}) does not hold, our dynamic sampling policy is designed as shown in Algorithm~\ref{Algo_DSCO} in the e-companion. 
If there exists a ($r,q$) such that $V(\mathcal{E}_{t};(r,q))>V(\mathcal{E}_{t})$, then sample allocation is determined by equation~(\ref{normal policy}); otherwise, let
\begin{gather}
W(\mathcal{E}_{t};(r,q))=\underset{j=1,\ldots,m}{\min}\ \underset{k_{1:n}\in\mathcal{K},\ell_{j}\in\mathcal{L}}{\min}\text{APCS}(k_{1:n},\ell_{j},\mathcal{E}_{(t,(r,q)_{E})}),\label{VFA_W}
\end{gather}
and then sample allocation is determined by
\begin{gather}\label{special policy}
A_{t+1}(\mathcal{E}_{t})=\left\{(r^{*},q^{*})\Bigg|W(\mathcal{E}_{t};(r^{*},q^{*}))=\underset{(r,q)}{\max}\ W(\mathcal{E}_{t};(r,q))\right\}
\end{gather}
until the following condition is met:
\begin{gather}\label{termination condition}
W(\mathcal{E}_{t};(r^{*},q^{*}))>V(\mathcal{E}_{t}).
\end{gather}
Note that there must exist a design-context pair ($r,q$) such that $$\underset{t\to+\infty}{\lim} W(\mathcal{E}_{t};(r,q))-W(\mathcal{E}_{t})>0,\ a.s.,$$ which will be shown in the proof of Theorem~\ref{thm_consistency}. It implies that the sampling rule in equation~(\ref{special policy}) can guarantee $W(\mathcal{E}_{t};(r^{*},q^{*}))$ is strictly increasing as $t$ goes to infinity. Apparently, $W(\mathcal{E}_{t};(r,q))$ considers the smallest $\text{APCS}(k_{1:n},\ell_{j},\mathcal{E}_{(t,(r,q)_{E})})$ among all possible clustering situations, and thus is a lower bound of $V(\mathcal{E}_{t};(r,q))$ which is the expectation of $\text{APCS}(k_{1:n},\ell_{j},\mathcal{E}_{(t,(r,q)_{E})})$ over the entire clustering situation space. Consequently, $V(\mathcal{E}_{t};(r^{*},q^{*}))>V(\mathcal{E}_{t})$ holds when the condition~(\ref{termination condition}) is met. Therefore, although the sampling rule in equation~(\ref{special policy}) is conservative, it can achieve the condition~(\ref{condition}) when meeting the termination condition~(\ref{termination condition}), and thus the asymptotic properties remain the same as those when the condition~(\ref{condition}) holds.
\end{remark}

The asymptotically optimal sampling ratio is interpreted from a large deviations perspective~\citep{glynn2004large}. Note that $t$ is the total sampling budget (number of simulation replications), and $t_{ij}$ is the number of simulation replications allocated to design $i$ and context $j$. Define $r_{ij}=t_{ij}/t$ and $\bm{r}$ is the vector of $r_{ij}$. Then the probability of false selection $\text{PFS}_\text{W}=1-\text{PCS}_\text{W}$ is proved to converge exponentially with a rate function of $\bm{r}$~\citep{gao2019selecting}, i.e.,
$$\underset{t\to+\infty}{\lim}\frac{1}{t}\log\text{PFS}_\text{W}=-\mathcal{R}(\bm{r}).$$ 
We prove that our proposed DSCO can achieve an asymptotically optimal sampling ratio that optimizes large deviations rate $\mathcal{R}(\bm{r})$ as $t$ goes to infinity.

\begin{theorem}\label{thm_consistency}
The proposed DSCO is consistent, i.e., $\forall j=1,\ldots,m$,
$$\lim_{t\to+\infty}\langle1\rangle_{jt}=\langle1\rangle_{j}\ \ a.s.$$
In addition, the sampling ratio of each design-context pair asymptotically achieves the optimal convergence rate of $\text{PCS}_{\text{W}}$ in~\cite{gao2019selecting}, i.e.,
$$\lim_{t\to+\infty} r_{ij}^{(t)}=r_{ij}^{*},\ a.s.,\ i=1,\ldots,n,\ j=1,\ldots,m,$$
where $r_{ij}^{(t)}\triangleq t_{ij}/t$, $\sum_{i=1}^{n}\sum_{j=1}^{m}r_{ij}^{*}=1,\ r_{ij}^{*}\geq0$, and
\begin{gather}
\frac{(r_{\langle1\rangle_{j} j}^{*})^{2}}{\sigma_{\langle1\rangle_{j}}^{2}(\bm{x}_{j})}=\sum_{i=2}^{n}\frac{(r_{\langle i\rangle_{j} j}^{*})^{2}}{\sigma_{\langle i\rangle_{j}}^{2}(\bm{x}_{j})},\ j=1,\ldots,m,\label{ratio_1}
\end{gather}
\begin{gather}
\frac{(y_{\langle 1\rangle_{j}}(\bm{x}_{j})-y_{\langle i\rangle_{j}}(\bm{x}_{j}))^{2}}{\sigma_{\langle 1\rangle_{j}}^{2}(\bm{x}_{j})/r_{\langle1\rangle_{j} j}^{*}+\sigma_{\langle i\rangle_{j}}^{2}(\bm{x}_{j})/r_{\langle i\rangle_{j} j}^{*}}=\frac{(y_{\langle 1\rangle_{j}}(\bm{x}_{j})-y_{\langle i'\rangle_{j}}(\bm{x}_{j}))^{2}}{\sigma_{\langle 1\rangle_{j}}^{2}(\bm{x}_{j})/r_{\langle1\rangle_{j} j}^{*}+\sigma_{\langle i'\rangle_{j}}^{2}(\bm{x}_{j})/r_{\langle i'\rangle_{j} j}^{*}},\ i,i'=2,\ldots,n,\ j=1,\ldots,m,\label{ratio_2}
\end{gather}
\begin{gather}
\frac{(y_{\langle 1\rangle_{j}}(\bm{x}_{j})-y_{\langle i\rangle_{j}}(\bm{x}_{j}))^{2}}{\sigma_{\langle 1\rangle_{j}}^{2}(\bm{x}_{j})/r_{\langle1\rangle_{j} j}^{*}+\sigma_{\langle i\rangle_{j}}^{2}(\bm{x}_{j})/r_{\langle i\rangle_{j} j}^{*}}=\frac{(y_{\langle 1\rangle_{j'}}(\bm{x}_{j'})-y_{\langle i'\rangle_{j'}}(\bm{x}_{j'}))^{2}}{\sigma_{\langle 1\rangle_{j'}}^{2}(\bm{x}_{j'})/r_{\langle1\rangle_{j'} j'}^{*}+\sigma_{\langle i'\rangle_{j'}}^{2}(\bm{x}_{j'})/r_{\langle i'\rangle_{j'} j'}^{*}},\ i,i'=2,\ldots,n,\ j,j'=1,\ldots,m,\label{ratio_3}
\end{gather}
\end{theorem}

The proof is for the dynamic sampling policy in Remark~\ref{special} and applies to the simplified policy in equation~(\ref{normal policy}) if the condition~(\ref{condition}) holds. Equation~(\ref{ratio_1}) and Equation~(\ref{ratio_2}) are the total balance condition and individual balance condition under a certain context, which is consistent with the optimal large deviations conditions in R\&S with single context~\citep{glynn2004large}. Equation~(\ref{ratio_3}) is the balance condition between design-context pairs from different contexts, which reflects effective sampling switching among different contexts due{} to the worst-case PCS considered in our study.

\section{Numerical Experiment}\label{sec:NE}
In this section, we conduct numerical experiments to test the performance of different sampling procedures for context-dependent simulation optimization problems. The proposed DSCO is compared with the equal allocation (EA), two-stage indifference-zone (IZ) procedure in~\cite{shen2017ranking}, the contextual optimal computing budget allocation (C-OCBA) for contextual R\&S in~\cite{gao2019selecting}, and sequential UCB-style algorithm (SUCB) for in~\cite{han2020sequential}. Specifically, EA equally allocates sampling budget to estimate the performance of each design-context pair $y_{i}(\bm{x}_{j}),\ i=1,\ldots,n,\ j=1,\ldots,m$ (roughly $t/(n\times m)$ samples for each $y_{i}(\bm{x}_{j})$); IZ takes $n_{0}$ independent samples of each design-context pair and calculates sample variances $S_{ij}^{2}$ at first stage, and then takes $\max\{\lceil h^{2}S_{ij}^{2}/\delta^{2} \rceil-n_{0},0\}$ additional independent samples of design $i$ in context $\bm{x}_{j}$ at second stage, where $h$ and $\delta$ are IZ parameters; C-OCBA allocates samples based on the optimality conditions (\ref{ratio_1})-(\ref{ratio_3}), where each replication should be allocated to a certain design-context pair in order to balance the equations; SUCB sequentially allocates each sample to design-context pair ($i,j$) such that
$\underset{j}{\max}\ \underset{i}{\min}\ \bm{x}_{j}^{\top}\hat{\theta}_{i,t}+\gamma\sqrt{\bm{x}_{j}^{\top}A^{-1}\bm{x}_{j}}$
where $\gamma>0$ is tuning parameter, $A=I_{d}+\sum_{j=1}^{m}\bm{x}_{j}\bm{x}_{j}^{\top}$, and $\hat{\theta}_{i,t}=A^{-1}\sum_{j=1}^{m}\bar{Y}_{i,t}(\bm{x}_{j})\bm{x}_{j}$ is sequentially updated. Both IZ and SUCB assume a linear dependency between the responses or rewards of a design and the contexts, and they utilize context parameters in addition to sample information; C-OCBA only utilizes the information in the posterior means and variances of the context-dependent performances, but it does not consider the information in clustering among designs and contexts. In all numerical examples, the statistical efficiency of the sampling procedures is measured by the $\text{PCS}_{\text{W}}$ estimated by 10,000 independent experiments. The $\text{PCS}_{\text{W}}$ is reported as a function of the sampling budget in each experiment. The codes for implementing the experiments can be found in GitHub (\url{https://github.com/mmpku1105/code-for-DSCO}).

\subsection{Synthetic Case}
\subsubsection{Example 1: $10\times10$ design-context pair}
\begin{figure}[htbp]
	\begin{center}
		\includegraphics[width=4.5in]{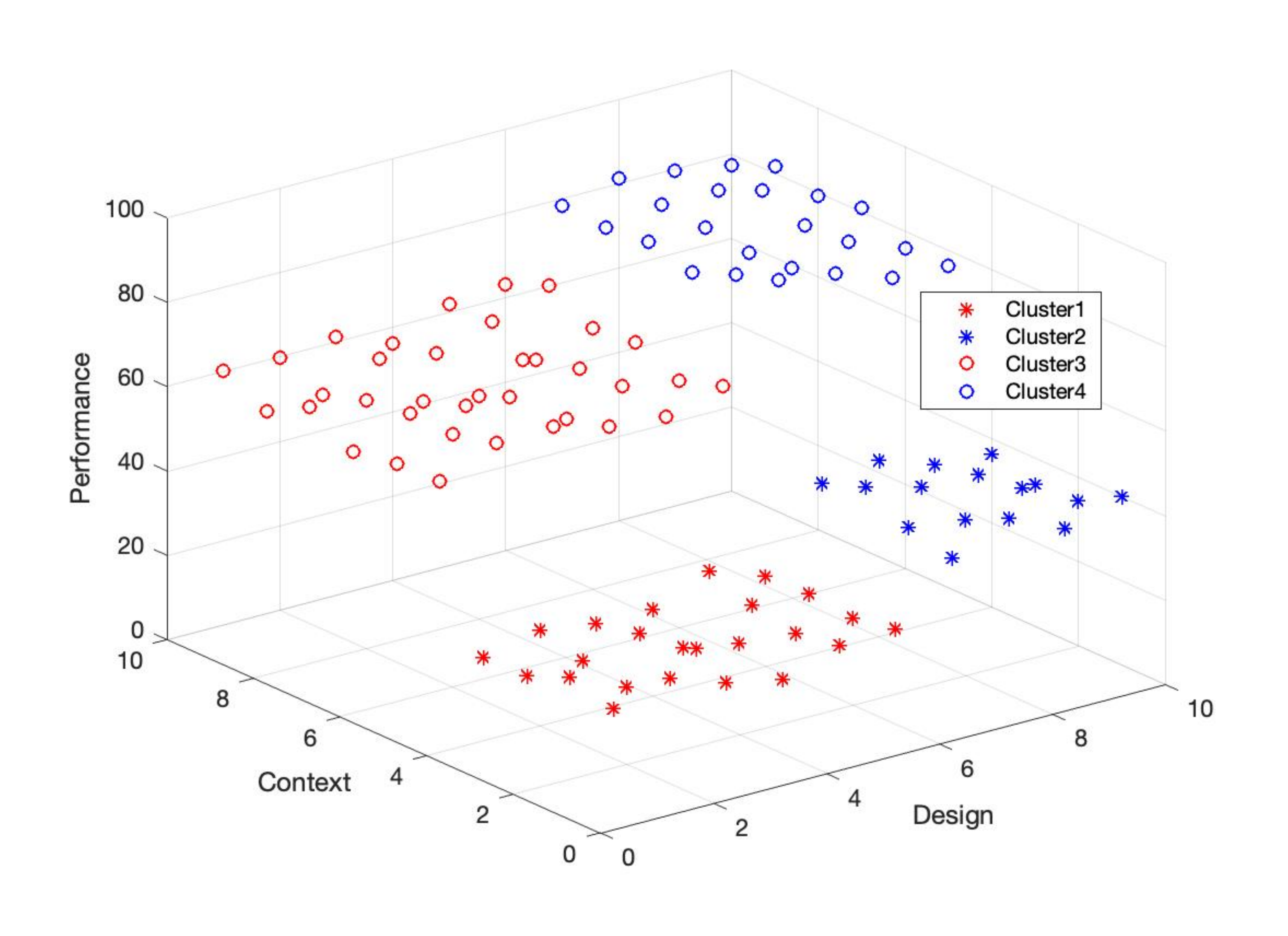}\\
		\caption{Performance clustering in multiple clusters case.}\label{performance_clustering1}
	\end{center}
\end{figure}

\begin{figure}[htbp]
	\begin{center}
		\includegraphics[width=3in]{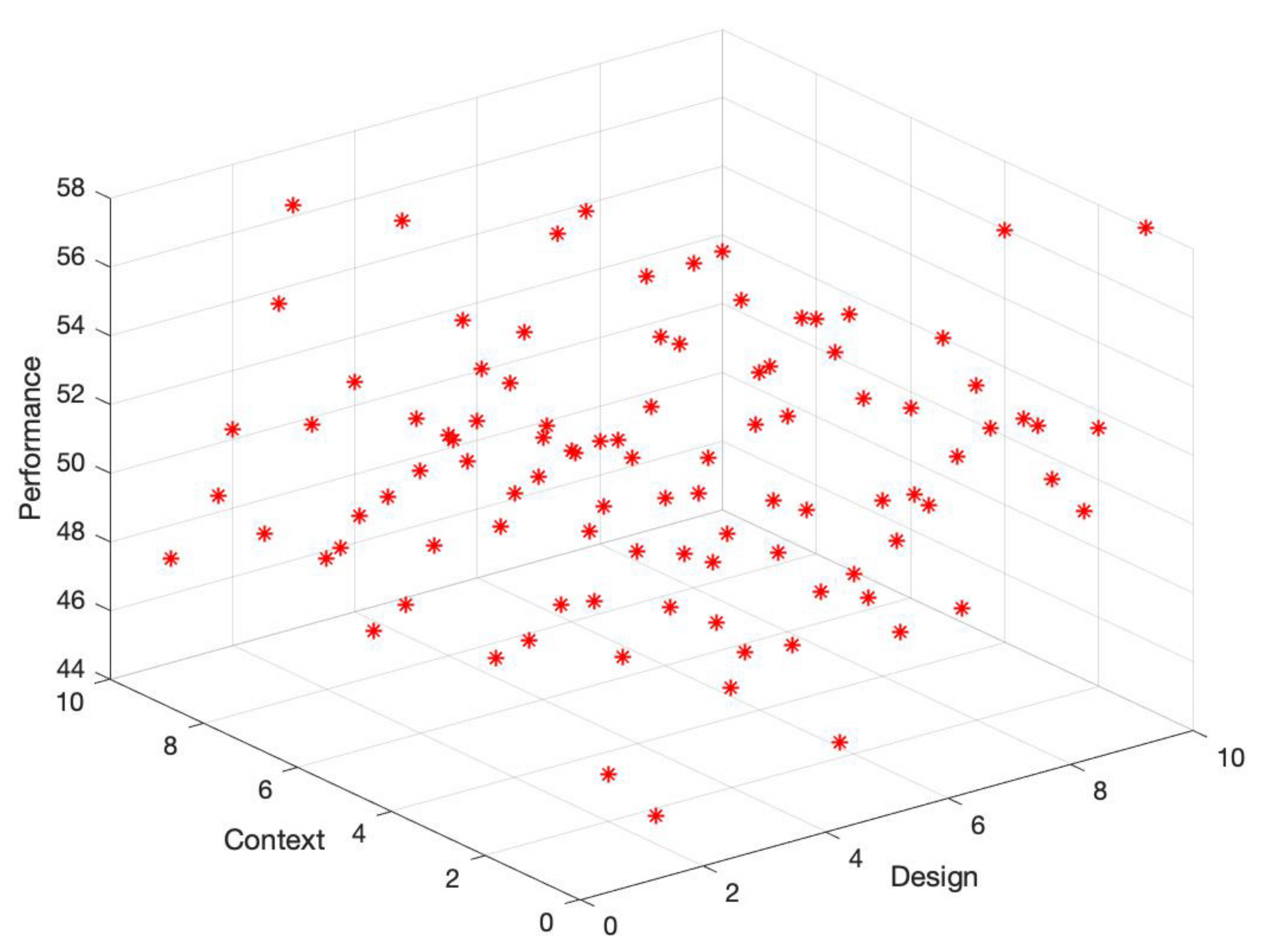}\\
		\caption{Performance clustering in one cluster case.}\label{performance_clustering2}
	\end{center}
\end{figure}

We test our proposed DSCO in a synthetic case with 10 designs and 10 contexts. In order to test the robustness for the performance of DSCO under different performance clustering phenomena, we consider two cases: one cluster case and multiple clusters case. In one cluster case, the performances of each design for each context are generated as follows:
\begin{eqnarray*}
y_{i}(\bm{x}_{j})\sim N(50,3^2),\ i=1,\ldots,10,\ j=1,\ldots,10,
\end{eqnarray*}
which means there is no clear performance clustering structure and all design-context pairs belong to a common cluster as considered by~\cite{shen2017ranking} and~\cite{gao2019selecting}. Context parameters $\bm{x}_{j}$ are set as single-dimensional variables drawn from $N(5,1^2)$. As for multiple clusters case,
the performances of each design for each context are generated as follows:
\begin{eqnarray*}
&y_{i}(\bm{x}_{j})\sim N(20,3^2),\ i=1,\ldots,6,\ j=1,\ldots,4;&y_{i}(\bm{x}_{j})\sim N(40,3^2),\ i=7,\ldots,10,\ j=1,\ldots,4;\\
&y_{i}(\bm{x}_{j})\sim N(60,3^2),\ i=1,\ldots,6,\ j=5,\ldots,10;&y_{i}(\bm{x}_{j})\sim N(80,3^2),\ i=7,\ldots,10,\ j=5,\ldots,10.
\end{eqnarray*}
That is to say both design dimension and context dimension have two clusters respectively. For design $i$ and context $j$, samples are drawn independently from a normal distribution $N(y_{i}(\bm{x}_{j}),\sigma_{i}^{2}(\bm{x}_{j}))$, where $\sigma_{i}(\bm{x}_{j})\sim U(8,12)$. Considering the linear dependency assumption in IZ and SUCB, we set context parameters $\bm{x}_{j}$ as single-dimensional variables generated as follows:
\begin{eqnarray*}
\bm{x}_{j}\sim N(4,1^2),\ j=1,\ldots,4;\ \bm{x}_{j}\sim N(6,1^2),\ j=5,\ldots,10.
\end{eqnarray*}
We set the number of initial replications as $n_{0}=5$ for each design-context pair. The other parameters involved in IZ are specified as follows: $\delta=0.1$ and the constant $h$ is computed by the numerical method in~\cite{shen2017ranking} when the target $\text{PCS}_{\text{W}}$ is 95\%. The tuning parameter $\gamma$ in SUCB is set as 1. The numbers of clusters ($K$ and $L$) are determined based on these initial replications, and the performance clustering phenomenon can been seen in Figure~\ref{performance_clustering2} and Figure~\ref{performance_clustering1}. 

\begin{figure}[htbp]
	\begin{center}
		\includegraphics[width=5in]{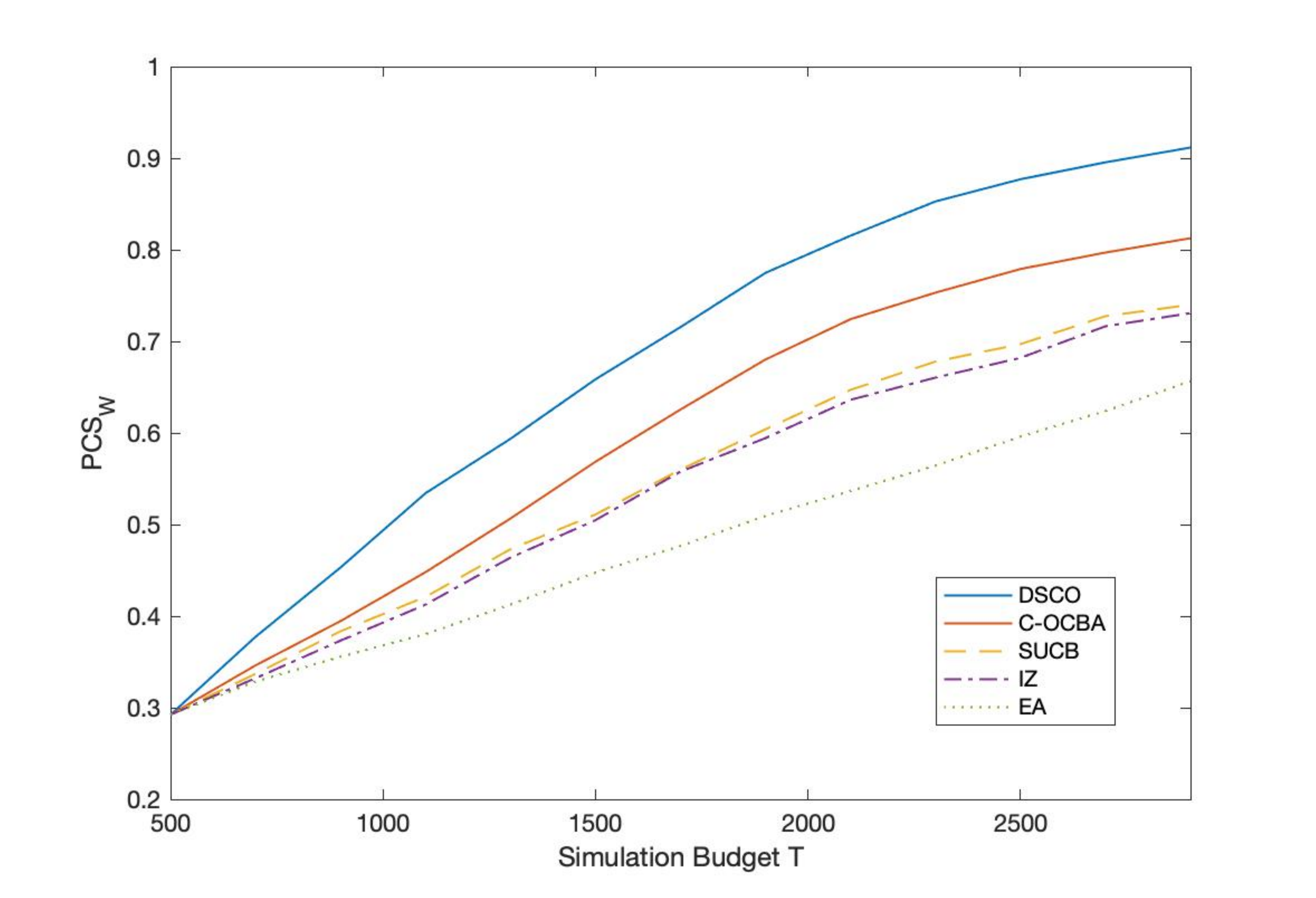}\\
		\caption{$\text{PCS}_{\text{W}}$ of the five sampling policies in one cluster case of Example 1.}\label{example1_1}
	\end{center}
\end{figure}

In Figure~\ref{example1_1}, we can see that DSCO and C-OCBA perform better than IZ, SUCB and EA, which could be attributed to the reason that EA utilizes no sample information, IZ only utilizes sample variances, and SUCB only utilizes sample means while the other two sampling policies utilize the information in the posterior means and variances. In order to attain $\text{PCS}_{\text{W}}=80\%$, the number of samples consumed by DSCO is 2000, while EA, IZ, SUCB and C-OCBA require more than 2800 samples. That is to say DSCO reduces the sampling budget by more than 28\%. Note that there is no clear performance clustering structure in one cluster case. The performance enhancement of DSCO could be attributed to its stochastic dynamic programming framework which formulates the optimal decision under finite sampling budget. On the contrary, the asymptotically optimal sampling ratio in C-OCBA does not have a
theoretical support for the finite-sample performance.

\begin{figure}[htbp]
	\begin{center}
		\includegraphics[width=5in]{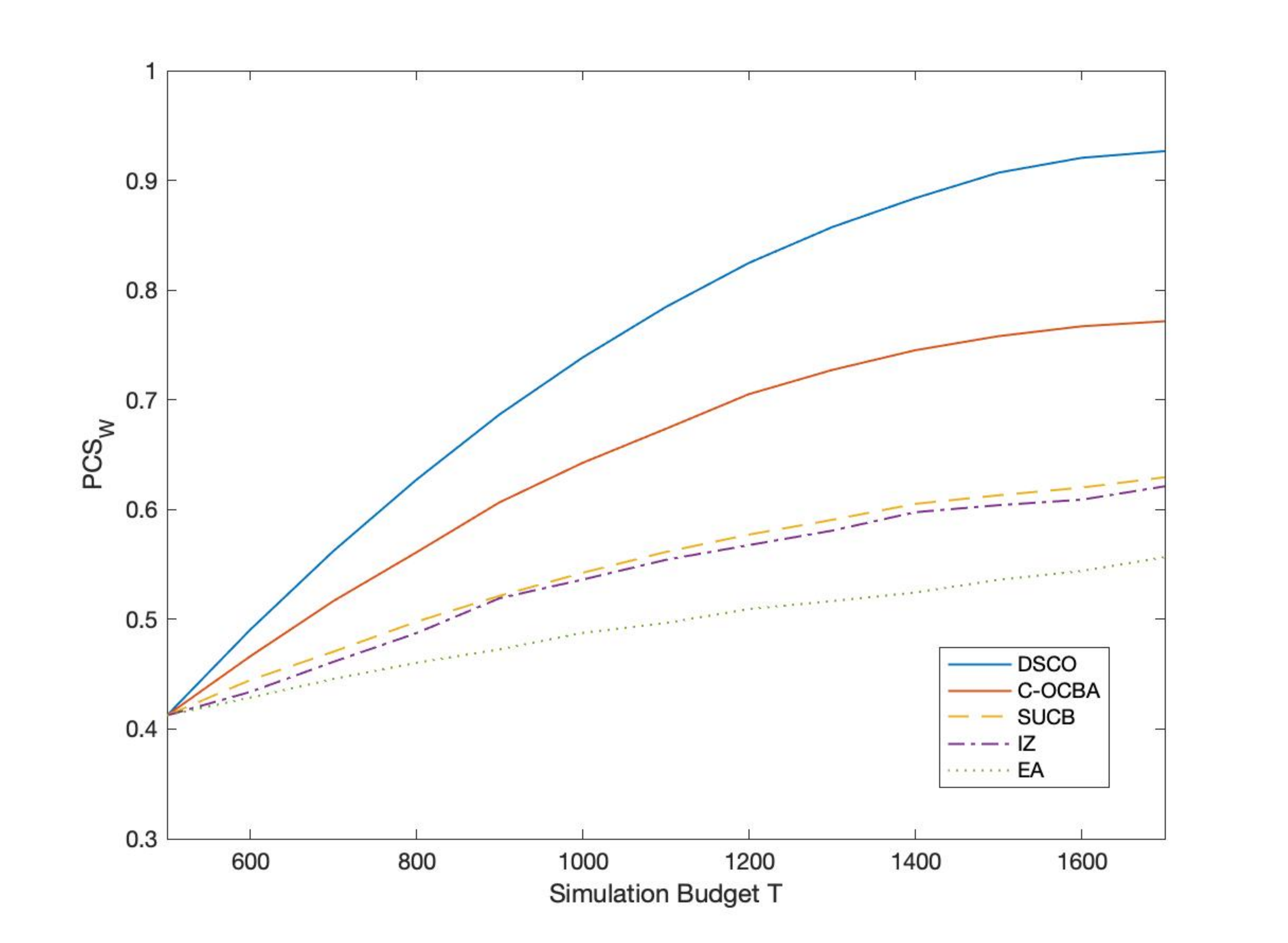}\\
		\caption{$\text{PCS}_{\text{W}}$ of the five sampling policies in multiple clusters case of Example 1.}\label{example1_2}
	\end{center}
\end{figure}

In Figure~\ref{example1_2}, we can see that DSCO performs significantly better than the other four sampling policies in multiple clusters case. DSCO needs 1400 samples to attain $\text{PCS}_{\text{W}}=90\%$, whereas EA, IZ, SUCB and C-OCBA cannot achieve the same $\text{PCS}_{\text{W}}$ even when simulation budget is 1700. Compared with one cluster case, the advantage of DSCO over C-OCBA increases as performance clustering phenomena becomes apparent. Besides stochastic dynamic programming framework, the performance enhancement of DSCO could be attributed to the benefit of using clustering information which provides useful global information for learning the best designs. DSCO allocates more samples to clusters 2 and 4 where designs have better performances and thus are expected to competitors of the best designs.

\subsubsection{Example 2: $30\times30$ design-context pair}
\begin{figure}[htbp]
	\begin{center}
		\includegraphics[width=3in]{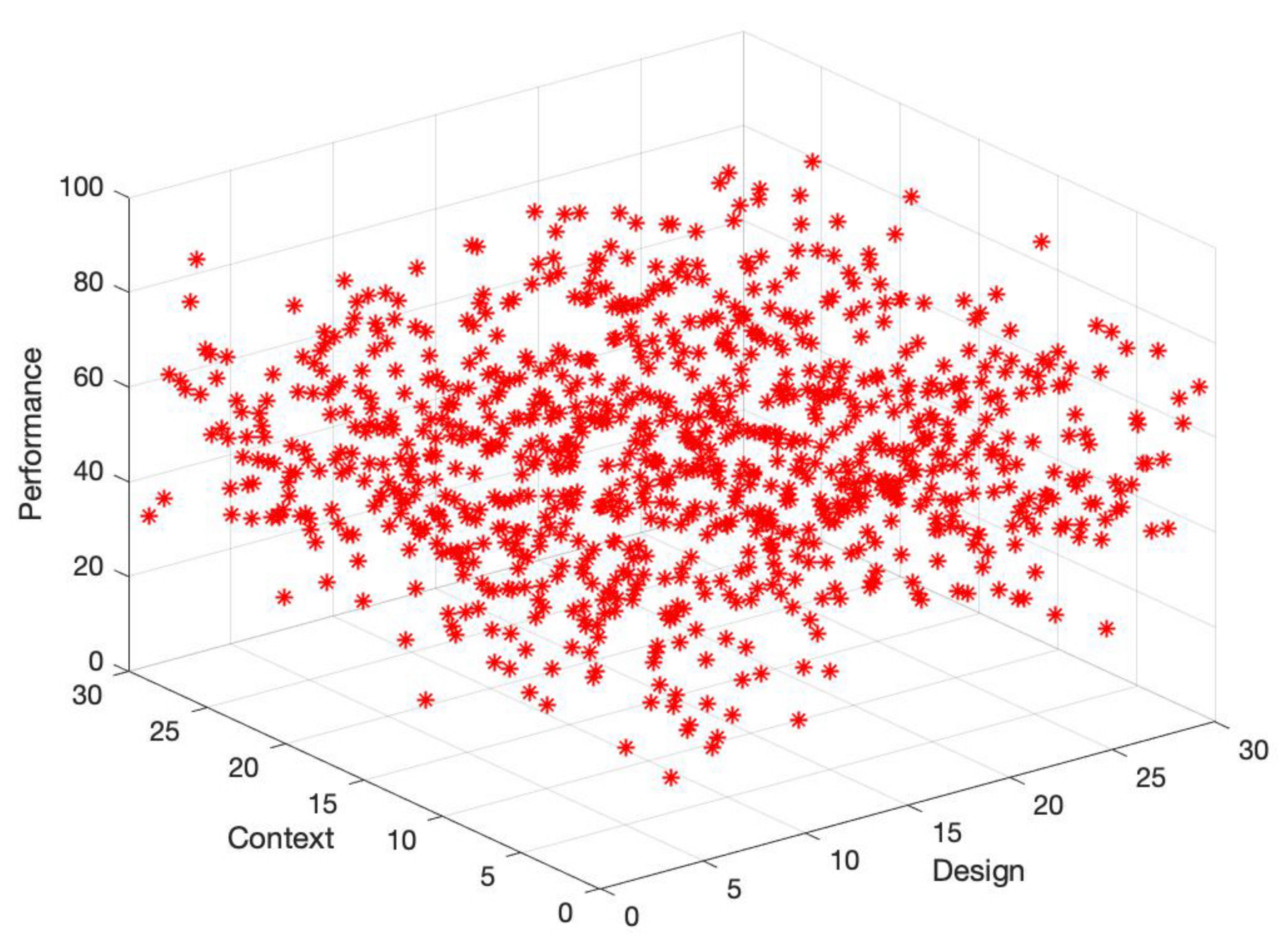}\\
		\caption{Performance clustering in one cluster case.}\label{performance_clustering4}
	\end{center}
\end{figure}

\begin{figure}[htbp]
	\begin{center}
		\includegraphics[width=3in]{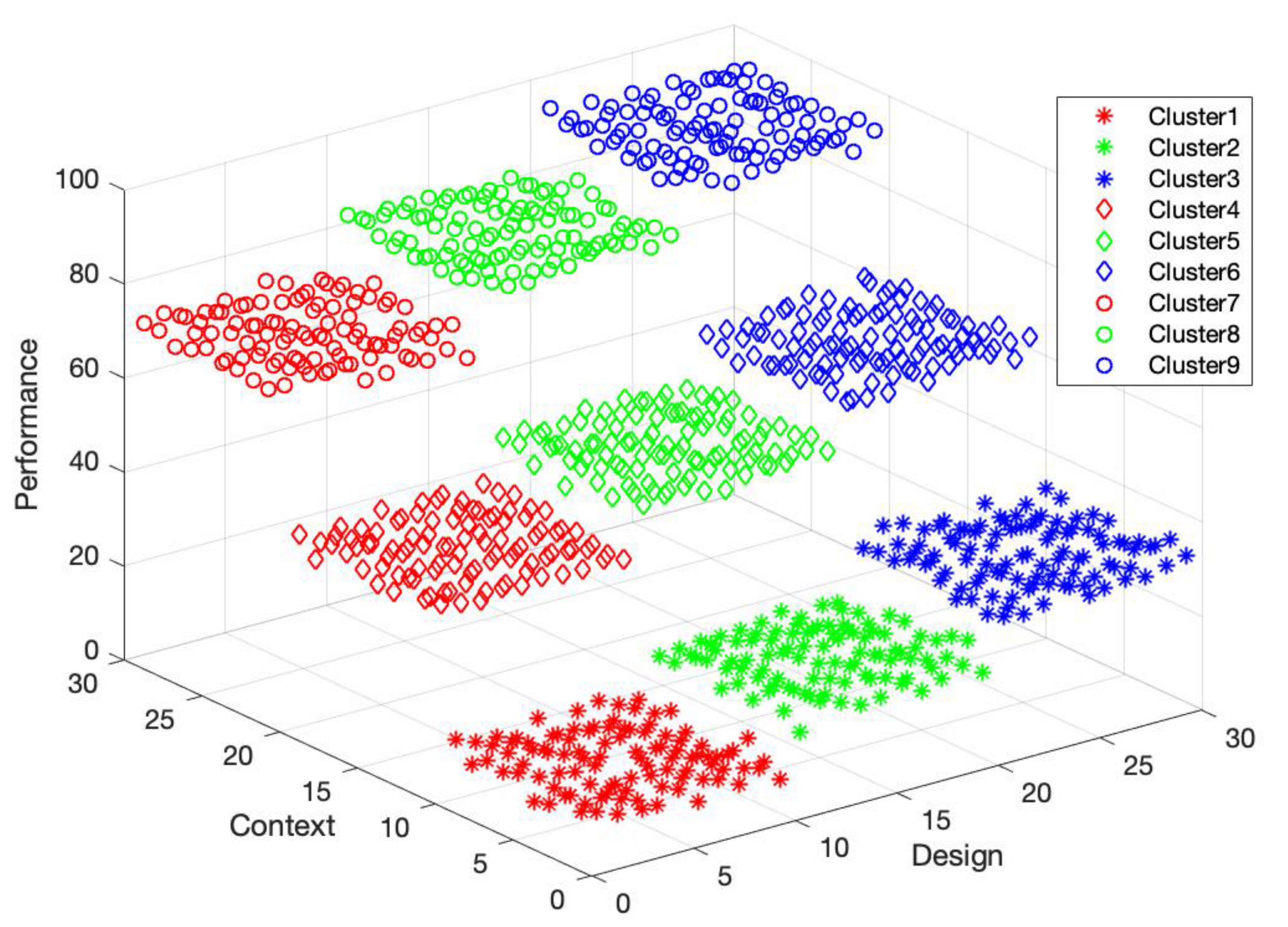}\\
		\caption{Performance clustering in multiple clusters case.}\label{performance_clustering3}
	\end{center}
\end{figure}

In this example, our proposed DSCO is tested in a larger synthetic case with 30 designs and 30 contexts. In one cluster case, the performances of each design for each context are generated as follows:
\begin{gather*}
y_{i}(\bm{x}_{j})\sim N(50,15^2),\ i=1,\ldots,30,\ j=1,\ldots,30.
\end{gather*}
Context parameters $\bm{x}_{j}$ are set as single-dimensional variables drawn from $N(5,1^2)$. As for multiple clusters case,
the performances of each design for each context are generated as follows:
\begin{gather*}
y_{i}(\bm{x}_{j})\sim N(10,1.5^2),\ i=1,\ldots,10,\ j=1,\ldots,10;\\
y_{i}(\bm{x}_{j})\sim N(20,1.5^2),\ i=11,\ldots,20,\ j=1,\ldots,10;\ y_{i}(\bm{x}_{j})\sim N(30,1.5^2),\ i=21,\ldots,30,\ j=1,\ldots,10;\\
y_{i}(\bm{x}_{j})\sim N(40,1.5^2),\ i=1,\ldots,10,\ j=11,\ldots,20;\ y_{i}(\bm{x}_{j})\sim N(50,1.5^2),\ i=11,\ldots,20,\ j=11,\ldots,20;\\
y_{i}(\bm{x}_{j})\sim N(60,1.5^2),\ i=21,\ldots,30,\ j=11,\ldots,20;\ y_{i}(\bm{x}_{j})\sim N(70,1.5^2),\ i=1,\ldots,10,\ j=21,\ldots,30;\\
y_{i}(\bm{x}_{j})\sim N(80,1.5^2),\ i=11,\ldots,20,\ j=21,\ldots,30;\ y_{i}(\bm{x}_{j})\sim N(90,1.5^2),\ i=21,\ldots,30,\ j=21,\ldots,30.
\end{gather*}
That is to say both design dimension and context dimension have 3 clusters respectively. For design $i$ and context $j$, samples are drawn independently from a normal distribution $N(y_{i}(\bm{x}_{j}),\sigma_{i}^{2}(\bm{x}_{j}))$, where $\sigma_{i}(\bm{x}_{j})\sim U(4,6)$. Considering the linear dependency assumption in IZ and SUCB, we set context parameters $\bm{x}_{j}$ as single-dimensional variables generated as follows:
\begin{eqnarray*}
\bm{x}_{j}\sim N(2,1^2),\ j=1,\ldots,10;\ \bm{x}_{j}\sim N(5,1^2),\ j=11,\ldots,20;\ \bm{x}_{j}\sim N(8,1^2),\ j=21,\ldots,30.
\end{eqnarray*}
The other parameters ($n_{0},\delta,h,\gamma$) are the same as the last one. The numbers of clusters ($K$ and $L$) are determined based on these initial replications, and the performance clustering phenomenon can been seen in Figure~\ref{performance_clustering4} and Figure~\ref{performance_clustering3}.

\begin{figure}[htbp]
	\begin{center}
		\includegraphics[width=5in]{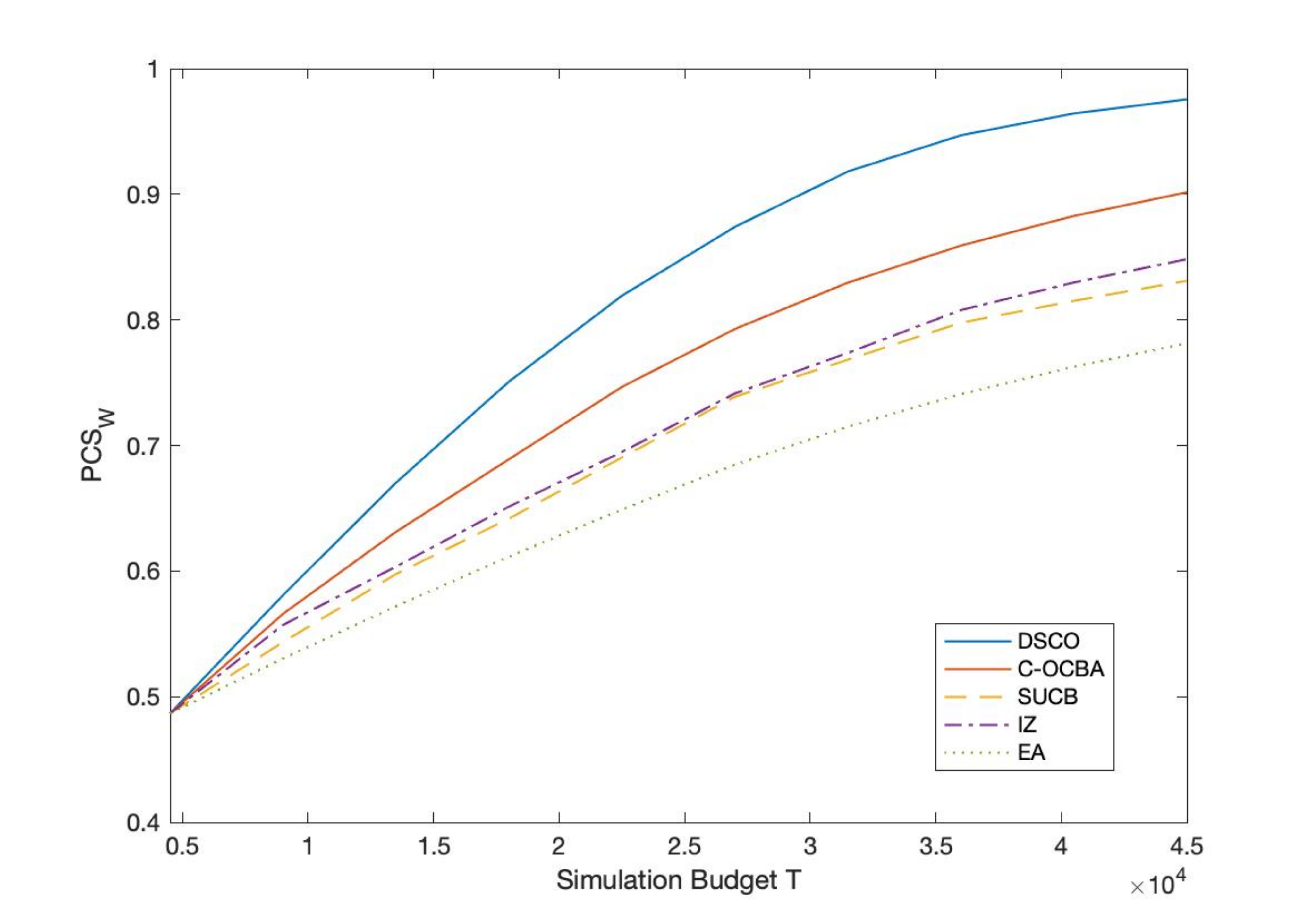}\\
		\caption{$\text{PCS}_{\text{W}}$ of the five sampling policies in one cluster case of Example 2.}\label{example2_1}
	\end{center}
\end{figure}

\begin{figure}[htbp]
	\begin{center}
		\includegraphics[width=5in]{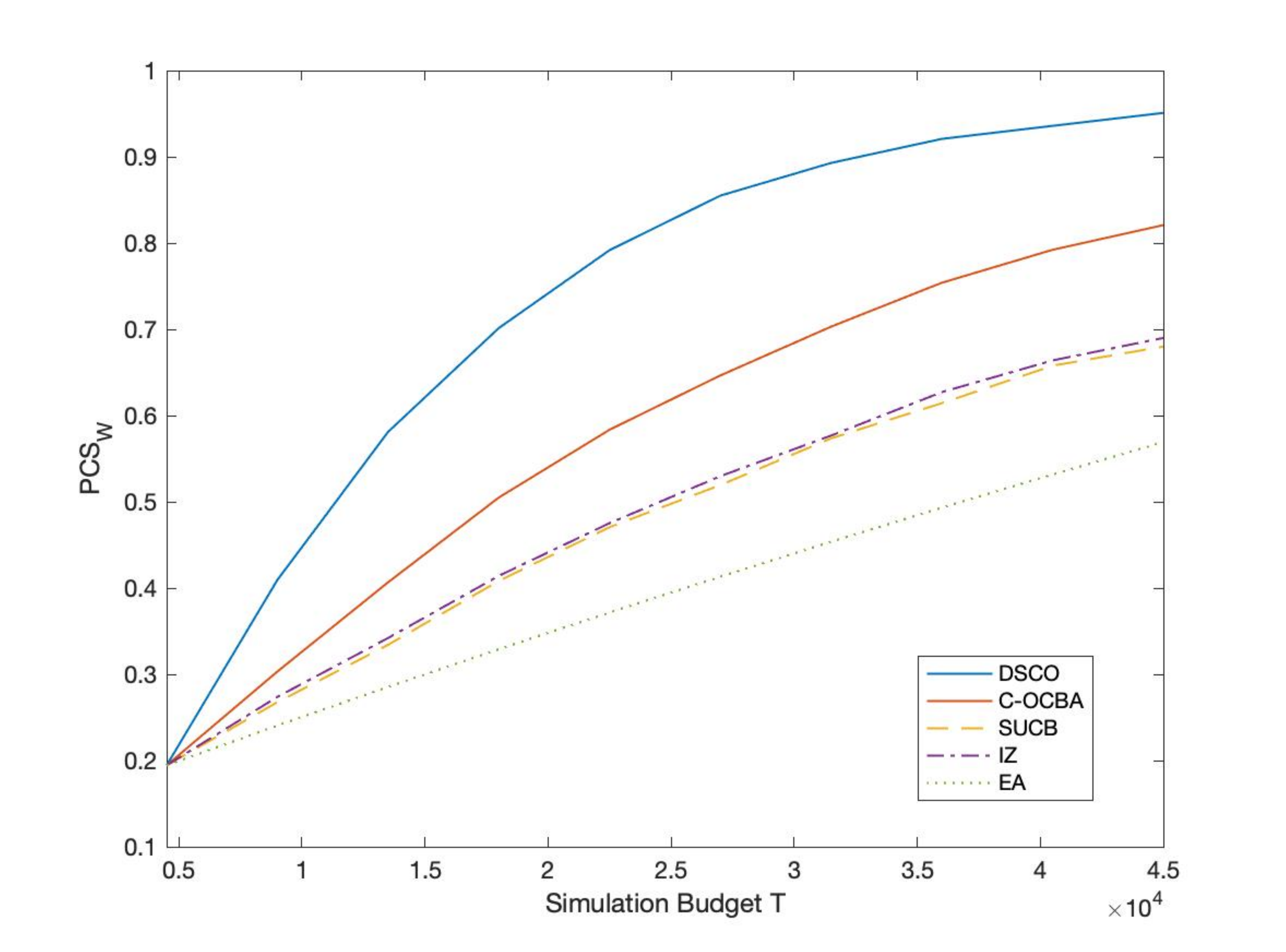}\\
		\caption{$\text{PCS}_{\text{W}}$ of the five sampling policies in multiple clusters case of Example 2.}\label{example2_2}
	\end{center}
\end{figure}

Figures~\ref{example2_1} and~\ref{example2_2} illustrate the performance of the five sampling policies. Similar to Example 1, DSCO remains as the most efficient sampling policy among the five, and C-OCBA is better than EA, IZ, and SUCB. However, it can be noticed that the advantage of DSCO is more significant when the scale of the problem (the number of designs, contexts, and clusters) grows large. In order to attain
$\text{PCS}_{\text{W}}=90\%$ in one cluster case, DSCO consumes less than 30,000 samples, while EA, IZ, SUCB and C-OCBA require more than 45,000 samples. That is to say DSCO reduces the sampling budget by more than 33\%. In multiple clusters case, DSCO needs 33,000 samples to attain $\text{PCS}_{\text{W}}=90\%$, whereas EA, IZ, SUCB and C-OCBA cannot achieve the same $\text{PCS}_{\text{W}}$ even when simulation budget is 45,000. DSCO allocates more samples to clusters 3, 6, and 9 where designs have better performances and thus are expected to competitors of the best designs.

\subsection{Cancer Prevention Treatment Example}
\begin{figure}[htbp]
	\begin{center}
		\includegraphics[width=6in]{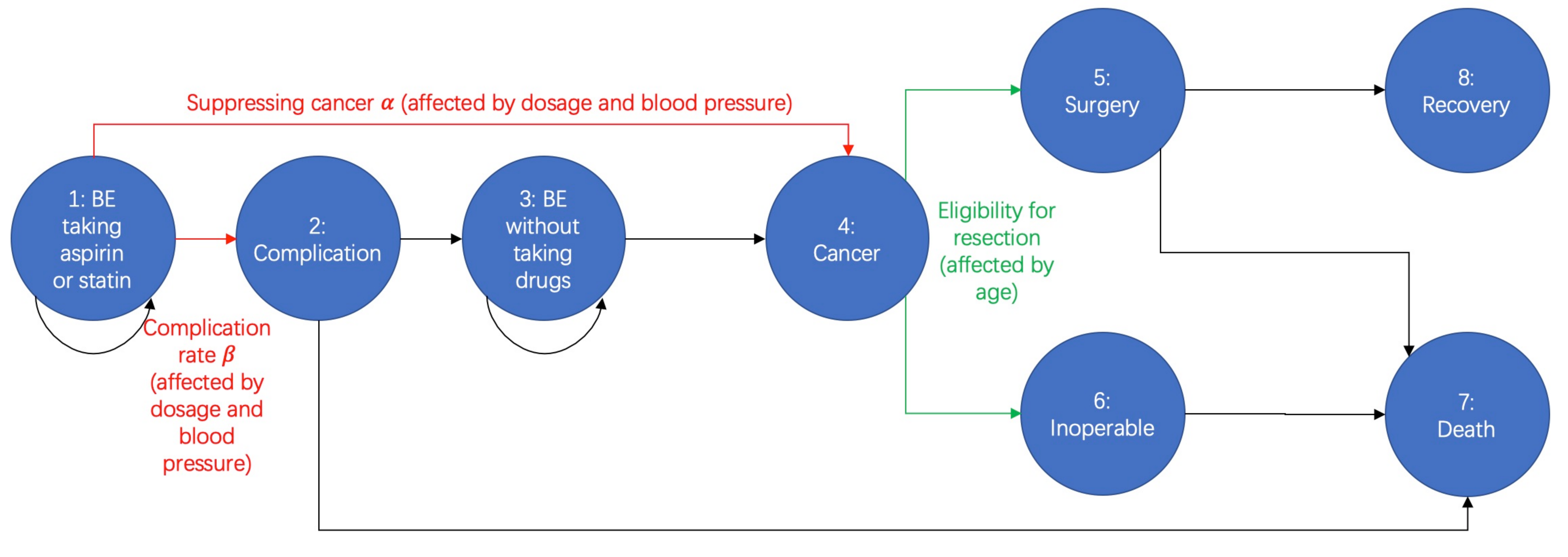}\\
		\caption{State of the Markov chain.\protect\footnotemark[1]}\label{markov}
	\end{center}
\end{figure}

Non-steroidal anti-inflammatory drugs such as aspirin and statin can prevent the progression of Barrett’s esophagus (BE) to adenocarcinoma, which is a main sub-type of esophageal cancer. However, use of such drugs is associated with numerous potential complications, including gastrointestinal bleeding and hemorrhagic strokes. In this example, we use a Markov chain as shown in Figure~\ref{markov} to capture the state transition in cancer prevention. Inputs of the simulation model contain design parameters (drug dosages) and context parameters (patient's characteristics), and some transition probabilities are dependent on these inputs, e.g., red colored transitions in Figure~\ref{markov} are context-dependent. For patients clustered by certain characteristic, we aim to determine the optimal drug dosage for their cancer prevention treatment. This example has also been considered in~\cite{shen2017ranking} and~\cite{gao2019selecting}. Parameters in the probability transition matrix are set based on~\cite{hur2004cost} and~\cite{choi2014statins}.
\footnotetext[1]{(1) The transitions from each state to death, which is due to all-cause mortality, are omitted in this illustration. (2) The time duration between state transitions is one month. (3) The details of the transition probabilities depends on whether aspirin chemoprevention or statin chemoprevention is used.}

Drug can reduce the probability of canceration, i.e., in Figure~\ref{markov}, the transition probability from state 1 to 4 is less than that from state 3 to 4. Then we denote $\alpha=P_{1,4}/P_{3,4}$ as the drug effect. However, drug can also cause complication, and complication rate $\beta$ is defined as the transition probability from state 1 to 2 in Figure~\ref{markov}. Different drug dosages have different drug effects and complication rates, and their relationship is set as follows:
\begin{eqnarray*}
\alpha_{\text{aspirin}}&=&0.5+(a-75)\times0.003-(b-120)\times0.005,\ a\in[50,150],\ b\in[110,150],\\
\beta_{\text{aspirin}}&=&0.025+(a-75)\times0.0005-(b-120)\times0.001,\ a\in[50,150],\ b\in[110,150],\\
\alpha_{\text{statin}}&=&0.5+(a-9)\times0.0417-(b-120)\times0.0025,\ a\in[6,18],\ b\in[110,150],\\
\beta_{\text{statin}}&=&0.04+(a-9)\times0.01+(b-120)\times0.001,\ a\in[6,18],\ b\in[110,150],
\end{eqnarray*}
where $a$ is drug dosage (mg) and $b$ is systolic blood pressure (mmHg). In the above formula, based on~\cite{hur2004cost}, we set the standard $\alpha$ and $\beta$ of aspirin as 0.5 and 0.025 under standard dosage $a=75$mg and normal pressure $b=120$mmHg, and set the standard $\alpha$ and $\beta$ of statin as 0.5 and 0.04 under standard dosage $a=9$mg and normal pressure $b=120$mmHg. Other coefficients in the above formula are determined by linear regression and (linearly interpolating) few observations already available in the literature. In this example, we consider 40 feasible candidate treatment strategies (designs). Half of them use aspirin, and their dosages are set to be $\{52.5,57.5,\ldots,142.5,147.5\}$; the other half use statin, and theirs dosages are set to be $\{6.2,6.8,\ldots,17.0,17.6\}$.

Patients' characteristics are denoted by ($x_{1},x_{2}$), where $x_{1}\in[45,80]$ is the starting age of a treatment and $x_{2}\in[110,150]$ is systolic blood pressure. All-cause mortality $\lambda$ denotes the transition probability from each state to death, which is age-related and set as
$$\lambda=\frac{1}{12\times(85-x_{1})}.$$
This formula is derived from a geometric distribution and a life expectancy of 85 years. Eligibility for resection determines the transitions from state 4 to state 5 or 6, which is affected by $x_{1}$ as follows:
$$P_{4,5}=(1-\lambda)\times[1-(x_{1}-45)\times0.00225].$$
The coefficients in the above formula are determined by the results of~\cite{hur2004cost}, stating that 100\% of patients at age 45 are eligible for resection while only 91.9\% of patients at age 81 are eligible. In general, an older age has a lower eligibility for resection and a higher death rate from all-cause mortality. As for $x_{2}$, it will affect drug effect and complication rate as $b$ in the expressions of $\alpha$ and $\beta$. We expect to find some clusters in patients' characteristics. We take 60 possible values of ($x_{1},x_{2}$) as contexts of interest. The performance of a treatment strategy is measured by quality-adjusted life years,
$$y_{i}(\bm{x}_{j})=\underset{N\to+\infty}{\lim}\mathbb{E}[\sum_{t=1}^{N}Q_{t}(a_{i},\bm{x}_{j})],$$
where $Q_{t}(\cdot)$ is the quality of life at time period $t$.
The quality of life $Q_{t}(\cdot)$ takes 1 as the initial value, then will have a 50\% discount after the development into cancer and a 3\% extra discount after surgery, and takes 0 for death. The initial number of simulation replications $n_{0}$ is set to be 10. The other parameters involved in IZ are specified as follows: $\delta=0.2$ and the constant $h$ is computed when the target $\text{PCS}_{\text{W}}$ is 95\%. The tuning parameter $\gamma$ in SUCB is set as 1.

After conducting DSCO, we obtain 4 clusters of designs and 6 clusters of contexts by the posterior probabilities of clustering $\widehat{z}$ and $\widehat{v}$. Tables~\ref{design} and~\ref{context} summarize common statistics (mean and standard deviation) of drug dosage, age, and blood pressure in each cluster. As shown in Table~\ref{design}, each design cluster only contains one type of drug, that is, DSCO can 
distinguish different drugs. Treatment strategies using the same drug are classified by DSCO into two levels of dosage. As shown in Table~\ref{context}, DSCO classifies all contexts into three clusters of age and two clusters of blood pressure. Specifically, Context Clusters 1 and 2 contain the age group of fifties, Context Clusters 3 and 4 contain the age group of sixties, and Context Clusters 5 and 6 contain the age group of seventies; Context Clusters 1, 3, and 5 contain patients with normal blood pressure while Context Clusters 2, 4, and 6 contain patients with
hypertension.

\begin{figure}[htbp]
	\begin{center}
		\includegraphics[width=5in]{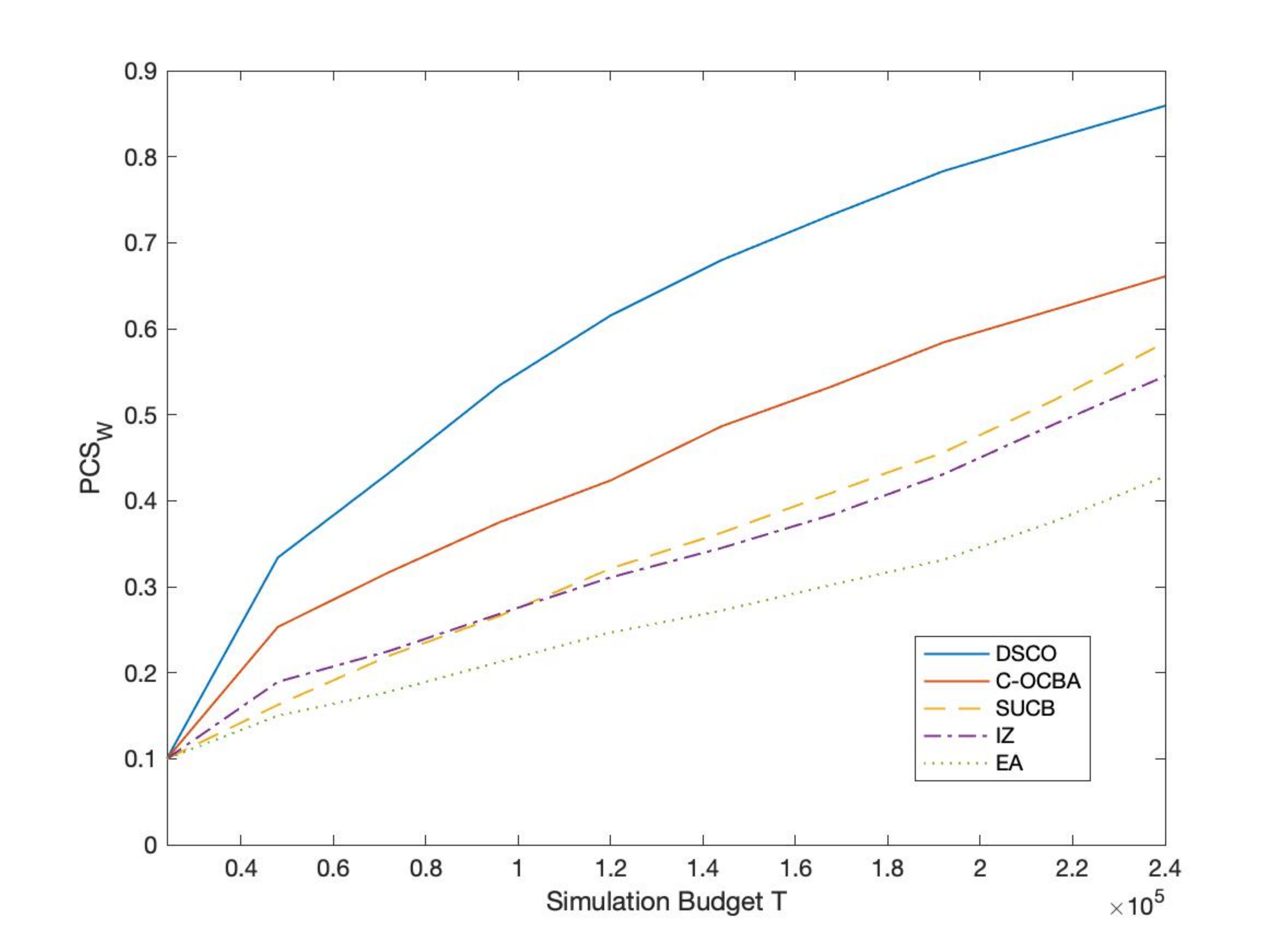}\\
		\caption{$\text{PCS}_{\text{W}}$ of the five sampling policies for cancer prevention treatment.}\label{example3}
	\end{center}
\end{figure}

DSCO allocates more samples to design-context pairs in Design Cluster 1 (relatively-high-dose aspirin) and Context Clusters 2, 4, and 6 (hypertension) where designs have better performances, which is in accord with the fact that aspirin significantly reduces major cardiovascular events with the greatest benefit seen in all myocardial infarction~\citep{hansson1998effects}. The performance comparison is shown in Figure~\ref{example3}. DSCO outperforms EA, IZ, SUCB and C-OCBA. The relative performances of the five compared sampling procedures are similar to those in the two synthetic cases.

\section{Conclusions}\label{sec:conclusion}
This paper studies a sample allocation problem for context-dependent R\&S. We take the performance cluster information
into consideration, and utilize a Gaussian mixture model as a priori. Under a Bayesian framework, we update model parameters and posterior estimates, and formulate the sequential sampling decision as a stochastic dynamic programming problem. We propose an efficient sampling policy named DSCO, which simultaneously learns the global clustering information and local performance information in design-context pairs. The proposed sampling policy is proved to be consistent and achieve the asymptotically optimal sampling ratio.
Numerical experiments demonstrate that DSCO can significantly enhance the efficiency for learning the best designs in all contexts by using performance cluster information.
\ACKNOWLEDGMENT{This work was supported in part by the National Science Foundation of China (NSFC) under Grants 71901003 and 72022001, by the National Science Foundation under Awards ECCS-1462409, CMMI-1462787, CAREER CMMI-1834710 and IIS-1849280, and by the scholarship from China Scholarship Council (CSC) under the Grant CSC. A preliminary version of this work has been published in Proceedings of 2020 Winter Simulation Conference (\citealp{li2020context}).}


\bibliographystyle{informs2014} 
\bibliography{ref}
\ECSwitch


\ECHead{Proofs and Supplementary Materials}

\section{Proof of Theorem~\ref{thm_EM}}
\proof{Proof of Theorem~\ref{thm_EM}}
The log-likelihood of the complete state variables has the following form:
\begin{eqnarray}
&&\log\mathcal{L}(\mathcal{E}_{t},\bm{y},Z,V;\theta)\nonumber\\
&=&\left(\sum_{i=1}^{n}\sum_{k=1}^{K}z_{i,k}\log\tau_{k}\right)+\left(\sum_{j=1}^{m}\sum_{\ell=1}^{L}v_{j,\ell}\log\omega_{\ell}\right)+\sum_{i=1}^{n}\sum_{j=1}^{m}\left\{-\sum_{h=1}^{t_{ij}}\left[\frac{1}{2}\log(2\pi\sigma_{i}^{2}(\bm{x}_{j}))+\frac{(Y_{i,h}(\bm{x}_{j})-y_{i}(\bm{x}_{j}))^{2}}{2\sigma_{i}^{2}(\bm{x}_{j})}\right]\right.\nonumber\\
&&\left.-\sum_{k=1}^{K}\sum_{\ell=1}^{L}\left[\frac{1}{2}z_{i,k}v_{j,\ell}\log(2\pi\sigma_{k\ell}^{2})+z_{i,k}v_{j,\ell}\frac{(y_{i}(\bm{x}_{j})-\mu_{k\ell})^{2}}{2\sigma_{k\ell}^{2}}\right]\right\}.\label{eq.log-likelihood}
\end{eqnarray}

Given $\widehat{\theta}^{(t,s)}$, an analytical form for the likelihood of the observations can be obtained by integrating out $\bm{y}$, $Z$, and $V$ as follow:
\begin{eqnarray*}
&&\mathcal{L}(\mathcal{E}_{t};\widehat{\theta}^{(t,s)})\\
&=&\sum_{k_{1:n}\in\mathcal{K}}\sum_{\ell_{1:m}\in\mathcal{L}}\left(\prod_{i=1}^{n}\widehat{\tau}_{k_{i}}^{(t,s)}\right)\left(\prod_{j=1}^{m}\widehat{\omega}_{\ell_{j}}^{(t,s)}\right)\prod_{i=1}^{n}\prod_{j=1}^{m}\left[\int_{\mathbb{R}}\phi\left(y_{i}(\bm{x}_{j})|\widehat{\mu}_{k_{i}\ell_{j}}^{(t,s)},(\widehat{\sigma}_{k_{i}\ell_{j}}^{2})^{(t,s)}\right)\prod_{h=1}^{t_{ij}}\phi\left(Y_{i,h}(\bm{x}_{j})|y_{i}(\bm{x}_{j}),\sigma_{i}^{2}(\bm{x}_{j})\right)dy_{i}(\bm{x}_{j})\right]\\
&=&\sum_{k_{1:n}\in\mathcal{K}}\sum_{\ell_{1:m}\in\mathcal{L}}\left(\prod_{i=1}^{n}\widehat{\tau}_{k_{i}}^{(t,s)}\right)\left(\prod_{j=1}^{m}\widehat{\omega}_{\ell_{j}}^{(t,s)}\right)\prod_{i=1}^{n}\prod_{j=1}^{m}\left[\frac{1}{\sqrt{2\pi(\widehat{\sigma}_{k_{i}\ell_{j}}^{2})^{(t,s)}}}\prod_{h=1}^{t_{ij}}\frac{1}{\sqrt{2\pi\sigma_{i}^{2}(\bm{x}_{j})}}\int_{\mathbb{R}}\exp\left\{-\frac{1}{2}\left[\frac{(y_{i}(\bm{x}_{j})-\widehat{\mu}_{k_{i}\ell_{j}}^{(t,s)})^{2}}{(\widehat{\sigma}_{k_{i}\ell_{j}}^{2})^{(t,s)}}\right.\right.\right.\\
&&\left.\left.\left.+\sum_{h=1}^{t_{ij}}\frac{(Y_{i,h}(\bm{x}_{j})-y_{i}(\bm{x}_{j}))^{2}}{\sigma_{i}^{2}(\bm{x}_{j})}\right]\right\}dy_{i}(\bm{x}_{j})\right]\\
&=&\sum_{k_{1:n}\in\mathcal{K}}\sum_{\ell_{1:m}\in\mathcal{L}}\left(\prod_{i=1}^{n}\widehat{\tau}_{k_{i}}^{(t,s)}\right)\left(\prod_{j=1}^{m}\widehat{\omega}_{\ell_{j}}^{(t,s)}\right)\prod_{i=1}^{n}\prod_{j=1}^{m}\left[C_{ij,k_{i}\ell_{j}}^{(t,s)}\int_{\mathbb{R}}\phi(y_{i}(\bm{x}_{j})|\mu_{ij,k_{i}\ell_{j}}^{(t,s)},(\sigma_{ij,k_{i}\ell_{j}}^{2})^{(t,s)})dy_{i}(\bm{x}_{j})\right]\\
&=&\sum_{k_{1:n}\in\mathcal{K}}\sum_{\ell_{1:m}\in\mathcal{L}}\left(\prod_{i=1}^{n}\widehat{\tau}_{k_{i}}^{(t,s)}\right)\left(\prod_{j=1}^{m}\widehat{\omega}_{\ell_{j}}^{(t,s)}\right)\prod_{i=1}^{n}\prod_{j=1}^{m}C_{ij,k_{i}\ell_{j}}^{(t,s)},
\end{eqnarray*}
where
$$(\sigma_{ij,k_{i}\ell_{j}}^{2})^{(t,s)}=1\Big/\left[\frac{t_{ij}}{\sigma_{i}^{2}(\bm{x}_{j})}+\frac{1}{(\widehat{\sigma}_{k_{i}\ell_{j}}^{2})^{(t,s)}}\right],$$
$$\mu_{ij,k_{i}\ell_{j}}^{(t,s)}=(\sigma_{ij,k_{i}\ell_{j}}^{2})^{(t,s)}\left[\frac{\sum_{h=1}^{t_{ij}}Y_{i,h}(\bm{x}_{j})}{\sigma_{i}^{2}(\bm{x}_{j})}+\frac{\widehat{\mu}_{k_{i}\ell_{j}}^{(t,s)}}{(\widehat{\sigma}_{k_{i}\ell_{j}}^{2})^{(t,s)}}\right],$$
and $$C_{ij,k_{i}\ell_{j}}^{(t,s)}=\left(\frac{1}{2\pi\sigma_{i}^{2}(\bm{x}_{j})}\right)^{\frac{t_{ij}}{2}}\sqrt{\frac{(\sigma_{ij,k_{i}\ell_{j}}^{2})^{(t,s)}}{(\widehat{\sigma}_{k_{i}\ell_{j}}^{2})^{(t,s)}}}\exp\left\{\frac{1}{2}\left[\frac{(\mu_{ij,k_{i}\ell_{j}}^{(t,s)})^{2}}{(\sigma_{ij,k_{i}\ell_{j}}^{2})^{(t,s)}}-\frac{\sum_{h=1}^{t_{ij}}Y_{i,h}^{2}(\bm{x}_{j})}{\sigma_{i}^{2}(\bm{x}_{j})}-\frac{(\widehat{\mu}_{k_{i}\ell_{j}}^{(t,s)})^{2}}{(\widehat{\sigma}_{k_{i}\ell_{j}}^{2})^{(t,s)}}\right]\right\}.$$
By the Bayes' rule, the posterior distribution of $\{z_{i,k}=1\}$ conditional on $\mathcal{E}_{t}$ and given $\widehat{\theta}^{(t,s)}$ is
$$\widehat{z}_{i,k}^{(t,s)}=\frac{\sum_{k_{1:n}\in\mathcal{K},k_{i}=k}\sum_{\ell_{1:m}\in\mathcal{L}}\left(\prod_{i=1}^{n}\widehat{\tau}_{k_{i}}^{(t,s)}\right)\left(\prod_{j=1}^{m}\widehat{\omega}_{\ell_{j}}^{(t,s)}\right)\prod_{i=1}^{n}\prod_{j=1}^{m}C_{ij,k_{i}\ell_{j}}^{(t,s)}}{\sum_{k_{1:n}\in\mathcal{K}}\sum_{\ell_{1:m}\in\mathcal{L}}\left(\prod_{i=1}^{n}\widehat{\tau}_{k_{i}}^{(t,s)}\right)\left(\prod_{j=1}^{m}\widehat{\omega}_{\ell_{j}}^{(t,s)}\right)\prod_{i=1}^{n}\prod_{j=1}^{m}C_{ij,k_{i}\ell_{j}}^{(t,s)}},$$
the posterior distribution of $\{v_{j,\ell}=1\}$ conditional on $\mathcal{E}_{t}$ and given $\widehat{\theta}^{(t,s)}$ is
$$\widehat{v}_{j,\ell}^{(t,s)}=\frac{\sum_{k_{1:n}\in\mathcal{K}}\sum_{\ell_{1:m}\in\mathcal{L},\ell_{j}=\ell}\left(\prod_{i=1}^{n}\widehat{\tau}_{k_{i}}^{(t,s)}\right)\left(\prod_{j=1}^{m}\widehat{\omega}_{\ell_{j}}^{(t,s)}\right)\prod_{i=1}^{n}\prod_{j=1}^{m}C_{ij,k_{i}\ell_{j}}^{(t,s)}}{\sum_{k_{1:n}\in\mathcal{K}}\sum_{\ell_{1:m}\in\mathcal{L}}\left(\prod_{i=1}^{n}\widehat{\tau}_{k_{i}}^{(t,s)}\right)\left(\prod_{j=1}^{m}\widehat{\omega}_{\ell_{j}}^{(t,s)}\right)\prod_{i=1}^{n}\prod_{j=1}^{m}C_{ij,k_{i}\ell_{j}}^{(t,s)}},$$
and given $\widehat{\theta}^{(t,s)}$, the posterior distribution of $y_{i}(\bm{x}_{j})$ conditional on $\{z_{i,k}=1\}$, $\{v_{j,\ell}=1\}$, and $\mathcal{E}_{t}$ is
$$\phi(y_{i}(\bm{x}_{j})|\mu_{ij,k\ell}^{(t,s)},(\sigma_{ij,k\ell}^{2})^{(t,s)}).$$

From the log-likelihood of complete state variables given by~(\ref{eq.log-likelihood}), we have
\begin{eqnarray*}
&&\mathcal{Q}(\theta|\widehat{\theta}^{(t,s)})\\
&=&\mathbb{E}\left[\log\mathcal{L}(\mathcal{E}_{t},\bm{y},Z,V;\theta)|\mathcal{E}_{t},\widehat{\theta}^{(t,s)}\right]\\
&=&\mathbb{E}\left[\left(\sum_{i=1}^{n}\sum_{k=1}^{K}z_{i,k}\log\tau_{k}\right)+\left(\sum_{j=1}^{m}\sum_{\ell=1}^{L}v_{j,\ell}\log\omega_{\ell}\right)-\sum_{i=1}^{n}\sum_{j=1}^{m}\sum_{k=1}^{K}\sum_{\ell=1}^{L}\left[\frac{1}{2}z_{i,k}v_{j,\ell}\log(2\pi\sigma_{k\ell}^{2})\right.\right.\\
&&\left.\left.+z_{i,k}v_{j,\ell}\frac{(y_{i}(\bm{x}_{j})-\mu_{k\ell})^{2}}{2\sigma_{k\ell}^{2}}\right]\Bigg|\mathcal{E}_{t},\widehat{\theta}^{(t,s)}\right]+C^{(t,s)},
\end{eqnarray*}
where $C^{(t,s)}$ is a constant independent of $\theta$. The estimate $\widehat{\tau}_{k}^{(t,s+1)}$ in the ($s+1$)-th iteration of the EM algorithm is obtained by solving the following optimization problem:
$$\underset{\tau}{\max}\sum_{i=1}^{n}\sum_{k=1}^{K}\widehat{z}_{i,k}^{(t,s)}\log\tau_{k},\ \text{s.t.}\sum_{k=1}^{K}\tau_{k}=1,\ \tau_{k}\geq0,$$
which is given by
$$\widehat{\tau}_{k}^{(t,s+1)}=\frac{\sum_{i=1}^{n}\widehat{z}_{i,k}^{(t,s)}}{n}.$$
Similarly,
$$\widehat{\omega}_{\ell}^{(t,s+1)}=\frac{\sum_{j=1}^{m}\widehat{v}_{j,\ell}^{(t,s)}}{m}.$$
Posterior estimate $\widehat{\mu}_{k\ell}^{(t,s+1)}$ is the solution of the follow equation:
$$\mathbb{E}\left[\sum_{i=1}^{n}\sum_{j=1}^{m}z_{i,k}v_{j,\ell}(y_{i}(\bm{x}_{j})-\mu_{k\ell})\Bigg|\mathcal{E}_{t},\widehat{\theta}^{(t,s)}\right]=\sum_{i=1}^{n}\sum_{j=1}^{m}\widehat{z}_{i,k}^{(t,s)}\widehat{v}_{j,\ell}^{(t,s)}\mathbb{E}\left[y_{i}(\bm{x}_{j})-\mu_{k\ell}\Big|z_{i,k}=1,v_{j,\ell}=1,\mathcal{E}_{t},\widehat{\theta}^{(t,s)}\right]=0,$$
which yields
$$\widehat{\mu}_{k\ell}^{(t,s+1)}=\frac{\sum_{i=1}^{n}\sum_{j=1}^{m}\widehat{z}_{i,k}^{(t,s)}\widehat{v}_{j,\ell}^{(t,s)}\mu_{ij,k\ell}^{(t,s)}}{\sum_{i=1}^{n}\sum_{j=1}^{m}\widehat{z}_{i,k}^{(t,s)}\widehat{v}_{j,\ell}^{(t,s)}}.$$
To calculate $(\widehat{\sigma}_{k\ell}^{2})^{(t,s+1)}$, we solve the follow equation:
\begin{eqnarray*}
\sum_{i=1}^{n}\sum_{j=1}^{m}\widehat{z}_{i,k}^{(t,s)}\widehat{v}_{j,\ell}^{(t,s)}\mathbb{E}\left[1-\frac{(y_{i}(\bm{x}_{j})-\widehat{\mu}_{k\ell}^{(t,s+1)})^{2}}{\sigma_{k\ell}^{2}}\Bigg|z_{i,k}=1,v_{j,\ell}=1,\mathcal{E}_{t},\widehat{\theta}^{(t,s)}\right]=0,
\end{eqnarray*}
which leads to
$$(\widehat{\sigma}_{k\ell}^{2})^{(t,s+1)}=\frac{\sum_{i=1}^{n}\sum_{j=1}^{m}\widehat{z}_{i,k}^{(t,s)}\widehat{v}_{j,\ell}^{(t,s)}\left[(\sigma_{ij,k\ell}^{2})^{(t,s)}+\left(\mu_{ij,k\ell}^{(t,s)}-\widehat{\mu}_{k\ell}^{(t,s+1)}\right)^{2}\right]}{\sum_{i=1}^{n}\sum_{j=1}^{m}\widehat{z}_{i,k}^{(t,s)}\widehat{v}_{j,\ell}^{(t,s)}}.$$\Halmos
\endproof

\section{The asymptotic analysis of Theorem~\ref{thm_EM}}
\begin{proposition}\label{proposition_lim}
The posterior probability of $\{z_{i,k}=1\}$ conditional on $\bm{y}$ and given $\widehat{\theta}^{(s)}$ is
$$\widehat{z}_{i,k}^{(s)}=\frac{\sum_{k_{1:n}\in\mathcal{K},k_{i}=k}\sum_{\ell_{1:m}\in\mathcal{L}}f_{\tau}^{(s)}(k_{1:n})f_{\omega}^{(s)}(\ell_{1:m})f_{\bm{y}}^{(s)}(k_{1:n},\ell_{1:m})}{\sum_{k_{1:n}\in\mathcal{K}}\sum_{\ell_{1:m}\in\mathcal{L}}f_{\tau}^{(s)}(k_{1:n})f_{\omega}^{(s)}(\ell_{1:m})f_{\bm{y}}^{(s)}(k_{1:n},\ell_{1:m})},$$
the posterior probability of $\{v_{j,\ell}=1\}$ conditional on $\bm{y}$ and given $\widehat{\theta}^{(s)}$ is
$$\widehat{v}_{j,\ell}^{(s)}=\frac{\sum_{k_{1:n}\in\mathcal{K}}\sum_{\ell_{1:m}\in\mathcal{L},\ell_{j}=\ell}f_{\tau}^{(s)}(k_{1:n})f_{\omega}^{(s)}(\ell_{1:m})f_{\bm{y}}^{(s)}(k_{1:n},\ell_{1:m})}{\sum_{k_{1:n}\in\mathcal{K}}\sum_{\ell_{1:m}\in\mathcal{L}}f_{\tau}^{(s)}(k_{1:n})f_{\omega}^{(s)}(\ell_{1:m})f_{\bm{y}}^{(s)}(k_{1:n},\ell_{1:m})},$$
where $f_{\tau}^{(s)}(k_{1:n})\triangleq\prod_{i=1}^{n}\widehat{\tau}_{k_{i}}^{(s)}$, $f_{\omega}^{(s)}(\ell_{1:m})\triangleq\prod_{j=1}^{m}\widehat{\omega}_{\ell_{j}}^{(s)}$, and $f_{\bm{y}}^{(s)}(k_{1:n},\ell_{1:m})\triangleq\prod_{i=1}^{n}\prod_{j=1}^{m}\phi\left(y_{i}(\bm{x}_{j})|\widehat{\mu}_{k_{i}\ell_{j}}^{(s)},(\widehat{\sigma}_{k_{i}\ell_{j}}^{2})^{(s)}\right)$. The estimates of the parameters in the $(s+1)$-th iteration of the EM algorithm are given by
$$\widehat{\tau}_{k}^{(s+1)}=\frac{\sum_{i=1}^{n}\widehat{z}_{i,k}^{(s)}}{n},\ \widehat{\omega}_{\ell}^{(s+1)}=\frac{\sum_{j=1}^{m}\widehat{v}_{j,\ell}^{(s)}}{m},$$
$$\widehat{\mu}_{k\ell}^{(s+1)}=\frac{\sum_{i=1}^{n}\sum_{j=1}^{m}\widehat{z}_{i,k}^{(s)}\widehat{v}_{j,\ell}^{(s)}y_{i}(\bm{x}_{j})}{\sum_{i=1}^{n}\sum_{j=1}^{m}\widehat{z}_{i,k}^{(s)}\widehat{v}_{j,\ell}^{(s)}},$$
and
$$(\widehat{\sigma}_{k\ell}^{2})^{(s+1)}=\frac{\sum_{i=1}^{n}\sum_{j=1}^{m}\widehat{z}_{i,k}^{(s)}\widehat{v}_{j,\ell}^{(s)}\left(y_{i}(\bm{x}_{j})-\widehat{\mu}_{k\ell}^{(s+1)}\right)^{2}}{\sum_{i=1}^{n}\sum_{j=1}^{m}\widehat{z}_{i,k}^{(s)}\widehat{v}_{j,\ell}^{(s)}}.$$
\end{proposition}

\begin{proposition}\label{proposition_con}
Suppose each design-context pair is sampled infinitely often as $t$ goes to infinity. Then
$$\underset{t\to+\infty}{\lim} \left[\frac{C_{ij,k_{i}\ell_{j}}^{(t,s)}}{C_{ij,k\ell}^{(t,s)}}-\frac{\phi\left(y_{i}(\bm{x}_{j})|\widehat{\mu}_{k_{i}\ell_{j}}^{(s)},(\widehat{\sigma}_{k_{i}\ell_{j}}^{2})^{(s)}\right)}{\phi\left(y_{i}(\bm{x}_{j})|\widehat{\mu}_{k\ell}^{(s)},(\widehat{\sigma}_{k\ell}^{2})^{(s)}\right)}\right]=0,\ a.s.$$
where
$$C_{ij,k\ell}^{(t,s)}\triangleq\left(\frac{1}{2\pi\sigma_{i}^{2}(\bm{x}_{j})}\right)^{\frac{t_{ij}}{2}}\sqrt{\frac{(\sigma_{ij,k\ell}^{2})^{(t,s)}}{(\widehat{\sigma}_{k\ell}^{2})^{(t,s)}}}\exp\left\{\frac{1}{2}\left[\frac{(\mu_{ij,k\ell}^{(t,s)})^{2}}{(\sigma_{ij,k\ell}^{2})^{(t,s)}}-\frac{\sum_{h=1}^{t_{ij}}Y_{i,h}^{2}(\bm{x}_{j})}{\sigma_{i}^{2}(\bm{x}_{j})}-\frac{(\widehat{\mu}_{k\ell}^{(t,s)})^{2}}{(\widehat{\sigma}_{k\ell}^{2})^{(t,s)}}\right]\right\}.$$ 
\end{proposition}
\proof{Proof of Proposition~\ref{proposition_con}}
We have
\begin{eqnarray*}
&&\frac{(\mu_{ij,k_{i}\ell_{j}}^{(t,s)})^{2}}{(\sigma_{ij,k_{i}\ell_{j}}^{2})^{(t,s)}}\\
&=&(\sigma_{ij,k_{i}\ell_{j}}^{2})^{(t,s)}\left[\frac{\sum_{h=1}^{t_{ij}}Y_{i,h}(\bm{x}_{j})}{\sigma_{i}^{2}(\bm{x}_{j})}+\frac{\widehat{\mu}_{k_{i}\ell_{j}}^{(t,s)}}{(\widehat{\sigma}_{k_{i}\ell_{j}}^{2})^{(t,s)}}\right]^{2}\\
&=&\frac{t_{ij}\left[\frac{1}{t_{ij}}\sum_{h=1}^{t_{ij}}Y_{i,h}(\bm{x}_{j})\right]^{2}}{\sigma_{i}^{2}(\bm{x}_{j})}-\frac{\left[\frac{1}{t_{ij}}\sum_{h=1}^{t_{ij}}Y_{i,h}(\bm{x}_{j})\right]^{2}}{(\widehat{\sigma}_{k_{i}\ell_{j}}^{2})^{(t,s)}}+2\frac{\left[\frac{1}{t_{ij}}\sum_{h=1}^{t_{ij}}Y_{i,h}(\bm{x}_{j})\right]\widehat{\mu}_{k_{i}\ell_{j}}^{(t,s)}}{(\widehat{\sigma}_{k_{i}\ell_{j}}^{2})^{(t,s)}}+o_{a.s.}(1),
\end{eqnarray*}
where $o_{a.s.}(1)$ means a term that goes to zero as $t$ goes to infinity a.s., by observing $\underset{t\to+\infty}{\lim}t_{ij}(\sigma_{ij,k_{i}\ell_{j}}^{2})^{(t,s)}=\sigma_{i}^{2}(\bm{x}_{j})$, a.s. Therefore, the conclusion of the proposition can be obtained straightforwardly by observing $\underset{t\to+\infty}{\lim}\frac{1}{t_{ij}}\sum_{h=1}^{t_{ij}}Y_{i,h}(\bm{x}_{j})=y_{i}(\bm{x}_{j})$, a.s., and canceling the terms independent of $k_{i}$ and $\ell_{j}$ in $C_{ij,k_{i}\ell_{j}}^{(t,s)}$.  
\Halmos
\endproof

The above proposition implies $C_{ij,k_{i}\ell_{j}}^{(t,s)}$ corresponds to $\phi\left(y_{i}(\bm{x}_{j})|\widehat{\mu}_{k_{i}\ell_{j}}^{(s)},(\widehat{\sigma}_{k_{i}\ell_{j}}^{2})^{(s)}\right)$. Further, $f_{Y}^{(t,s)}(\mathcal{E}_{t}|k_{1:n},\ell_{1:m})$
corresponds to $f_{\bm{y}}^{(s)}(k_{1:n},\ell_{1:m})$ in the classic results. Therefore, Corollary~\ref{corollary_con} is a direct conclusion from Proposition~\ref{proposition_con}.
\begin{corollary}\label{corollary_con}
Suppose each design-context pair is sampled infinitely often as $t$ goes to infinity. Then
$$\underset{t\to+\infty}{\lim} \left[\widehat{z}_{i,k}^{(t,s)}-\widehat{z}_{i,k}^{(s)}\right]=0\ \text{and}\ \underset{t\to+\infty}{\lim} \left[\widehat{v}_{j,\ell}^{(t,s)}-\widehat{v}_{j,\ell}^{(s)}\right]=0,\ a.s.$$
\end{corollary}

\section{Proposition~\ref{proposition_zero}}
\begin{proposition}\label{proposition_zero}
As $t_{ij}$ goes to infinity, $C_{ij,k_{i}\ell_{j}}^{(t,s)}$ approaches zero when $\sigma_{i}^{2}(x_{j})\geq1/(2e\pi)$; otherwise $C_{ij,k_{i}\ell_{j}}^{(t,s)}$ goes to infinity.
\end{proposition}
\proof{Proof of Proposition~\ref{proposition_zero}}
Notice that we have $\underset{t\to+\infty}{\lim}\widehat{\mu}_{k_{i}\ell_{j}}^{(t,s)}=\mu_{k_{i}\ell_{j}}^{(s)}\ \text{and}\ \underset{t\to+\infty}{\lim}(\widehat{\sigma}_{k_{i}\ell_{j}}^{2})^{(t,s)}=(\sigma_{k_{i}\ell_{j}}^{2})^{(s)}>0$. Therefore, when $2e\pi\sigma_{i}^{2}(x_{j})\geq1$,
\begin{eqnarray*}
&&\underset{t_{ij}\to+\infty}{\lim}C_{ij,k_{i}\ell_{j}}^{(t,s)}\\
&=&\underset{t_{ij}\to+\infty}{\lim}\left(\frac{1}{2\pi\sigma_{i}^{2}(\bm{x}_{j})}\right)^{\frac{t_{ij}}{2}}\sqrt{\frac{(\sigma_{ij,k_{i}\ell_{j}}^{2})^{(t,s)}}{(\widehat{\sigma}_{k_{i}\ell_{j}}^{2})^{(t,s)}}}\exp\left\{\frac{1}{2}\left[\frac{(\mu_{ij,k_{i}\ell_{j}}^{(t,s)})^{2}}{(\sigma_{ij,k_{i}\ell_{j}}^{2})^{(t,s)}}-\frac{\sum_{h=1}^{t_{ij}}Y_{i,h}^{2}(\bm{x}_{j})}{\sigma_{i}^{2}(\bm{x}_{j})}-\frac{(\widehat{\mu}_{k_{i}\ell_{j}}^{(t,s)})^{2}}{(\widehat{\sigma}_{k_{i}\ell_{j}}^{2})^{(t,s)}}\right]\right\}\\
&=&\underset{t_{ij}\to+\infty}{\lim}\left(\frac{1}{2\pi\sigma_{i}^{2}(\bm{x}_{j})}\right)^{\frac{t_{ij}}{2}}\sqrt{\frac{\sigma_{i}^{2}(x_{j})}{t_{ij}(\sigma_{k_{i}\ell_{j}}^{2})^{(s)}}}\exp\left\{\frac{1}{2}\left[\frac{t_{ij}\left(E[Y_{i,h}(\bm{x}_{j})]\right)^{2}}{\sigma_{i}^{2}(\bm{x}_{j})}-\frac{t_{ij}E[Y_{i,h}^{2}(\bm{x}_{j})]}{\sigma_{i}^{2}(\bm{x}_{j})}-o(t_{ij})\right]\right\}\\
&=&\underset{t_{ij}\to+\infty}{\lim}\left(\frac{1}{2\pi\sigma_{i}^{2}(\bm{x}_{j})}\right)^{\frac{t_{ij}}{2}}\sqrt{\frac{\sigma_{i}^{2}(x_{j})}{t_{ij}(\sigma_{k_{i}\ell_{j}}^{2})^{(s)}}}\exp\left\{\frac{1}{2}\left[-t_{ij}-o(t_{ij})\right]\right\}\\
&=&0,
\end{eqnarray*}
where the second equation follows from the law of large numbers and the last equation is a result of $2e\pi\sigma_{i}^{2}(x_{j})\geq1$. Similarly, when $2e\pi\sigma_{i}^{2}(x_{j})<1$, $\underset{t_{ij}\to+\infty}{\lim}C_{ij,k_{i}\ell_{j}}^{(t,s)}=+\infty$ since the exponential function grows faster than any power function.
\Halmos
\endproof

\section{Algorithm~\ref{Algo_trans}}
\begin{minipage}{14.8cm}
\begin{algorithm}[H]
\caption{Equivalent transformation for $\widehat{z}_{i,k}^{(t,s)}$ and $\widehat{v}_{j,\ell}^{(t,s)}$}\label{Algo_trans}
Log transformation: For each $k_{1:n}\in\mathcal{K}, \ell_{1:m}\in\mathcal{L}$, $$\log f^{(t,s)}(k_{1:n},\ell_{1:m})\triangleq \sum_{i=1}^{n}\log \widehat{\tau}_{k_{i}}^{(t,s)}+\sum_{j=1}^{m}\log \widehat{\omega}_{\ell_{j}}^{(t,s)}+\sum_{i=1}^{n}\sum_{j=1}^{m}\log C_{ij,k_{i}\ell_{j}}^{(t,s)}.$$\\
Magnification: $$g^{(t,s)}(k_{1:n},\ell_{1:m})\triangleq\log f^{(t,s)}(k_{1:n},\ell_{1:m})-\underset{k_{1:n}\in\mathcal{K},\ell_{1:m}\in\mathcal{L}}{\max}\log f^{(t,s)}(k_{1:n},\ell_{1:m}).$$\\
Equivalent transformation: $$\widehat{z}_{i,k}^{(t,s)}=\frac{\sum_{k_{1:n}\in\mathcal{K},k_{i}=k}\sum_{\ell_{1:m}\in\mathcal{L}}\exp(g^{(t,s)}(k_{1:n},\ell_{1:m}))}{\sum_{k_{1:n}\in\mathcal{K}}\sum_{\ell_{1:m}\in\mathcal{L}}\exp(g^{(t,s)}(k_{1:n},\ell_{1:m}))},$$ $$\widehat{v}_{j,\ell}^{(t,s)}=\frac{\sum_{k_{1:n}\in\mathcal{K}}\sum_{\ell_{1:m}\in\mathcal{L},\ell_{j}=\ell}\exp(g^{(t,s)}(k_{1:n},\ell_{1:m}))}{\sum_{k_{1:n}\in\mathcal{K}}\sum_{\ell_{1:m}\in\mathcal{L}}\exp(g^{(t,s)}(k_{1:n},\ell_{1:m}))}.$$\\
\Return $\widehat{z}_{i,k}^{(t,s)}$ and $\widehat{v}_{j,\ell}^{(t,s)}$.
\end{algorithm}
\end{minipage}

\section{Proposition~\ref{proposition_trans}}
\begin{proposition}\label{proposition_trans}
The denominator satisfies $$1\leq \sum_{k_{1:n}\in\mathcal{K}}\sum_{\ell_{1:m}\in\mathcal{L}}\exp(g^{(t,s)}(k_{1:n},\ell_{1:m}))\leq K^{n}\times L^{m}~.$$ 
\end{proposition}
\proof{Proof of Proposition~\ref{proposition_trans}}
Note that each $g^{(t,s)}(k_{1:n},\ell_{1:m})$ is not greater than zero, and there must exist a clustering situation $(k_{1:n},\ell_{1:m})$ such that $g^{(t,s)}(k_{1:n},\ell_{1:m})=0$. Therefore, the denominator $\sum_{k_{1:n}\in\mathcal{K}}\sum_{\ell_{1:m}\in\mathcal{L}}\exp(g^{(t,s)}(k_{1:n},\ell_{1:m}))$ is not less than one and is bounded by $K^{n}\times L^{m}$.\Halmos
\endproof

\section{Exponential decay of approximation error}
\begin{proposition}\label{pro_exp}
The error between the integral of multivariate standard normal density over a region $\Omega$ and that over the maximal tangent inner ball in $\Omega$ decreases to zero at least in an order of $O(t^{(n-1)/2}e^{-t/2})$ as $t\to+\infty$.
\end{proposition}
\proof{Proof of Proposition~\ref{pro_exp}}
We have
\begin{eqnarray*}
&&\int\cdots\int_{\Omega} \frac{1}{\sqrt{(2\pi)^{n}}}\exp(-\frac{1}{2}\sum_{i=1}^{n}z_{i}^{2})dz_{1}\ldots dz_{n}-\int\cdots\int_{\sum_{i=1}^{n}z_{i}^{2}\leq R^{2}} \frac{1}{\sqrt{(2\pi)^{n}}}\exp(-\frac{1}{2}\sum_{i=1}^{n}z_{i}^{2})dz_{1}\ldots dz_{n}\\
&\leq&1-\int\cdots\int_{\sum_{i=1}^{n}z_{i}^{2}\leq R^{2}} \frac{1}{\sqrt{(2\pi)^{n}}}\exp(-\frac{1}{2}\sum_{i=1}^{n}z_{i}^{2})dz_{1}\ldots dz_{n}\\
&=&\int_{R}^{+\infty}\int_{0}^{\pi}\cdots\int_{0}^{\pi}\int_{0}^{2\pi} \frac{1}{\sqrt{(2\pi)^{n}}}\exp(-\frac{1}{2}r^{2})r^{n-1}(\sin\varphi_{1})^{n-2}\cdots\sin\varphi_{n-2}\ dr d\varphi_{1}\ldots d\varphi_{n-1}\\
&=&C(n)\int_{R}^{+\infty}\exp(-\frac{1}{2}r^{2})r^{n-1}\ dr
\end{eqnarray*}
where $R$ is the radius of the inner ball, and $$C(n)=\frac{1}{\sqrt{(2\pi)^{n}}}\int_{0}^{\pi}\cdots\int_{0}^{\pi}\int_{0}^{2\pi} (\sin\varphi_{1})^{n-2}\cdots\sin\varphi_{n-2}\ d\varphi_{1}\ldots d\varphi_{n-1}$$
is a constant depending only on $n$.

Given by integration by parts,
\begin{gather*}
\int_{R}^{+\infty}\exp(-\frac{1}{2}r^{2})r^{n-1}\ dr=\frac{1}{n}\left(\left[\exp(-\frac{1}{2}r^{2})r^{n}\right]_{R}^{+\infty}-\int_{R}^{+\infty}\exp(-\frac{1}{2}r^{2})(-r)r^{n-1}\ dr\right).
\end{gather*}
In addition, we have $$\int_{R}^{+\infty}\exp(-\frac{1}{2}r^{2})r^{n}\ dr\geq R\int_{R}^{+\infty}\exp(-\frac{1}{2}r^{2})r^{n-1}\ dr.$$
Therefore,
\begin{eqnarray*}
\exp(-\frac{1}{2}R^{2})R^{n}+n\int_{R}^{+\infty}\exp(-\frac{1}{2}r^{2})r^{n-1}\ dr\geq R\int_{R}^{+\infty}\exp(-\frac{1}{2}r^{2})r^{n-1}\ dr
\end{eqnarray*}
which concludes $$\int_{R}^{+\infty}\exp(-\frac{1}{2}r^{2})r^{n-1}\ dr\leq\frac{1}{R-n}\exp(-\frac{1}{2}R^{2})R^{n}.$$

Note that $$R^2=\underset{i\neq 1}{\min}\frac{\left(\mu_{\langle1\rangle_{jt}j,k_{\langle1\rangle_{jt}}\ell_{j}}^{(t)}-\mu_{\langle i\rangle_{jt}j,k_{\langle i\rangle_{jt}}\ell_{j}}^{(t)}\right)^{2}}{\left(\sigma_{\langle1\rangle_{jt}j,k_{\langle1\rangle_{jt}}\ell_{j}}^{2}\right)^{(t)}+\left(\sigma_{\langle i\rangle_{jt}j,k_{\langle i\rangle_{jt}}\ell_{j}}^{2}\right)^{(t)}}=O(t),$$ which concludes the proposition.\Halmos
\endproof

\section{Approximations of $\widehat{z}_{i,k}^{(t;(r,q)_{E})}$ and $\widehat{v}_{j,\ell}^{(t;(r,q)_{E})}$}
In order to reduce the complexity of
$$\widehat{z}_{i,k}^{(t;(r,q)_{E})}=\frac{\sum_{k_{1:n}\in\mathcal{K},k_{i}=k}\sum_{\ell_{1:m}\in\mathcal{L}}f_{\tau}^{(t;(r,q)_{E})}(k_{1:n})f_{\omega}^{(t;(r,q)_{E})}(\ell_{1:m})f_{Y}^{(t;(r,q)_{E})}(\mathcal{E}_{t}|k_{1:n},\ell_{1:m})}{\sum_{k_{1:n}\in\mathcal{K}}\sum_{\ell_{1:m}\in\mathcal{L}}f_{\tau}^{(t;(r,q)_{E})}(k_{1:n})f_{\omega}^{(t;(r,q)_{E})}(\ell_{1:m})f_{Y}^{(t;(r,q)_{E})}(\mathcal{E}_{t}|k_{1:n},\ell_{1:m})},$$
we focus on the change in the numerator of $\widehat{z}_{i,k}^{(t;(r,q)_{E})}$ and make the following approximation by considering the denominators of $\widehat{z}_{i,k}^{(t;(r,q)_{E})}$ and $\widehat{z}_{i,k}^{(t)}$ as a same constant:
$$\frac{\widehat{z}_{i,k}^{(t;(r,q)_{E})}}{\widehat{z}_{i,k}^{(t)}}\approx\frac{\sum_{k_{1:n}\in\mathcal{K},k_{i}=k}\sum_{\ell_{1:m}\in\mathcal{L}}f_{\tau}^{(t;(r,q)_{E})}(k_{1:n})f_{\omega}^{(t;(r,q)_{E})}(\ell_{1:m})f_{Y}^{(t;(r,q)_{E})}(\mathcal{E}_{t}|k_{1:n},\ell_{1:m})}{\sum_{k_{1:n}\in\mathcal{K},k_{i}=k}\sum_{\ell_{1:m}\in\mathcal{L}}f_{\tau}^{(t)}(k_{1:n})f_{\omega}^{(t)}(\ell_{1:m})f_{Y}^{(t)}(\mathcal{E}_{t}|k_{1:n},\ell_{1:m})}.$$

Note that $k_{i}^{*}=\underset{k=1,\ldots,K}{\arg\max}\ \widehat{z}_{i,k}^{(t)}$ and $\ell_{j}^{*}=\underset{\ell=1,\ldots,L}{\arg\max}\ \widehat{v}_{j,\ell}^{(t)}$ denote the indexes of the optimal posterior probabilities of clustering for each design and context. If there indeed exists obvious clustering phenomenon in designs and contexts, then we would tend to have $\phi\left(y_{i}(\bm{x}_{j})|\widehat{\mu}_{k_{i}^{*}\ell_{j}^{*}}^{(t,s)},(\widehat{\sigma}_{k_{i}^{*}\ell_{j}^{*}}^{2})^{(t,s)}\right)\gg\phi\left(y_{i}(\bm{x}_{j})|\widehat{\mu}_{k\ell}^{(t,s)},(\widehat{\sigma}_{k\ell}^{2})^{(t,s)}\right),\ \forall (k,\ell)\neq(k_{i}^{*},\ell_{j}^{*})$. According to the results of Proposition~\ref{proposition_con}, we have $C_{ij,k_{i}^{*}\ell_{j}^{*}}^{(t,s)}\gg C_{ij,k\ell}^{(t,s)}$ when $t$ is relatively large. Therefore, by ignoring the events with non-optimal posterior probabilities of clustering, we have
\begin{eqnarray*}
&&\sum_{k_{1:n}\in\mathcal{K},k_{i}=k}\sum_{\ell_{1:m}\in\mathcal{L}}f_{\tau}^{(t)}(k_{1:n})f_{\omega}^{(t)}(\ell_{1:m})f_{Y}^{(t)}(\mathcal{E}_{t}|k_{1:n},\ell_{1:m})\\
&\approx&f_{\tau}^{(t)}([k_{1:(i-1)}^{*},k,k_{(i+1):n}^{*}])f_{\omega}^{(t)}(\ell_{1:m}^{*})f_{Y}^{(t)}(\mathcal{E}_{t}|[k_{1:(i-1)}^{*},k,k_{(i+1):n}^{*}],\ell_{1:m}^{*})\\
&=&\widehat{\tau}_{k}^{(t)}\left(\prod_{i'=1,i'\neq i}^{n}\widehat{\tau}_{k_{i'}^{*}}^{(t)}\right)\left(\prod_{j=1}^{m}\widehat{\omega}_{\ell_{j}^{*}}^{(t)}\right)\left(\prod_{j=1}^{m}C_{ij,k\ell_{j}^{*}}^{(t)}\right)\left(\prod_{i'=1,i'\neq i}^{n}\prod_{j=1}^{m}C_{i'j,k_{i'}^{*}\ell_{j}^{*}}^{(t)}\right),
\end{eqnarray*}
and then $$\frac{\widehat{z}_{i,k}^{(t;(r,q)_{E})}}{\widehat{z}_{i,k}^{(t)}}\approx\frac{\left(\prod_{j=1}^{m}C_{ij,k\ell_{j}^{*}}^{(t;(r,q)_{E})}\right)\left(\prod_{i'=1,i'\neq i}^{n}\prod_{j=1}^{m}C_{i'j,k_{i'}^{*}\ell_{j}^{*}}^{(t;(r,q)_{E})}\right)}{\left(\prod_{j=1}^{m}C_{ij,k\ell_{j}^{*}}^{(t)}\right)\left(\prod_{i'=1,i'\neq i}^{n}\prod_{j=1}^{m}C_{i'j,k_{i'}^{*}\ell_{j}^{*}}^{(t)}\right)}.$$

The following proposition indicates sampling a design-context pair $(r,q)$ would become less likely to change the likelihood of observations for other design-context pair $(i,j)$ as the number of samples grows large.
\begin{proposition}\label{appro_1}
Suppose each design-context pair is sampled infinitely often as t goes to infinity. For each $(i,j)\neq(r,q)$, $$\underset{t\to+\infty}{\lim} \dfrac{C_{ij,k\ell}^{(t;(r,q)_{E})}}{C_{ij,k\ell}^{(t)}}=1,\ a.s.$$
\end{proposition}
\proof{Proof of Proposition~\ref{appro_1}}
Note that for $(i,j)\neq(r,q)$,
\begin{eqnarray*}
\dfrac{C_{ij,k\ell}^{(t;(r,q)_{E})}}{C_{ij,k\ell}^{(t)}}=\sqrt{\frac{(\sigma_{ij,k\ell}^{2})^{\left(t;(r,q)_{E}\right)}}{(\widehat{\sigma}_{k\ell}^{2})^{\left(t;(r,q)_{E}\right)}}\frac{(\widehat{\sigma}_{k\ell}^{2})^{(t)}}{(\sigma_{ij,k\ell}^{2})^{(t)}}}\exp\left\{\frac{1}{2}\left[\left(\frac{(\mu_{ij,k\ell}^{(t)})^{2}}{(\sigma_{ij,k\ell}^{2})^{\left(t;(r,q)_{E}\right)}}-\frac{(\mu_{ij,k\ell}^{(t)})^{2}}{(\sigma_{ij,k\ell}^{2})^{\left(t\right)}}\right)-\left(\frac{(\widehat{\mu}_{k\ell}^{(t)})^{2}}{(\widehat{\sigma}_{k\ell}^{2})^{\left(t;(r,q)_{E}\right)}}-\frac{(\widehat{\mu}_{k\ell}^{(t)})^{2}}{(\widehat{\sigma}_{k\ell}^{2})^{\left(t\right)}}\right)\right]\right\}.
\end{eqnarray*}
In addition,
\begin{gather*}
\underset{t\to+\infty}{\lim} \frac{1}{(\sigma_{ij,k\ell}^{2})^{\left(t;(r,q)_{E}\right)}}-\frac{1}{(\sigma_{ij,k\ell}^{2})^{\left(t\right)}}=\underset{t\to+\infty}{\lim} \frac{1}{(\widehat{\sigma}_{k\ell}^{2})^{\left(t;(r,q)_{E}\right)}}-\frac{1}{(\widehat{\sigma}_{k\ell}^{2})^{\left(t\right)}}=0\ a.s.
\end{gather*}
where the first equation is due to $(i,j)\neq(r,q)$ and the second equation is due to that both $(\widehat{\sigma}_{k\ell}^{2})^{\left(t;(r,q)_{E}\right)}$ and $(\widehat{\sigma}_{k\ell}^{2})^{\left(t\right)}$ converge to a same positive value. Therefore, $\underset{t\to+\infty}{\lim}\dfrac{C_{ij,k\ell}^{(t;(r,q)_{E})}}{C_{ij,k\ell}^{(t)}}=1\ a.s.$
\Halmos
\endproof

The proposition suggests us to make the following approximation:
$$\frac{\widehat{z}_{r,k}^{(t;(r,q)_{E})}}{\widehat{z}_{r,k}^{(t)}}\approx\frac{C_{rq,k\ell_{q}^{*}}^{(t;(r,q)_{E})}}{C_{rq,k\ell_{q}^{*}}^{(t)}}$$
and
$$\frac{\widehat{z}_{i,k}^{(t;(r,q)_{E})}}{\widehat{z}_{i,k}^{(t)}}\approx\frac{C_{rq,k_{r}^{*}\ell_{q}^{*}}^{(t;(r,q)_{E})}}{C_{rq,k_{r}^{*}\ell_{q}^{*}}^{(t)}},\ i\neq r.$$
Further, the following proposition provides the asymptotic results for the change rate of the likelihood of observations for design-context pair $(r,q)$ after allocating one more sample to this design-context pair.
\begin{proposition}\label{appro_2}
Suppose each design-context pair is sampled infinitely often as $t$ goes to infinity. Then
\begin{eqnarray*}
\underset{t\to+\infty}{\lim} \left[\frac{C_{rq,k\ell_{q}^{*}}^{(t;(r,q)_{E})}}{C_{rq,k\ell_{q}^{*}}^{(t)}}-\sqrt{\frac{1}{2\pi\sigma_{r}^{2}(\bm{x}_{q})}}\exp\left\{\frac{1}{2}\left[-\frac{(\mu_{rq,k\ell_{q}^{*}}^{(t)}-\widehat{\mu}_{k\ell_{q}^{*}}^{(t)})^{2}(\sigma_{rq,k\ell_{q}^{*}}^{4})^{(t)}}{\sigma_{r}^{2}(\bm{x}_{q})(\widehat{\sigma}_{k\ell_{q}^{*}}^{4})^{(t)}}\right]\right\}\right]=0\ a.s.
\end{eqnarray*}
\end{proposition}
\proof{Proof of Proposition~\ref{appro_2}}
Note that
\begin{eqnarray*}
&\dfrac{C_{rq,k\ell_{q}^{*}}^{(t;(r,q)_{E})}}{C_{rq,k\ell_{q}^{*}}^{(t)}}=&\sqrt{\frac{1}{2\pi\sigma_{r}^{2}(\bm{x}_{q})}}\sqrt{\frac{(\sigma_{rq,k\ell_{q}^{*}}^{2})^{\left(t;(r,q)_{E}\right)}}{(\widehat{\sigma}_{k\ell_{q}^{*}}^{2})^{\left(t;(r,q)_{E}\right)}}\frac{(\widehat{\sigma}_{k\ell_{q}^{*}}^{2})^{(t)}}{(\sigma_{rq,k\ell_{q}^{*}}^{2})^{(t)}}}\exp\left\{\frac{1}{2}\left[\dfrac{\left[\dfrac{\sum_{h=1}^{t_{rq}}Y_{r,h}(\bm{x}_{q})+\mu_{rq,k\ell_{q}^{*}}^{(t)}}{\sigma_{r}^{2}(\bm{x}_{q})}+\dfrac{\widehat{\mu}_{k\ell_{q}^{*}}^{(t)}}{\left(\widehat{\sigma}_{k\ell_{q}^{*}}^{2}\right)^{(t)}}\right]^{2}}{\dfrac{t_{rq}+1}{\sigma_{r}^{2}(\bm{x}_{q})}+\dfrac{1}{\left(\widehat{\sigma}_{k\ell_{q}^{*}}^{2}\right)^{(t)}}}\right.\right.\\
&&\left.\left.-\dfrac{\left[\dfrac{\sum_{h=1}^{t_{rq}}Y_{r,h}(\bm{x}_{q})}{\sigma_{r}^{2}(\bm{x}_{q})}+\dfrac{\widehat{\mu}_{k\ell_{q}^{*}}^{(t)}}{(\widehat{\sigma}_{k\ell_{q}^{*}}^{2})^{(t)}}\right]^{2}}{\dfrac{t_{rq}}{\sigma_{r}^{2}(\bm{x}_{q})}+\dfrac{1}{\left(\widehat{\sigma}_{k\ell_{q}^{*}}^{2}\right)^{(t)}}}-\dfrac{\left(\mu_{rq,k\ell_{q}^{*}}^{(t)}\right)^{2}}{\sigma_{r}^{2}(\bm{x}_{q})}\right]\right\}.
\end{eqnarray*}
In addition, $\underset{t\to+\infty}{\lim}\dfrac{(\sigma_{rq,k\ell_{q}^{*}}^{2})^{\left(t;(r,q)_{E}\right)}}{(\widehat{\sigma}_{k\ell_{q}^{*}}^{2})^{\left(t;(r,q)_{E}\right)}}\dfrac{(\widehat{\sigma}_{k\ell_{q}^{*}}^{2})^{(t)}}{(\sigma_{rq,k\ell_{q}^{*}}^{2})^{(t)}}=1\ a.s.$ and $\underset{t\to+\infty}{\lim}\dfrac{1}{t_{rq}}\sum_{h=1}^{t_{rq}}Y_{r,h}(\bm{x}_{q})-\mu_{rq,k\ell_{q}^{*}}^{(t)}=0\ a.s.$,
which means that the effect of prior information on the posterior estimate vanishes as the number of samples goes to infinity. Therefore, the conclusion of the proposition can be obtained.  
\Halmos
\endproof

Summarizing the discussions above, we have
\begin{eqnarray*}
\frac{\widehat{z}_{r,k}^{(t;(r,q)_{E})}}{\widehat{z}_{r,k}^{(t)}}\approx\sqrt{\frac{1}{2\pi\sigma_{r}^{2}(\bm{x}_{q})}}\exp\left\{\frac{1}{2}\left[-\frac{(\mu_{rq,k\ell_{q}^{*}}^{(t)}-\widehat{\mu}_{k\ell_{q}^{*}}^{(t)})^{2}(\sigma_{rq,k\ell_{q}^{*}}^{4})^{(t)}}{\sigma_{r}^{2}(\bm{x}_{q})(\widehat{\sigma}_{k\ell_{q}^{*}}^{4})^{(t)}}\right]\right\}
\end{eqnarray*} 
and similarly
\begin{eqnarray*}
\frac{\widehat{v}_{q,\ell}^{(t;(r,q)_{E})}}{\widehat{v}_{q,\ell}^{(t)}}\approx\sqrt{\frac{1}{2\pi\sigma_{r}^{2}(\bm{x}_{q})}}\exp\left\{\frac{1}{2}\left[-\frac{(\mu_{rq,k_{r}^{*}\ell}^{(t)}-\widehat{\mu}_{k_{r}^{*}\ell}^{(t)})^{2}(\sigma_{rq,k_{r}^{*}\ell}^{4})^{(t)}}{\sigma_{r}^{2}(\bm{x}_{q})(\widehat{\sigma}_{k_{r}^{*}\ell}^{4})^{(t)}}\right]\right\}.
\end{eqnarray*} 
Notice that the formulas on the right-hand side of the approximations can be calculated efficiently.
Due to constraints $\sum_{k=1}^{K}\widehat{z}_{i,k}^{(t;(r,q)_{E})}=1$ and $\sum_{\ell=1}^{L}\widehat{v}_{j,\ell}^{(t;(r,q)_{E})}=1$, we obtain efficient approximations for the one-step-ahead looking posterior probabilities of clustering $\widehat{z}_{i,k}^{(t;(r,q)_{E})}$ and $\widehat{v}_{j,\ell}^{(t;(r,q)_{E})}$ by normalization as shown in~\eqref{onestep_z} and \eqref{onestep_v}.

\section{Algorithm~\ref{Algo_DSCO}}
\begin{algorithm}
\small
\caption{Dynamic Sampling Policy for Context-Dependent Optimization}\label{Algo_DSCO}
\textbf{Inputs}: Number of designs $n$, number of contexts $m$, total sampling budget $T$, common initial sample size $n_{0}$.\\
Generate $n_{0}$ samples from each design $i$ in each context $j$, and $t\leftarrow n\times m\times n_{0}$.\\
Determine number of design clusters $K$ and number of context clusters $L$ by BIC \eqref{eq.BIC} and EM algorithm \eqref{eq.E-step}-\eqref{eq.M-step}.\\
Initialize parameter estimates $\widehat{\tau}_{k}^{(t)}$, $\widehat{\omega}_{\ell}^{(t)}$, $\widehat{\mu}_{k\ell}^{(t)}$, $(\widehat{\sigma}_{k\ell}^{2})^{(t)}$, and posterior estimates $\widehat{z}_{i,k}^{(t)}$, $\widehat{v}_{j,\ell}^{(t)}$, $\widehat{\mu}_{ij,k\ell}^{(t)}$, $(\widehat{\sigma}_{ij,k\ell}^{2})^{(t)}$ by Theorem~\ref{thm_EM} \eqref{thm1.1}-\eqref{thm1.7}.\\
\While{$t\leq T$}
{
	\eIf{there exists a ($r,q$) such that $V(\mathcal{E}_{t};(r,q))>V(\mathcal{E}_{t})$}
	{
	  Choose $A_{t+1}(\mathcal{E}_{t})=\left\{(r^{*},q^{*})\Bigg|V(\mathcal{E}_{t};(r^{*},q^{*}))=\underset{(r,q)}{\max}\ V(\mathcal{E}_{t};(r,q))\right\}$, where $(\sigma^{2}_{ij,k_{i}\ell_{j}})^{(t,(r,q)_{E})}$, $(\widehat{\sigma}^{2}_{k\ell})^{(t,(r,q)_{E})}$, $\widehat{z}_{i,k}^{(t,(r,q)_{E})}$, and $\widehat{v}_{j,\ell}^{(t,(r,q)_{E})}$ are calculated by \eqref{onestep_sigma}-\eqref{onestep_v} and $V(\mathcal{E}_{t};(r,q))$ is calculated by \eqref{final_VFA}.\\
	  Take a sample $Y_{r^{*},t_{r^{*}q^{*}}+1}(\bm{x}_{q^{*}})$ from design $r^{*}$ in context $q^{*}$.\\
	  Update parameter estimates $\widehat{\tau}_{k}^{(t)}$, $\widehat{\omega}_{\ell}^{(t)}$, $\widehat{\mu}_{k\ell}^{(t)}$, $(\widehat{\sigma}_{k\ell}^{2})^{(t)}$, and posterior estimates $\widehat{z}_{i,k}^{(t)}$, $\widehat{v}_{j,\ell}^{(t)}$, $\widehat{\mu}_{ij,k\ell}^{(t)}$, $(\widehat{\sigma}_{ij,k\ell}^{2})^{(t)}$ by Theorem~\ref{thm_EM} \eqref{thm1.1}-\eqref{thm1.7}.\\
	  Set $t\leftarrow t+1$.
	}
	{
      \While{$W(\mathcal{E}_{t};(r^{*},q^{*}))>V(\mathcal{E}_{t})$ and $t\leq T$}
      {
        Choose $A_{t+1}(\mathcal{E}_{t})=\left\{(r^{*},q^{*})\Bigg|W(\mathcal{E}_{t};(r^{*},q^{*}))=\underset{(r,q)}{\max}\ W(\mathcal{E}_{t};(r,q))\right\}$, where $(\sigma^{2}_{ij,k_{i}\ell_{j}})^{(t,(r,q)_{E})}$ and $(\widehat{\sigma}^{2}_{k\ell})^{(t,(r,q)_{E})}$ are calculated by \eqref{onestep_sigma}-\eqref{onestep_clustersigma} and $W(\mathcal{E}_{t};(r,q))$ is calculated by \eqref{VFA_W}.\\
        Take a sample $Y_{r^{*},t_{r^{*}q^{*}}+1}(\bm{x}_{q^{*}})$ from design $r^{*}$ in context $q^{*}$.\\
	    Update parameter estimates $\widehat{\tau}_{k}^{(t)}$, $\widehat{\omega}_{\ell}^{(t)}$, $\widehat{\mu}_{k\ell}^{(t)}$, $(\widehat{\sigma}_{k\ell}^{2})^{(t)}$, and posterior estimates $\widehat{z}_{i,k}^{(t)}$, $\widehat{v}_{j,\ell}^{(t)}$, $\widehat{\mu}_{ij,k\ell}^{(t)}$, $(\widehat{\sigma}_{ij,k\ell}^{2})^{(t)}$ by Theorem~\ref{thm_EM} \eqref{thm1.1}-\eqref{thm1.7}.\\
	    Set $t\leftarrow t+1$.
      }
	}
}
\Return Select $\arg\max_{i}\ \widehat{\mu}_{ij,k_{i}^{*}\ell_{j}^{*}}^{(T)}$ as the bests.
\end{algorithm}

\section{Proof of Theorem~\ref{thm_consistency}}
\proof{Proof of Theorem~\ref{thm_consistency}}
We only need to prove that each $y_{i}(\bm{x}_{j})$ will be sampled infinitely often a.s. following DSCO policy, and the consistency will follow by the law of large numbers. Suppose $y_{i}(\bm{x}_{j})$ is only sampled finitely often and $y_{r}(\bm{x}_{q})$ is sampled infinitely often. Therefore, there exists a finite number $N_{0}$ such that $y_{i}(\bm{x}_{j})$ will stop receiving replications after the sampling number $t$ exceeds $N_{0}$. Thus we have
$$\lim_{t\to+\infty}(\sigma_{ij,k\ell}^{2})^{(t)}>0,\ \lim_{t\to+\infty}(\sigma_{rq,k\ell}^{2})^{(t)}=0,\ \forall k=1,\ldots,K,\ \ell=1,\ldots,L.$$

By noticing that
$$\lim_{t\to+\infty}\left[(\sigma_{rq,k\ell}^{2})^{(t)}-(\sigma_{rq,k\ell}^{2})^{(t;(r,q)_{E})}\right]=0,$$
$$\lim_{t\to+\infty}\left[(\sigma_{k\ell}^{2})^{(t)}-(\sigma_{k\ell}^{2})^{(t;(r,q)_{E})}\right]=0,$$
and
$$\lim_{t\to+\infty}\left[\widehat{z}_{r,k}^{(t)}-\widehat{z}_{r,k}^{(t;(r,q)_{E})}\right]=0,$$
$$\lim_{t\to+\infty}\left[\widehat{v}_{q,\ell}^{(t)}-\widehat{v}_{q,\ell}^{(t;(r,q)_{E})}\right]=0,$$
we have
$$\lim_{t\to+\infty}\left[V(\mathcal{E}_{t};(r,q))-V(\mathcal{E}_{t})\right]=0\ \ a.s.$$
If there exists a design-context pair ($i,j$) whose performance $y_{i}(\bm{x}_{j})$ is only sampled finitely often such that
$$\lim_{t\to+\infty}\left[V(\mathcal{E}_{t};(i,j))-V(\mathcal{E}_{t})\right]>0\ \ a.s.,$$
then it contradicts with the sampling rule in equation (\ref{normal policy}) that the design-context pair with the largest $V(\mathcal{E}_{t};(i,j))$ is sampled. 

Therefore, $$\lim_{t\to+\infty}\left[V(\mathcal{E}_{t};(i,j))-V(\mathcal{E}_{t})\right]\leq0$$ holds for all design-context pairs, and thus sample allocation is determined by the sampling rule in equation~(\ref{special policy}). By noticing that
$$\lim_{t\to+\infty}\left[(\sigma_{ij,k\ell}^{2})^{(t)}-(\sigma_{ij,k\ell}^{2})^{(t;(i,j)_{E})}\right]>0,$$
and
$$\lim_{t\to+\infty}\left[(\sigma_{k_{i}^{*},\ell_{j}^{*}}^{2})^{(t)}-(\sigma_{k_{i}^{*},\ell_{j}^{*}}^{2})^{(t;(i,j)_{E})}\right]>0,$$
there must exist a design-context pair ($i,j$) which is only sampled finitely often such that
$$\lim_{t\to+\infty}\left[W(\mathcal{E}_{t};(i,j))-W(\mathcal{E}_{t})\right]>0\ \ a.s.,$$
which contradicts with the sampling rule in equation (\ref{special policy}) that the design-context pair with the largest $W(\mathcal{E}_{t};(i,j))$ is sampled. Therefore, the proposed DSCO policy must be consistent.

By the law of large numbers, $\underset{t\to+\infty}{\lim}\mu_{ij,k\ell}^{(t)}=y_{i}(\bm{x}_{j})$. For simplicity of analysis, we can replace $\mu_{ij,k\ell}^{(t)}$ and $(\sigma_{ij,k\ell}^{2})^{(t)}$ with $y_{i}(\bm{x}_{j})$ and $\sigma_{i}^{2}(\bm{x}_{j})/t_{ij}$ in $V(\mathcal{E}_{t};(r,q))$. Then when $t\to+\infty$, both $V(\mathcal{E}_{t};(r,q))$ and $W(\mathcal{E}_{t};(r,q))$ are reduced to $$\underset{j=1,\ldots,m}{\min}\ \underset{i\neq 1}{\min}\frac{\left(y_{\langle1\rangle_{j}}(\bm{x}_{j})-y_{\langle i\rangle_{j}}(\bm{x}_{j})\right)^{2}}{\dfrac{\sigma_{\langle1\rangle_{j}}^{2}(\bm{x}_{j})}{t_{\langle1\rangle_{j} j}+\mathbbm{1}\{(\langle1\rangle_{j},j)=(r,q)\}}+\dfrac{\sigma_{\langle i\rangle_{j}}^{2}(\bm{x}_{j})}{t_{\langle i\rangle_{j} j}+\mathbbm{1}\{(\langle i\rangle_{j},j)=(r,q)\}}}.$$ 

Let $r_{ij}^{(t)}\triangleq t_{ij}/t$, $i=1,\ldots,n,\ j=1,\ldots,m$. By the Bolzano-Weierstrass theorem~\citep{rudin1964principles}, there exists a subsequence of $\{r_{ij}^{(t)}\}$ converging to $\{r_{ij}\}$ such that $\sum_{i=1}^{n}\sum_{j=1}^{m}r_{ij}=1,\ r_{ij}\geq0$. Without loss of generality, we can assume $\{r_{ij}^{(t)}\}$ converges to $\{r_{ij}\}$; otherwise, the following argument is made over a subsequence. We claim $r_{ij}>0,\ i=1,\ldots,n,\ j=1,\ldots,m$; otherwise, there exist $r_{\langle i\rangle_{j} j}=0$ and $r_{\langle i'\rangle_{j'} j'}>0$. Notice that
\begin{eqnarray*}
&&\underset{t\to+\infty}{\lim} \left[\frac{\left(y_{\langle1\rangle_{j}}(\bm{x}_{j})-y_{\langle i\rangle_{j}}(\bm{x}_{j})\right)^{2}}{\sigma_{\langle1\rangle_{j}}^{2}(\bm{x}_{j})/t_{\langle1\rangle_{j} j}+\sigma_{\langle i\rangle_{j}}^{2}(\bm{x}_{j})/(t_{\langle i\rangle_{j} j}+1)}-\frac{\left(y_{\langle1\rangle_{j}}(\bm{x}_{j})-y_{\langle i\rangle_{j}}(\bm{x}_{j})\right)^{2}}{\sigma_{\langle1\rangle_{j}}^{2}(\bm{x}_{j})/t_{\langle1\rangle_{j} j}+\sigma_{\langle i\rangle_{j}}^{2}(\bm{x_{j}})/t_{\langle i\rangle_{j} j}}\right]\\
&=&\underset{t\to+\infty}{\lim} t\left[\frac{\left(y_{\langle1\rangle_{j}}(\bm{x}_{j})-y_{\langle i\rangle_{j}}(\bm{x}_{j})\right)^{2}}{\sigma_{\langle1\rangle_{j}}^{2}(\bm{x}_{j})/r_{\langle1\rangle_{j} j}^{(t)}+\sigma_{\langle i\rangle_{j}}^{2}(\bm{x}_{j})/(r_{\langle i\rangle_{j} j}^{(t)}+1/t)}-\frac{\left(y_{\langle1\rangle_{j}}(\bm{x}_{j})-y_{\langle i\rangle_{j}}(\bm{x}_{j})\right)^{2}}{\sigma_{\langle1\rangle_{j}}^{2}(\bm{x}_{j})/r_{\langle1\rangle_{j} j}^{(t)}+\sigma_{\langle i\rangle_{j}}^{2}(\bm{x}_{j})/r_{\langle i\rangle_{j} j}^{(t)}}\right]\\
&=&\underset{t\to+\infty}{\lim} \left(\frac{\sigma_{\langle i\rangle_{j}}(\bm{x}_{j})}{r_{\langle i\rangle_{j} j}^{(t)}}\right)^{2}\frac{\left(y_{\langle1\rangle_{j}}(\bm{x}_{j})-y_{\langle i\rangle_{j}}(\bm{x}_{j})\right)^{2}}{\left(\sigma_{\langle1\rangle_{j}}^{2}(\bm{x}_{j})/r_{\langle1\rangle_{j} j}^{(t)}+\sigma_{\langle i\rangle_{j}}^{2}(\bm{x}_{j})/r_{\langle i\rangle_{j} j}^{(t)}\right)^{2}}
\end{eqnarray*}
and
\begin{eqnarray*}
&&\underset{t\to+\infty}{\lim} \left[\frac{\left(y_{\langle1\rangle_{j}}(\bm{x}_{j})-y_{\langle i\rangle_{j}}(\bm{x}_{j})\right)^{2}}{\sigma_{\langle1\rangle_{j}}^{2}(\bm{x}_{j})/(t_{\langle1\rangle_{j} j}+1)+\sigma_{\langle i\rangle_{j}}^{2}(\bm{x}_{j})/t_{\langle i\rangle_{j} j}}-\frac{\left(y_{\langle1\rangle_{j}}(\bm{x}_{j})-y_{\langle i\rangle_{j}}(\bm{x}_{j})\right)^{2}}{\sigma_{\langle1\rangle_{j}}^{2}(\bm{x}_{j})/t_{\langle1\rangle_{j} j}+\sigma_{\langle i\rangle_{j}}^{2}(\bm{x}_{j})/t_{\langle i\rangle_{j} j}}\right]\\
&=&\underset{t\to+\infty}{\lim} t\left[\frac{\left(y_{\langle1\rangle_{j}}(\bm{x}_{j})-y_{\langle i\rangle_{j}}(\bm{x}_{j})\right)^{2}}{\sigma_{\langle1\rangle_{j}}^{2}(\bm{x}_{j})/(r_{\langle1\rangle_{j} j}^{(t)}+1/t)+\sigma_{\langle i\rangle_{j}}^{2}(\bm{x}_{j})/r_{\langle i\rangle_{j} j}^{(t)}}-\frac{\left(y_{\langle1\rangle_{j}}(\bm{x}_{j})-y_{\langle i\rangle_{j}}(\bm{x}_{j})\right)^{2}}{\sigma_{\langle1\rangle_{j}}^{2}(\bm{x}_{j})/r_{\langle1\rangle_{j} j}^{(t)}+\sigma_{\langle i\rangle_{j}}^{2}(\bm{x}_{j})/r_{\langle i\rangle_{j} j}^{(t)}}\right]\\
&=&\underset{t\to+\infty}{\lim} \left(\frac{\sigma_{\langle 1\rangle_{j}}(\bm{x}_{j})}{r_{\langle 1\rangle_{j} j}^{(t)}}\right)^{2}\frac{\left(y_{\langle1\rangle_{j}}(\bm{x}_{j})-y_{\langle i\rangle_{j}}(\bm{x}_{j})\right)^{2}}{\left(\sigma_{\langle1\rangle_{j}}^{2}(\bm{x}_{j})/r_{\langle1\rangle_{j} j}^{(t)}+\sigma_{\langle i\rangle_{j}}^{2}(\bm{x}_{j})/r_{\langle i\rangle_{j} j}^{(t)}\right)^{2}}.
\end{eqnarray*}
If $r_{\langle 1\rangle_{j} j}=0$ and $r_{\langle i\rangle_{j} j}>0$, then $$\underset{t\to+\infty}{\lim} \left(\frac{\sigma_{\langle 1\rangle_{j}}(\bm{x}_{j})}{r_{\langle 1\rangle_{j} j}^{(t)}}\right)^{2}\frac{\left(y_{\langle1\rangle_{j}}(\bm{x}_{j})-y_{\langle i\rangle_{j}}(\bm{x}_{j})\right)^{2}}{\left(\sigma_{\langle1\rangle_{j}}^{2}(\bm{x}_{j})/r_{\langle1\rangle_{j} j}^{(t)}+\sigma_{\langle i\rangle_{j}}^{2}(\bm{x}_{j})/r_{\langle i\rangle_{j} j}^{(t)}\right)^{2}}>0$$ and $$\underset{t\to+\infty}{\lim} \left(\frac{\sigma_{\langle i\rangle_{j}}(\bm{x}_{j})}{r_{\langle i\rangle_{j} j}^{(t)}}\right)^{2}\frac{\left(y_{\langle1\rangle_{j}}(\bm{x}_{j})-y_{\langle i\rangle_{j}}(\bm{x}_{j})\right)^{2}}{\left(\sigma_{\langle1\rangle_{j}}^{2}(\bm{x}_{j})/r_{\langle1\rangle_{j} j}^{(t)}+\sigma_{\langle i\rangle_{j}}^{2}(\bm{x}_{j})/r_{\langle i\rangle_{j} j}^{(t)}\right)^{2}}=0$$
which contradicts with the sampling rules in equations (\ref{normal policy}) and (\ref{special policy}). If $r_{\langle 1\rangle_{j} j}>0$ and $r_{\langle i\rangle_{j} j}=0$, then $$\underset{t\to+\infty}{\lim} \left(\frac{\sigma_{\langle 1\rangle_{j}}(\bm{x}_{j})}{r_{\langle 1\rangle_{j} j}^{(t)}}\right)^{2}\frac{\left(y_{\langle1\rangle_{j}}(\bm{x}_{j})-y_{\langle i\rangle_{j}}(\bm{x}_{j})\right)^{2}}{\left(\sigma_{\langle1\rangle_{j}}^{2}(\bm{x}_{j})/r_{\langle1\rangle_{j} j}^{(t)}+\sigma_{\langle i\rangle_{j}}^{2}(\bm{x}_{j})/r_{\langle i\rangle_{j} j}^{(t)}\right)^{2}}=0$$ and $$\underset{t\to+\infty}{\lim} \left(\frac{\sigma_{\langle i\rangle_{j}}(\bm{x}_{j})}{r_{\langle i\rangle_{j} j}^{(t)}}\right)^{2}\frac{\left(y_{\langle1\rangle_{j}}(\bm{x}_{j})-y_{\langle i\rangle_{j}}(\bm{x}_{j})\right)^{2}}{\left(\sigma_{\langle1\rangle_{j}}^{2}(\bm{x}_{j})/r_{\langle1\rangle_{j} j}^{(t)}+\sigma_{\langle i\rangle_{j}}^{2}(\bm{x}_{j})/r_{\langle i\rangle_{j} j}^{(t)}\right)^{2}}>0$$
which contradicts with the sampling rules in equations (\ref{normal policy}) and (\ref{special policy}). If $r_{\langle i\rangle_{j} j}=0$, $i=1,\ldots,n$ and $r_{\langle i\rangle_{j'} j'}>0$, $i=1,\ldots,n$, then by replacing $\mu_{ij,k\ell}^{(t)}$ and $(\sigma_{ij,k\ell}^{2})^{(t)}$ with $y_{i}(\bm{x}_{j})$ and $\sigma_{i}^{2}(\bm{x}_{j})/t_{ij}$ when $t\to+\infty$, we have
$$\underset{t\to+\infty}{\lim}\sum_{k_{1:n}\in\mathcal{K},\ell_{j}\in\mathcal{L}}p_{z}(k_{1:n},\mathcal{E}_{t})p_{v}(\ell_{j},\mathcal{E}_{t})\text{APCS}(k_{1:n},\ell_{j},\mathcal{E}_{t})-\underset{i\neq 1}{\min}\frac{\left(y_{\langle1\rangle_{j}}(\bm{x}_{j})-y_{\langle i\rangle_{j}}(\bm{x}_{j})\right)^{2}}{\sigma_{\langle1\rangle_{j}}^{2}(\bm{x}_{j})/t_{\langle1\rangle_{j} j}+\sigma_{\langle i\rangle_{j}}^{2}(\bm{x}_{j})/t_{\langle i\rangle_{j} j}}=0,$$
$$\underset{i\neq 1}{\min}\frac{\left(y_{\langle1\rangle_{j}}(\bm{x}_{j})-y_{\langle i\rangle_{j}}(\bm{x}_{j})\right)^{2}}{\sigma_{\langle1\rangle_{j}}^{2}(\bm{x}_{j})/t_{\langle1\rangle_{j} j}+\sigma_{\langle i\rangle_{j}}^{2}(\bm{x}_{j})/t_{\langle i\rangle_{j} j}}=o(t),$$
and
$$\underset{t\to+\infty}{\lim}\sum_{k_{1:n}\in\mathcal{K},\ell_{j'}\in\mathcal{L}}p_{z}(k_{1:n},\mathcal{E}_{t})p_{v}(\ell_{j'},\mathcal{E}_{t})\text{APCS}(k_{1:n},\ell_{j'},\mathcal{E}_{t})-\underset{i\neq 1}{\min}\frac{\left(y_{\langle1\rangle_{j'}}(\bm{x}_{j'})-y_{\langle i\rangle_{j'}}(\bm{x}_{j'})\right)^{2}}{\sigma_{\langle1\rangle_{j'}}^{2}(\bm{x}_{j'})/t_{\langle1\rangle_{j'} j'}+\sigma_{\langle i\rangle_{j'}}^{2}(\bm{x}_{j'})/t_{\langle i\rangle_{j'} j'}}=0,$$
$$\underset{i\neq 1}{\min}\frac{\left(y_{\langle1\rangle_{j'}}(\bm{x}_{j'})-y_{\langle i\rangle_{j'}}(\bm{x}_{j'})\right)^{2}}{\sigma_{\langle1\rangle_{j'}}^{2}(\bm{x}_{j'})/t_{\langle1\rangle_{j'} j'}+\sigma_{\langle i\rangle_{j'}}^{2}(\bm{x}_{j'})/t_{\langle i\rangle_{j'} j'}}=O(t),$$
which contradict with the definition of $\text{PCS}_{\text{W}}$. Therefore, $r_{ij}>0,\ i=1,\ldots,n,\ j=1,\ldots,m$.

Let $G_{ij}(r_{\langle1\rangle_{j} j},r_{\langle i\rangle_{j} j})\triangleq \dfrac{\left(y_{\langle1\rangle_{j}}(\bm{x}_{j})-y_{\langle i\rangle_{j}}(\bm{x}_{j})\right)^{2}}{\sigma_{\langle1\rangle_{j}}^{2}(\bm{x}_{j})/r_{\langle1\rangle_{j} j}+\sigma_{\langle i\rangle_{j}}^{2}(\bm{x}_{j})/r_{\langle i\rangle_{j} j}},\ i=2,\ldots,n,\ j=1,\ldots,m$. If $\{r_{ij}\}$ does not satisfy equation (\ref{ratio_2}), there exist $i\neq i',\ i,i'=2,\ldots,n$ such that $$G_{ij}(r_{\langle1\rangle_{j} j},r_{\langle i\rangle_{j} j})>G_{i'j}(r_{\langle1\rangle_{j} j},r_{\langle i'\rangle_{j} j}).$$ If the inequality above holds, there exists $T_{0}>0$ such that $\forall t>T_{0}$, $$G_{ij}(r_{\langle1\rangle_{j} j}^{(t)},r_{\langle i\rangle_{j} j}^{(t)})>G_{i'j}(r_{\langle1\rangle_{j} j}^{(t)},r_{\langle i'\rangle_{j} j}^{(t)}),$$ due to continuity of $G_{ij}$ on $(0,1)\times(0,1)$. By the sampling rules in equations (\ref{normal policy}) and (\ref{special policy}), $y_{\langle i'\rangle_{j}}(\bm{x}_{j})$ will be sampled and $y_{\langle i\rangle_{j}}(\bm{x}_{j})$ will stop receiving replications before the inequality above reverses. This contradicts $\{r_{ij}^{(t)}\}$ converging to $\{r_{ij}\}$, so equation (\ref{ratio_2}) must hold. If $\{r_{ij}\}$ does not satisfy equation (\ref{ratio_3}), there exist $i,i'=2,\ldots,n$ and $j\neq j',\ j,j'=1,\ldots,m$ such that $$G_{ij}(r_{\langle1\rangle_{j} j},r_{\langle i\rangle_{j} j})>G_{i'j'}(r_{\langle1\rangle_{j'} j'},r_{\langle i'\rangle_{j'} j'}).$$ If the inequality above holds, there exists $T_{0}>0$ such that $\forall t>T_{0}$, $$G_{ij}(r_{\langle1\rangle_{j} j}^{(t)},r_{\langle i\rangle_{j} j}^{(t)})>G_{i'j'}(r_{\langle1\rangle_{j'} j'}^{(t)},r_{\langle i'\rangle_{j'} j'}^{(t)}),$$ due to continuity of $G_{ij}$ on $(0,1)\times(0,1)$. By the definition of $\text{PCS}_{\text{W}}$, context $j'$ will be sampled and context $j$ will stop receiving replications before the inequality above reverses. This contradicts $\{r_{ij}^{(t)}\}$ converging to $\{r_{ij}\}$, so equation (\ref{ratio_3}) must hold. 

By the implicit function theorem~\citep{rudin1964principles}, equations (\ref{ratio_2}), (\ref{ratio_3}), and $\sum_{i=1}^{n}\sum_{j=1}^{m}r_{ij}=1$ determine implicit functions $r_{\langle i\rangle_{j} j}(x)\Big|_{x=(r_{\langle 1\rangle_{1} 1},\ldots,r_{\langle 1\rangle_{m} m})},\ i=2,\ldots,n,\ j=1,\ldots,m,$ because
\begin{gather*}
\text{det}(\Sigma)=\prod_{i=2}^{n}\prod_{j=1}^{m}\zeta_{ij}\left(\sum_{i=2}^{n}\sum_{j=1}^{m}\zeta_{ij}^{-1}\right)>0,
\end{gather*}
where 
\begin{gather*}
\zeta_{ij}\triangleq\frac{\partial G_{ij}(r_{\langle1\rangle_{j} j},x)}{\partial x}\Bigg|_{x=r_{\langle i\rangle_{j} j}},\ i=2,\ldots,n,\ j=1,\ldots,m,
\end{gather*}
\begin{gather*}
\Sigma\triangleq\left(\begin{matrix} \zeta_{2,1} & -\zeta_{3,1} & \cdots & 0 & 0 & \cdots & 0 & 0 \\ 0 & \zeta_{3,1} & \cdots & 0 & 0 & \cdots & 0 & 0 \\ \vdots & \vdots & \cdots & \vdots & \vdots & \cdots & \vdots & \vdots \\ 0 & 0 & \cdots & \zeta_{n,1} & -\zeta_{2,2} & \cdots & 0 & 0 \\ 0 & 0 & \cdots & 0 & \zeta_{2,2} & \cdots & 0 & 0 \\\vdots & \vdots & \cdots & \vdots & \vdots & \cdots & \vdots & \vdots\\ 0 & 0 & \cdots & 0 & 0 & \cdots & \zeta_{n-1,m} & -\zeta_{n,m} \\
1 & 1 & \cdots & 1 & 1 & \cdots & 1 & 1 \end{matrix}\right),
\end{gather*}
and $\Sigma R=-\Upsilon$, where
\begin{gather*}
R\triangleq\left(\begin{matrix} \frac{\partial r_{\langle2\rangle_{1} 1}(x)}{\partial x_1} & \cdots & \frac{\partial r_{\langle2\rangle_{1} 1}(x)}{\partial x_m}\\ \frac{r_{\langle3\rangle_{1} 1}(x)}{\partial x_1} & \cdots & \frac{\partial r_{\langle3\rangle_{1} 1}(x)}{\partial x_m}\\ \vdots & \cdots & \vdots \\ \frac{r_{\langle n\rangle_{1} 1}(x)}{\partial x_1} & \cdots & \frac{\partial r_{\langle n\rangle_{1} 1}(x)}{\partial x_m}\\ \frac{r_{\langle 2\rangle_{2} 2}(x)}{\partial x_1} & \cdots & \frac{\partial r_{\langle2\rangle_{2} 2}(x)}{\partial x_m}\\ \vdots & \cdots & \vdots \\ \frac{r_{\langle n\rangle_{m} m}(x)}{\partial x_1} & \cdots & \frac{\partial r_{\langle n\rangle_{m} m}(x)}{\partial x_m}\end{matrix}\right)_{x=(r_{\langle 1\rangle_{1} 1},\ldots,r_{\langle 1\rangle_{m} m})},
\end{gather*}
\begin{gather*}
\Upsilon\triangleq\left(\begin{matrix} \frac{\partial G_{2,1}(x_{1},r_{\langle2\rangle_{1} 1})-G_{3,1}(x_{1},r_{\langle3\rangle_{1} 1})}{\partial x_1} & \cdots & 0\\
\frac{\partial G_{3,1}(x_{1},r_{\langle3\rangle_{1} 1})-G_{4,1}(x_{1},r_{\langle4\rangle_{1} 1})}{\partial x_1} & \cdots & 0\\
\vdots & \cdots & \vdots \\ 
\frac{\partial G_{n,1}(x_{1},r_{\langle n\rangle_{1} 1})}{\partial x_1} & \cdots & 0\\
0 & \cdots & 0\\
\vdots & \cdots & \vdots \\ 
0 & \cdots & \frac{\partial G_{n-1,m}(x_{m},r_{\langle n-1\rangle_{m} m})-G_{n,m}(x_{m},r_{\langle n\rangle_{m} m})}{\partial x_m}\\
1 & \cdots & 1
\end{matrix}\right)_{x=(r_{\langle 1\rangle_{1} 1},\ldots,r_{\langle 1\rangle_{m} m})}.
\end{gather*}
In addition,
\begin{gather*}
\frac{\partial G_{i,j}(x_{j},r_{\langle i\rangle_{j} j})}{\partial x_{j}}\Bigg|_{x=(r_{\langle 1\rangle_{1} 1},\ldots,r_{\langle 1\rangle_{m} m})}+\zeta_{ij}\frac{\partial r_{\langle i\rangle_{j} j}(x)}{\partial x_{j}}\Bigg|_{x=(r_{\langle 1\rangle_{1} 1},\ldots,r_{\langle 1\rangle_{m} m})}=0,\ i=2,\ldots,n,\ j=1,\ldots,m; 
\end{gather*}
otherwise, there exist $i'\neq 1$ and $j'$ such that the equality above does not hold, say
\begin{gather*}
\frac{\partial G_{i',j'}(x_{j'},r_{\langle i'\rangle_{j'} j'})}{\partial x_{j'}}\Bigg|_{x=(r_{\langle 1\rangle_{1} 1},\ldots,r_{\langle 1\rangle_{m} m})}+\zeta_{i'j'}\frac{\partial r_{\langle i'\rangle_{j'} j'}(x)}{\partial x_{j'}}\Bigg|_{x=(r_{\langle 1\rangle_{1} 1},\ldots,r_{\langle 1\rangle_{m} m})}>0.
\end{gather*}
Following the sampling rules in equations (\ref{normal policy}) and (\ref{special policy}), $y_{\langle 1\rangle_{j'}}(\bm{x}_{j'})$ will be sampled and $y_{\langle i'\rangle_{j'}}(\bm{x}_{j'})$ will stop receiving replications before the inequality above reverses, which contradicts $\{r_{ij}^{(t)}\}$ converging to $\{r_{ij}\}$. Similarly,
\begin{gather*}
\frac{\partial G_{i,j}(x_{j},r_{\langle i\rangle_{j} j})}{\partial x_{j'}}\Bigg|_{x=(r_{\langle 1\rangle_{1} 1},\ldots,r_{\langle 1\rangle_{m} m})}+\zeta_{ij}\frac{\partial r_{\langle i\rangle_{j} j}(x)}{\partial x_{j'}}\Bigg|_{x=(r_{\langle 1\rangle_{1} 1},\ldots,r_{\langle 1\rangle_{m} m})}=0,\ i=2,\ldots,n,\ j\neq j'; 
\end{gather*}
otherwise, there exist $i\neq 1$ and $j'$ such that the equality above does not hold, say
\begin{eqnarray*}
&&\frac{\partial G_{i,j}(x_{j},r_{\langle i\rangle_{j} j})}{\partial x_{j'}}\Bigg|_{x=(r_{\langle 1\rangle_{1} 1},\ldots,r_{\langle 1\rangle_{m} m})}+\zeta_{ij}\frac{\partial r_{\langle i\rangle_{j} j}(x)}{\partial x_{j'}}\Bigg|_{x=(r_{\langle 1\rangle_{1} 1},\ldots,r_{\langle 1\rangle_{m} m})}\\
&>&\frac{\partial G_{i,j}(x_{j},r_{\langle i\rangle_{j} j})}{\partial x_{j}}\Bigg|_{x=(r_{\langle 1\rangle_{1} 1},\ldots,r_{\langle 1\rangle_{m} m})}+\zeta_{ij}\frac{\partial r_{\langle i\rangle_{j} j}(x)}{\partial x_{j}}\Bigg|_{x=(r_{\langle 1\rangle_{1} 1},\ldots,r_{\langle 1\rangle_{m} m})}=0
\end{eqnarray*}
without loss of generality. Following the sampling rules in equations (\ref{normal policy}) and (\ref{special policy}), context $j'$ will be sampled and context $j$ will stop receiving replications before the inequality above reverses, which contradicts $\{r_{ij}^{(t)}\}$ converging to $\{r_{ij}\}$. Further,
note that
\begin{gather*}
\frac{\partial G_{i,j}(x_{j},r_{\langle i\rangle_{j} j})}{\partial x_{j'}}\Bigg|_{x=(r_{\langle 1\rangle_{1} 1},\ldots,r_{\langle 1\rangle_{m} m})}=0,\ i=2,\ldots,n,\ j\neq j',
\end{gather*}
which is due to $x_{j}$ and $x_{j'}$ are independent, then we have
\begin{gather*}
\zeta_{ij}\frac{\partial r_{\langle i\rangle_{j} j}(x)}{\partial x_{j'}}\Bigg|_{x=(r_{\langle 1\rangle_{1} 1},\ldots,r_{\langle 1\rangle_{m} m})}=0,\ i=2,\ldots,n,\ j\neq j'.
\end{gather*}
Then, $HR=-G$, where
\begin{gather*}
G\triangleq\left(\begin{matrix} \frac{\partial G_{2,1}(x_{1},r_{\langle 2\rangle_{1} 1})}{\partial x_{1}} & \cdots & 0\\
\frac{\partial G_{3,1}(x_{1},r_{\langle 3\rangle_{1} 1})}{\partial x_{1}} & \cdots & 0\\
\vdots & \cdots & \vdots\\
\frac{\partial G_{n,1}(x_{1},r_{\langle n\rangle_{1} 1})}{\partial x_{1}} & \cdots & 0\\
0 & \cdots & 0\\
0 & \cdots & \frac{\partial G_{n,m}(x_{m},r_{\langle n\rangle_{m} m})}{\partial x_{m}}
\end{matrix}\right)_{x=(r_{\langle 1\rangle_{1} 1},\ldots,r_{\langle 1\rangle_{m} m})}
\end{gather*}
and
\begin{gather*}
H\triangleq\left(\begin{matrix}\zeta_{2,1} & 0 & \cdots & 0\\
0 & \zeta_{3,1} & \cdots & 0\\
\vdots & \vdots & \ddots & \vdots\\
0 & 0 & \cdots & \zeta_{n,m}
\end{matrix}\right).
\end{gather*}
Summarizing the above, we have
\begin{gather*}
\Upsilon=\Sigma H^{-1}G
\end{gather*}
which leads to
\begin{gather*}
\sum_{i=2}^{n} \frac{\partial G_{i,j}(x_{j},r_{\langle i\rangle_{j} j})/\partial x_{j}\Big|_{x=(r_{\langle 1\rangle_{1} 1},\ldots,r_{\langle 1\rangle_{m} m})}}{G_{i,j}(r_{\langle 1\rangle_{j} j},x)/\partial x\Big|_{x=r_{\langle i\rangle_{j} j}}}=1, \ j=1,\ldots,m \Leftrightarrow \text{Equation}~(\ref{ratio_1}).
\end{gather*}
Therefore, $\{r_{ij}^{(t)}\}$ converges to $\{r_{ij}^{*}\}$.
\Halmos
\endproof

\section{Performance clustering phenomenon}

\begin{table}[h!]
  \begin{center}
    \caption{Results for design clustering.}\label{design}
    \begin{tabular}{c|c|c|c|c}
      \hline  
            & \# of design & Type of drugs & Mean & Standard deviation \\
      \hline
      Design Cluster 1 & 11 & Aspirin & 77.7275 & 16.5825 \\
      \hline
      Design Cluster 2 & 9 & Aspirin & 127.7775 & 13.6925 \\
      \hline
      Design Cluster 3 & 7 & Statin & 8.0858 & 1.3152 \\
      \hline
      Design Cluster 4 & 13 & Statin & 14.1384 & 2.3432 \\
      \hline
    \end{tabular}
  \end{center}
\end{table}
\begin{table}[h!]
  \begin{center}
    \caption{Results for context clustering.}\label{context}
    \begin{tabular}{c|c|c|c|c} 
      \hline  
            & \# of context & Parameter & Mean & Standard deviation \\
      \hline
      \multirow{2}*{Context Cluster 1} & \multirow{2}*{12} & $x_{1}$ & 48.8417 & 3.1575 \\
      ~ & ~ & $x_{2}$ & 122.0083 & 7.2111 \\
      \hline
      \multirow{2}*{Context Cluster 2} & \multirow{2}*{9} & $x_{1}$ & 51.2333 & 3.1535 \\
      ~ & ~ & $x_{2}$ & 138.5667 & 11.3039 \\
      \hline
      \multirow{2}*{Context Cluster 3} & \multirow{2}*{10} & $x_{1}$ & 61.5100 & 3.0277 \\
      ~ & ~ & $x_{2}$ & 122.0100 & 6.0553 \\
      \hline
      \multirow{2}*{Context Cluster 4} & \multirow{2}*{13} & $x_{1}$ & 64.6231 & 4.7000 \\
      ~ & ~ & $x_{2}$ & 132.7000 & 13.7803 \\
      \hline
      \multirow{2}*{Context Cluster 5} & \multirow{2}*{9} & $x_{1}$ & 73.6778 & 3.1623 \\
      ~ & ~ & $x_{2}$ & 127.0111 & 5.4772 \\
      \hline
      \multirow{2}*{Context Cluster 6} & \multirow{2}*{7} & $x_{1}$ & 75.0143 & 2.1602 \\
      ~ & ~ & $x_{2}$ & 143.0143 & 4.3205 \\
      \hline  
    \end{tabular}
  \end{center}
\end{table}

%
%
%




\end{document}